\documentclass[aps,prd,reprint,superscriptaddress,nofootinbib,floatfix,notitlepage]{revtex4-2}

\usepackage{amssymb}
\usepackage{amsmath}
\usepackage[breaklinks,colorlinks=true]{hyperref}
\usepackage{graphicx}
\usepackage{mathtools}
\usepackage{slashed}
\usepackage[caption=false]{subfig}
\usepackage{xcolor}
\usepackage{upgreek}
\usepackage{mathrsfs}

\newcommand{\e}[0]{\hspace{0.1em}\textrm{e}}

\newcommand{\Oo}[0]{\mathcal{O}}

\newcommand{\D}[0]{\mathcal{D}}

\newcommand{\V}[0]{\mathcal{V}_4}
\newcommand{\Det}[0]{\textrm{Det}}
\newcommand{\Tr}[0]{\textrm{Tr}}
\newcommand{\tr}[0]{\textrm{tr}\,}
\newcommand{\Ds}{\slashed{D}}
\newcommand{\symb}{\textrm{Symb}}
\newcommand{\tcdot}[0]{\!\cdot\!}
\newcommand{\tp}[0]{\!+\!}
\newcommand{\tm}[0]{\!-\!}
\newcommand{\Del}{\underset{{}^\smile}{\Updelta}}
\newcommand{\dDel}{{}^{\bullet}\!\Del{}}

\newcommand{\dDeld}{{}^{\bullet}\!\Del\!{}^{\bullet}\!}

\newcommand{\oDel}{{}^{\circ}\!\Del{}}
\newcommand{\Delo}{\Del\!{}^{\circ}\!}
\newcommand{\oDelo}{{}^{\circ}\!\Del\!{}^{\circ}\!}

\newcommand{\dDelo}{{}^{\bullet}\!\Del\!{}^{\circ}\!}
\newcommand{\oDeld}{{}^{\circ}\!\Del\!{}^{\bullet}\!}
\newcommand{\G}{\mathcal{G}}
\newcommand{\Gd}{\dot{\G}}

\newcommand{\GF}{\mathcal{G}_{F}}

\newcommand{\oG}{{}^{\circ}\!\G}
\newcommand{\oGF}{{}^{\circ}\!\GF}
\newcommand{\GFo}{\GF^{\circ}}
\newcommand{\oGFo}{{}^{\circ}\!\GF^{\circ}}

\newcommand{\dXi}{{}^{\bullet}\,\!\Xi}
\newcommand{\Xid}{\Xi\,\!{}^{\bullet}\!}
\newcommand{\dXid}{{}^{\bullet}\,\!\Xi\,\!{}^{\bullet}\!}

\newcommand{\oXi}{{}^{\circ}\,\!\Xi}
\newcommand{\Xio}{{\Xi\,\!{}^{\circ}}\!}
\newcommand{\oXio}{{}^{\circ}\,\!\Xi\,\!{}^{\circ}\!}
\newcommand{\dXio}{{}^{\bullet}\,\!\Xi\,\!{}^{\circ}\!}
\newcommand{\oXid}{{}^{\circ}\,\!\Xi\,\!{}^{\bullet}\!}

\allowdisplaybreaks

\begin{document}

\title{Low-energy multi-photon scattering at tree-level and one-loop order in a homogeneous electromagnetic field}
\author{Ivan Ahumada}
\email{ivan.ahumadahernandez@plymouth.ac.uk}
\affiliation{Centre for Mathematical Sciences, University of Plymouth, Plymouth, PL4 8AA, UK}
\author{Patrick Copinger}
\email{patrick.copinger@plymouth.ac.uk}
\affiliation{Centre for Mathematical Sciences, University of Plymouth, Plymouth, PL4 8AA, UK}
\author{James P. Edwards}
\email{james.p.edwards@plymouth.ac.uk}
\affiliation{Centre for Mathematical Sciences, University of Plymouth, Plymouth, PL4 8AA, UK}

\begin{abstract}
We study low energy photons coupled to scalar and spinor matter in the presence of an arbitrary homogeneous electromagnetic field in a first-quantised (worldline) approach. Utilising a Fock-Schwinger gauge for both the scattering photons and homogeneous background, simple compact expressions are found for both the photon- and background-dressed effective action and propagator in scalar and spinor quantum electrodynamics. The low-energy limit allows identification of the coupling of the scattering photons as one of an effective homogeneous superposition of their field strengths, with amplitudes following from application of a suitable linearisation operator.

To treat the linearisation, several techniques are employed, including a functional expansion based on the proper time formalism and worldline Green functions, linearised vertex operators under a worldline path integral, and a matrix expansion in the field strengths. We find, in particular, that a replacement rule converting scalar amplitudes to spinor amplitudes at one-loop order can, surprisingly, be extended to tree level amplitudes in the low energy limit. Finally, we discuss a novel worldline representation of the momentum space matter propagators, obtaining a suitable worldline Green function for this path integral satisfying homogeneous Dirichlet boundary conditions and momentum space vertex operators representing the scattering photons already in momentum space.
\end{abstract}

\maketitle

%%%%%%%%%%%%%%%%%%%%%%%%%%%%%%%%%%%%%%%%%%%
\section{Introduction}
Quantum electrodynamics (QED) is recognised as a theory producing amongst the most precise predictions of physical phenomena -- well exemplified by the famous agreement of the perturbative calculation of the electron anomalous magnetic moment (known analytically to 4 loop order \cite{Laporta4} and studied numerically at 5 loop order \cite{Volkov5, Kinoshita5}) with experimental observation. The study of photon scattering amplitudes in QED goes back to its earliest days \cite{Euler:1935zz, Heisenberg:1936nmg, Akhieser, Karplus:1950zza, Karplus:1950zz, DeTollis:1964una, DeTollis:1965vna, Scharnhorst:2017wzh}, with revived interest due to the intentions of experimental groups at existing and upcoming high-intensity laser facilities and collider facilities~\cite{LUXE:2023crk, Clarke:2022rbd, PhysRevAccelBeams.22.101301, Ahmadiniaz:2024xob,Ahmadiniaz:2022nrv, Schutze:2024kzu, CMS, ATLAS, Sacla, Corels, ELI, SEL,dEnterria:2013zqi} to measure phenomena such as vacuum polarisation, vacuum birefringence and particle pair creation, either the perturbative Breit-Wheeler effect~\cite{PhysRevLett.127.052302} or the non-perturbative Schwinger effect (already anticipated by \cite{Sauter} and first calculated in \cite{Schwinger, Weisskopf}), which derive from (classically absent) light-by-light interactions (see also \cite{Dittrich:2000zu, Fradkin:1991zq, PhysRevLett.129.061802,Macleod:2023asi,Ahmadiniaz:2022mcy, Karbstein:2021otv, Karbstein:2018omb, Schubert:2024heu, HEINZL2006318}). These processes expose the nonlinear behaviour of the \textit{quantum} theory of electrodynamics \cite{Costantini:1971cj}, and are also relevant in various astrophysical scenarios \cite{10.1093mnrasstw2798, 10.10635.0215939, 2019arXiv190513439K, 2018MNRAS.481...36A, 2018Galax...6...76H, 2018Galax...6...57C, 2017JCAP...10..014S, Harding:2006qn, Heyl}.

Despite the effort that has gone in to calculating on-shell photon scattering amplitudes in QED, a general formula for arbitrary numbers of photons with generic momenta and massive fermions remains unknown even at one-loop order (the current state of the art is up to $N = 6$ photons at one-loop order for massless scalar QED \cite{1loop1, Binoth:2007ca, Badger} and $N = 4$ photons at two-loop level in spinor QED \cite{2loop1, 2loop2, 1loop2}; see also \cite{Badger}). There are also several examples involving both matter and (generally fewer) photons on external legs \cite{Fadin:2023phc, Anastasiou:2002zn, Dave:2024ewl}. One exception to this is the result that to one-loop the massless all-equal and one-unequal helicity amplitudes vanish for $N \neq 4$ scattering photons \cite{Mahlon}; results for the maximal-helicity-violating (e.g. $--+\ldots +$) cases are also known for arbitrary multiplicity (e.g. \cite{1loop1}) in this massless limit. For massive fermions, even the all-equal helicity amplitudes remain uncomputed beyond the 2-loop $N = 4$ point amplitude. Already the mathematical complexity of the Feynman integrals required to compute these amplitudes is high, and new techniques for their computation have had to be developed \cite{Henn:2024ngj, Forner:2024ojj}. The panorama is even more modest for amplitudes in the presence of strong background fields, such as in the context of ``Strong Field QED'' (SFQED). Here explicit, exact results in a plane wave are limited to $N = 2$ photons (``double Compton,'' ``double Breit-Wheeler'' or ``photon trident'' at tree-level \cite{Oleinik1, Oleinik2, FEDOROV2006413,Loetstedt:2009zz,Roshchupkin:2021yhd, PhysRevD.102.116012, DC1, DC2, Trident1, Trident2, Trident3, Trident4, Trident5, DCT,deVos:2023pen} and vacuum polarisation / helicity flip at one-loop order \cite{1975JETP...42..961B,Dittrich:2000zu, BeckerVP, PhysRevD.88.013007, Ilderton:2016qpj, VP, Meuren:2013oya, Dinu:2013gaa, Meuren:2014uia}), with integral representations for the one-loop photon splitting ($N = 3$) process available in \cite{DiPiazza:2007yx}. In a constant field, explicit results are known in fact to order $N = 3$ photons at one-loop order \cite{PhysRevLett.25.1061, PhysRevD.2.2341,Baier,PhysRevLett.77.1691,ADLER1971599,Adler:1996cja}(with generating functions known for the arbitrary multiplicity amplitudes -- see below).	

Significant simplifications occur, however, in the calculations for scattering of \textit{low-energy} photon amplitudes. Physically, this condition corresponds to $\omega_{i} / m \ll 1$, where $k_{i}^{\mu} = (\omega_{i}, \omega_{i} \hat{\mathbf{k}}_{i})$ is the momentum 4-vector of photon $i$, and $\frac{k_{i} \cdot k_{j} }{m^{2}} \ll 1$ for $i \neq j$. Formally this is implemented by projecting onto the contribution to the scattering amplitude that is multi-linear in the $k_{i}$. In this case, it is possible to obtain closed formulas for arbitrary multiplicity amplitudes at one- and two-loop \cite{Itzykson:1980rh, ChrisLow, Edwards:2018vjd} order in vacuum, for all photon helicity assignments (see also \cite{Dunne:2002qf, Baier:2018vso}). More recently, it was possible to obtain a ``Master Formula'' generating function for all-multiplicity amplitudes in a constant electromagnetic background to one-loop order \cite{Ahmadiniaz:2023jwd, MishaLow}. The former results can conveniently be obtained by expanding the effective action in a constant electromagnetic background to fixed order in the background field and following the recipe outlined in \cite{Itzykson:1980rh}. The latter were also obtained by direct calculation using the worldline formalism of quantum field theory \cite{UsRep, ChrisRev}, by expanding the worldline vertex operator (representing the external photon interactions) to linear order in momentum (see also \cite{Ahmadiniaz:2024rvi}). 

The first quantised worldline approach to QFT has proven to be particularly useful for calculating high-multiplicity amplitudes, since it has the advantage of combining multiple Feynman diagrams related by permutation of external legs \cite{Schubert:2023bed, Edwards:2021uif}. This, combined with the fact that virtual momenta are repackaged into global propertime integrals, leads to compact Master Formulae that better encapsulate physical information and gauge structure. Indeed, for the four-photon light-by-light scattering process, the worldline formalism naturally provides a particularly compact tensor decomposition \cite{Ahmadiniaz:2020jgo, Ahmadiniaz:2023vrk}. In recent years significant effort has been invested in developing worldline-specific techniques for calculating the parameter integrals that arise in this formalism without breaking them back into ordered sectors (essentially Feynman parameterisation) \cite{Ahmadiniaz:2024rvi, Edwards:2022dbd}. The worldline formalism has found particular success applied to study processes in constant electromagnetic backgrounds, such as the effective action \cite{Reuter:1996zm, Schmidt:1993rk, Shaisultanov:1995tm, Kors:1998ew, Fliegner:1997ra, Dunne:2002qf, Dunne:2002qg}, propagator \cite{Alexandrou:1998ia, Ahmad_2017, fppaper3} (where it confirmed and broadened the discovery of 1PR contributions in homogeneous backgrounds \cite{Edwards:2017bte, Ahmadiniaz:2017rrk}) and scattering amplitudes \cite{Ahmadiniaz:2019nhk, Dittrich:2000wz,Schubert:2000yt, McKeon:1994hd, Adler:1996cja} -- see also a proper time approach to studying the heat kernel expansion \cite{Franchino-Vinas:2023wea, Fecit:2025kqb}. More recently it has been extended to provide a fresh approach to calculations in plane wave backgrounds \cite{Ilderton:2016qpj, Ahmadiniaz:2019nhk, Edwards:2021vhg, Edwards:2021uif, Schubert:2023bed, Schubert:2023gsl} and shockwaves \cite{Copinger:2024twl, Tarasov:2019rfp}.

In the present manuscript we aim to develop a general approach to obtaining the low-energy limit of photon scattering amplitudes, including in the presence of an arbitrary homogeneous background electromagnetic field. First we will show how the expansion of the effective action with respect to a homogeneous component of a background electromagnetic field contains the full information on the low-energy photon amplitudes in the complement background. We will do this expansion in three ways: a direct expansion of the proper-time representation of effective action which simplifies the integral-structure of the resulting amplitudes, by employing the worldline representation to expand in the scattering photons' momenta (this simplifies the Lorentz structure instead), and by directly performing a power series expansion about the Lorentz matrix form of the homogeneous photon field strength. We also extend this technique to the previously overlooked case of tree-level amplitudes, both in vacuum and in a constant field.

Along the way, we will show how -- in the special case of low-energy scattering photons -- it is possible to derive a novel worldline representation of the \textit{momentum space} tree-level amplitudes that makes use of a coordinate space path integral with Dirichlet boundary conditions. This is, to the best of the authors' knowledge, the first time such a path integral representation has been presented (past efforts involve Fourier transforming the position space Master Formulae), and the resulting worldline theory is of intrinsic interest due to it having a non-diagonal kinetic operator.

This article is arranged as follows: in the following section we examine the low energy limit of $N$-photon scattering for the scalar effective action and propagator, and show how the sum over $N$-photon field strengths are represented by an effective homogeneous field. Further we perform the expansion with respect to the $N$-photon field strengths using the various techniques outlined above for the effective action and propagator in coordinate space. In Sec.~\ref{secSpinor} we do the same, however, for the case of spinors. Then last in Sec.~\ref{secmom} we look at the case of the momentum space propagators. Non-essential calculations and properties of operators and Green functions are included in an appendix.

We work in (3+1)-dimensions throughout with mostly minus signature, and natural units such that $\hbar=c=1$. A matrix form in Lorentz indices is used that can be taken to read $F\coloneqq F^\mu_{\hphantom{\mu}\nu}$; where ambiguous indices are written explicitly. Our Clifford algebra is $\{\gamma_{\mu},\gamma_{\nu}\}=2\eta_{\mu\nu}$ for Dirac equation $(i\slashed{\partial}_x-m)\psi(x)=0$.

%%%%%%%%%%%%%%%%%%%%%%%%%%%%%%%%%%%%%%%%%%%
\section{Master Formulae in Scalar SFQED}
\label{secScalar}
\subsection{Low-energy $\boldsymbol{N}$-photon in a Fock-Schwinger gauge}

To begin, let us first treat the case of a complex scalar particle dressed with $N$ photons in a background field with arbitrary strength, in which the propagator in coordinate space and one-loop effective action respectively are defined as~\cite{ChrisRev}
\begin{align}
	\label{eqDN}
    \mathcal{D}_{N}^{x'x}[A]&=(-ie)^N\int_{0}^{\infty}\!dT\!\int_{x(0)=x}^{x(T)=x'}\hspace{-2em}\mathcal{D}x(\tau)\,\e^{iS}\prod_{i=1}^{N}V [f_i]\,,\\
    \Gamma_{N}[A]&=(-ie)^N\int_{0}^{\infty}\!\frac{dT}{T}\!\oint \mathcal{D}x(\tau) \,\e^{iS}\prod_{i=1}^{N}V[f_i]\,,
    \label{eqGamN}
\end{align}
with worldline action 
\begin{equation}
	S= -\int_{0}^{T} d\tau\,\bigl[m^2 + \tfrac{1}{4}\dot{x}^{2}+eA \cdot \dot{x}\bigr]\,.
\end{equation}
The path integral for the effective action is taken over trajectories satisfying periodic boundary conditions, $x(0)=x(T)$, ($\text{PBC}$). For low-energy photons the vertices take on a simplified form for both the effective action (as is known~\cite{ChrisLow, Edwards:2018vjd}) and the propagator with a collective Fock-Schwinger gauge choice. In anticipation of this simplification we write for the scalar vertex, $V[f]$, as a functional over the field strength. A similar simplified form will indeed also hold for the case of spinors in the spacetime kernel for spinor QED in Sec.~\ref{secSpinor}.

Let us first illustrate the simplification for the effective action. A low-energy approximation is justified when the energy for each $i$ photon is much smaller than the scalar or spinor mass, $m$, i.e., $\omega_i\ll m \;\forall\; i \in [1,N]$~\cite{ChrisLow, Edwards:2018vjd}. This amounts to retaining only terms linear in photon momenta, $k_i$, on-shell. The effective action is gauge-independent, therefore for the vertex we may select the standard expansion of the photon gauge, in which
\begin{equation}
    V[f_i]=  \int^T_0 \!d\tau\, a_i\cdot\dot{x}=\int^T_0\!d\tau\, \varepsilon_i\cdot\dot{x}\,\e^{ik\cdot x}\,. \label{vertex}
\end{equation}
Then by virtue of the low-energy limit under \textit{periodic} boundary conditions one may find no boundary terms contribute to the vertex by halving the above integral, performing integration by parts on one half, and retaining only terms linear in $k_i$.
\begin{equation} \label{lowEvertex}
   V[f_i]=\!\int^T_0\!d\tau \,\frac{1}{2}x(\tau) \cdot f_i \cdot \dot{x}(\tau)+\mathcal{O}((k_i^\mu)^2)\,, 
\end{equation}
where $f^{\mu\nu}_i\coloneqq i[k^{\mu}_i\varepsilon^{\nu}_i-k^{\nu}_i\varepsilon^{\mu}_i]$. We emphasise $f_i$ is real in Minkowski space.

The path integral for the propagator is over trajectories with fixed endpoints. If we repeat the previous calculation using the same gauge choice in~\eqref{vertex} for these \textit{open} lines, we would find a boundary contribution already at \textit{zeroth} order in photon energy, being $\varepsilon \cdot (x' - x)$, which is inconvenient from the perspective of later calculations. However, we will later corroborate our results by comparing with an expansion of the propagator dressed to all orders in a homogeneous background field; since we shall take this background field in Fock-Schwinger gauge, it is convenient to adopt this gauge field choice also for the scattering photons:
\begin{align}
        \nonumber a_{\mu}(x) =& -\sum_{n=0}^{\infty}\frac{1}{n!(n\tp2)}
	(x\tm\hat{x})^\nu(x\tm\hat{x})^{\nu_1}\ldots(x\tm\hat{x})^{\nu_n}\\
        &ik_{\nu_1}\ldots ik_{\nu_n}\,f_{\mu\nu} \,,
\end{align}
where $\hat{x}$ is a chosen reference point. In the low-energy limit the gauge field reads
\begin{equation}\label{FS}
	a_{\mu}(x)=-\frac{1}{2} f_{\mu\nu} (x-\hat{x})^\nu \,,
\end{equation}
where for \textit{any} reference point we again see the vertex operator for the open line is linear in the field strength tensor:
\begin{equation} \label{lowEvertex_line}
	 V[f_i]=\!\int^T_0\!d\tau \,\frac{1}{2}(x(\tau)-\hat{x}) \cdot f_i \cdot \dot{x}(\tau)+\mathcal{O}((k_i^\mu)^2)\,.
\end{equation}
With PBC one can see the term with $\hat{x}$ vanishes making the connection to the loop vertex in~\eqref{lowEvertex}\footnote{Here we are providing the details behind the assertion that there are no boundary terms when the vertex operator is taken from the dynamical photon field in Fock-Schwinger gauge  that was first made in Chapter 3 of \cite{Corradini:2015tik}: since the boundary terms are pure gauge we may as well take the vertex operator with $\hat{x} = 0$ when calculating amplitudes.}. We remark that a Fock-Schwinger gauge choice will affect the functional form of the photon-dressed propagator due to gauge covariance; however, the gauge invariance of scattering amplitudes is, via the Ward identity, protected after LSZ reduction.

For both the scalar propagator and effective action (we will also find similarly for the case of spinors)  the low-energy limit of the $N$-photon vertices can be recast into an effective interaction resembling that of a homogeneous field. This is to be expected; in the low energy limit the photon wavefronts should vary infinitely slowly, projecting onto the constant crossed field limit of a plane wave. Indeed one may suggestively re-exponentiate the worldline vertices -- as is done in string theory -- via a linearisation operator to find a coupling to an effective homogeneous field coupling
\begin{equation}
    \prod_{i=1}^{N} (-ie)^NV[f_i]=\exp\Bigl[-ie\!\int_{0}^{T}\!d\tau\,a \cdot \dot{x}\Bigr]\Big|_{\text{lin \!\!}N}\,,
\end{equation}
where $a \coloneqq \sum_{i=1}^N a_i$ is the sum over all $N$-photon fields given by the respective Fock-Schwinger gauge choice,~\eqref{FS}, and $\text{lin }N$ is the linearisation operator serving to take only the multi-linear contribution in each of the $N$ $f_i$ (or $\varepsilon_i$), e.g. $f_1f_2|_{\text{lin}2}=f_1f_2$ but $f_{1,2}|_{\text{lin}2}=0$ or $1 |_{\text{lin}2}=0$. Thus the evaluation of the effective action and propagator to one-loop for an arbitrary background with $N$-photons in the low-energy limit is tantamount to evaluating one with both the background field plus a homogeneous field, $a$, and then applying the linearisation operator:
\begin{equation}\label{lin_op}
    \mathcal{D}_{N}^{x'x}[A]=\mathcal{D}^{x'x}[A\!+\!a]\Big|_{\text{lin}N}\,,\;
    \Gamma_{N}[A]=\Gamma[A\!+\!a]\Big|_{\text{lin}N}\,.
\end{equation}
(Here $\Gamma = \Gamma_{0}$ and $\mathcal{D}^{x'x} = \mathcal{D}^{x'x}_{0}$ are the effective action and propagator dressed by the external fields whose path integral representation omits the product of vertex operators). It is worth emphasising again that the above holds for \textit{arbitrary} $A$ coupled to any number of external $a_i$ photons, and that one need only evaluate the one-loop/line path integral without vertex insertions, and expand in accordance with the linearisation operator.

We will, however, now specialise to the case of a homogeneous background field, with similar Fock-Schwinger gauge,~\eqref{FS}, so that the combined field is homogeneous. This affords a simplification we use to highlight the merit of the low-energy photon approach, making otherwise intractable higher multiplicity calculations achievable.

\subsection{Effective action}
Now we wish to verify that the low-energy tree-level (resp. one-loop) amplitudes in the homogeneous background field can be obtained -- analogously to in vacuum -- from expanding the propagator (effective action) in a ``strong'' constant background, $A$, plus a ``weak'' constant fluctuation, $a$, in the fluctuation field. 

We start with the one-loop effective action in scalar QED. In a homogeneous background field the worldline action is quadratic in $x$ and thus the path integral (see \ref{eqGamN}) may be exactly evaluated as
\begin{equation}\label{Gamma_1}
    \Gamma[A\!+\!a]=-i\V\!\int^\infty_0\!\frac{dT}{(4\pi)^2T^3}\,\e^{-im^2T}\,K[A\!+\!a]\,,
\end{equation}
where $\V$ is a spacetime volume factor associated to the zero mode of the worldline kinetic operator. $K$ is a functional determinant factor arising from quadratic fluctuations. Taking also the background field in Fock-Schwinger gauge, so that $A_{\mu}(x) = -\frac{1}{2}F_{\mu\nu}(x - \hat{x})^{\nu}$, this functional determinant has a well-known solution under PBC orthogonal to the zero mode~\cite{Schmidt:1993rk} on $[0,T]$, being
\begin{align}
    K[F\tp f]&= \frac{\Det{}^{-\frac{1}{2}} [ \hat{\partial}_{\tau}^{2} 
    -2e(F\tp f)\hat{\partial}_{\tau}]  }{\Det{}^{-\frac{1}{2}} [ \hat{\partial}_{\tau}^{2}]}
    \label{Det_unexp}\\
    &=\det{}^{-\frac{1}{2}} \Big[ \frac{\sinh(Z\tp z)}{Z\tp z}\Big]\,,
\end{align}
where the matrices $Z\coloneqq eFT$ and $ z \coloneqq e f T$. 

In order to apply the linearisation operator in ~\eqref{lin_op}, it is convenient to first expand in powers of $z$. We can readily achieve this by expanding within the functional determinant form of~\eqref{Det_unexp}. To expand in the fluctuation we use the familiar ``ln Det = Tr ln'' relation and factorise the strong background part
\begin{align}
	K[F\tp f]&= K[F] \frac{ \Det{}^{-\frac{1}{2}} [ \hat{\partial}_{\tau}^{2} 
	-2e (F\tp f) \hat{\partial}_{\tau} ]}{ \Det{}^{-\frac{1}{2}} [ \hat{\partial}_{\tau}^{2} 
	- 2eF\hat{\partial}_{\tau} ] } \\
	&=K[F] \e^{ -\frac{1}{2} \Tr \ln\big[ 1 - 2e(\hat{\partial}_{\tau}^{2} 
	-2e F\hat{\partial}_{\tau})^{-1}  f \hat{\partial}_{\tau}  \big] }\,.
	\label{eff_log_form}
\end{align}
The logarithm supplies us with a suitable mechanism for expansion in $f$, where the Green function in functional form appears as
\begin{equation}\label{G_1}
	\G(\tau,\tau'|F)=\big\langle \tau \big| \frac{2}{\hat{\partial}_\tau^2-2eF\hat{\partial}_\tau} \big|\tau' \big\rangle\,.
\end{equation}
For $\G$ dependent only on the background field, $F$, we will further for brevity take $\G(\tau,\tau'|F)\eqqcolon\G(\tau,\tau')$. The inverse of the operator is to be taken on the space of periodic functions orthogonal to the zero mode (with $\G_{0j}=\G_{Tj}$, $\G_{i0}=\G_{iT}$ and $\int_{0}^{T}d\tau_{i}\, \G_{ij} = 0$, $\int_{0}^{T}d\tau_{j}\, \G_{ij} = 0$), so that it coincides with the worldline Green function in a homogeneous background field~\cite{Schmidt:1993rk}, and in Minkowski spacetime is given by
\begin{equation}
    \G_{ij}=
    \frac{T}{2Z^{2}}\Bigl(1-\frac{Z}{\sinh(Z)}\e^{-Z\dot{G}_{ij}}-Z\dot{G}_{ij}\Bigr)\,,
\end{equation}
where we have written for brevity $\G_{ij}\coloneqq \G(\tau_i,\tau_j)$, with Lorentz indices still implicit. The free Green function is $G_{ij}=|\tau_i-\tau_j|-T^{-1}(\tau_i-\tau_j)^2$. We remark that the Green functions are understood in the distributional sense, and this implies therein that $\mathrm{sgn}^2(\tau_i-\tau_j)=1$. The Green function satisfies  $(\partial_{i}^{2} -2e F\partial_{i} )\, \G_{ij} =2\delta_{ij}-2/T$. 

Using the PBC Green function, exponential corrections to the functional determinant can be written by expanding the logarithm in~\eqref{eff_log_form}
\begin{equation}
	K[F\tp f]= K[F]\e^{\sum_{n=1}^\infty \frac{(-ie)^n}{2n}\mathcal{I}_n} \,,\label{ExpDeltaF}
\end{equation}
where we have recovered the exponent in~\cite{Ahmadiniaz:2023jwd}:
\begin{equation}\label{Delta_F}
    \mathcal{I}^{(n)}=\!i^n\prod_{i=1}^n 
    \int^T_0\!d\tau_i \,\tr (\dot{\G}_{12} \cdot f \cdot \dot{\G}_{23} \cdot f  \!... \, \dot{\G}_{n1} \cdot f)\,.
\end{equation}
The multi-linear contribution to the effective action is then determined through application of the linearisation operator,~\eqref{lin_op}, onto just $K[F\tp f]$ as $K_N[F]=K[F\tp f] |_{\text{lin}N}$. We remark that although the logarithm has been expanded to infinite order, after acting with the linearisation operator, the summation will truncate as $\sum_{n=1}^\infty\to \sum_{n=1}^N$. Let us record the first two leading order corrections at order $N = 1$ and $N=2$.
\begin{align}\label{K_1}
	K_1[F]&=\frac{-ie}{2}K[F]\mathcal{I}^{(1)}\Big|_{\text{lin}1}\,,\\
	\label{K_2}
    K_2[F]&=\frac{(-ie)^2}{4}K[F] \Bigl [\frac{1}{2}(\mathcal{I}^{(1)})^2+\mathcal{I}^{(2)} \Bigr]\Big |_{\text{lin}2}\,,
\end{align}
where we have left the linearisation operator explicit. Its action which may be trivially determined by just summing over the symmetric combinations of $f_i$ (c.f. the following sub-section). 

At this point let us remark for the purpose of verification to previous literature~\cite{Ahmadiniaz:2023jwd}, that one may re-express the right hand side of the above if we now specialise $f = \sum_{k = 1}^{N}f_{k}$ and discard contributions not multi-linear in the $f_{k}$ following our low-energy prescription to obtain:
\begin{equation}
	\mathcal{I}^{(n)}=\sum_{\{i_1...i_n\}}\! ni^n\prod_{i=1}^n 
    \int^T_0\!d\tau_i \,\tr (\dot{\G}_{12} \cdot f_{i_2} \cdot \dot{\G}_{23} ... f_{i_n} \cdot \dot{\G}_{n1} \cdot f_{i_1})\,,
\end{equation}
in which the sum goes over trace-inequivalent orderings of the field strength tensors in the product.

\subsubsection{Worldline photon vertices}
Let us now demonstrate that the same structure follows from a formal worldline path integral evaluation. Although this has been shown for the effective action in \cite{Ahmadiniaz:2023jwd}, we will extend their work to the case of open lines and momentum space calculations later, so we take this opportunity to present the calculation in our notation. To do so let us expand path integral trajectory in the effective action,~\eqref{eqGamN}, about a string-inspired loop center, $x(\tau)=x_0+q(\tau)$, where $q(\tau)$ obeys PBC and as is usual $x_0=T^{-1}\int^T_0\!d\tau x(\tau)$. For the simple case of the effective action in a homogeneous field $x_0$ dependence disappears from the integrand leaving the volume factor, $\mathcal{V}_4$, and only periodic $q$ dependence in the path integral:
\begin{align}
    \Gamma_{N}[A]=&-i\mathcal{V}_4 \int_{0}^{\infty}\!dT\frac{\e^{-im^2T}}{(4\pi)^2T^3} K[F]\notag\\
    &(-ie)^N \Big\langle \prod_{i=1}^{N}V[f_i]\Big\rangle_{\text{PBC}}\,,
    \label{Gamma_vertex}
\end{align}
where we define for generic BCs the following worldline expectation value
\begin{equation}
    \langle \Oo \rangle =\frac{\int\! \D q(\tau)\, \Oo\, \e^{i\int^T_0\!d\tau\,[-\frac{1}{4}\dot{q}^{2} - q\cdot eF\cdot\dot{q}]}}{
    \int\! \D q(\tau)\, \e^{i\int^T_0\!d\tau\,[-\frac{1}{4}\dot{q}^{2} - q\cdot eF\cdot\dot{q}]}}\,.
\end{equation}
This implies that $\G_{ij}= -i\langle q(\tau_i)q(\tau_j)\rangle_{\text{PBC}}$. The vertices for the effective action are those defined in~\eqref{lowEvertex} with $x(\tau)\to q(\tau)$ owing the simplicity of the PBC. Indeed, comparing~\eqref{Gamma_vertex} with~\eqref{Gamma_1} we find
\begin{equation}
	K_N[F]=K[F](-ie)^N \Bigl \langle \prod_{i=1}^{N}V[f_i]\Bigr \rangle_{\text{PBC}}\,.
\end{equation}
We then verify the leading photon insertions at order $N = 1$ and $N=2$. We may readily confirm the $N=1$ contribution as
\begin{equation}
	K_1[F]=\frac{e}{2} K[F]\int^T_0\!d\tau_1\, \tr (\dot{\G}_{11} \cdot f_1)\,.
\end{equation}
For $N=2$, upon performing the Wick contraction we likewise find that 
\begin{align}
\hspace{-1.75em}    K_2[F]=\frac{e^2}{2}K[F]\int^T_0 \!d\tau_1d\tau_2\, &\Bigl[\frac{1}{2}\tr (\dot{\G}_{11} \cdot f_{1})\tr (\dot{\G}_{22} \cdot f_{2})\notag\\
\hspace{-1.75em}     &+\tr (\dot{\G}_{12} \cdot f_{2} \cdot \dot{\G}_{21} \cdot f_1)  \Bigr]\,.
\end{align}
These agree with the expressions for $K_1[F]$ and $K_2[F]$ given in~\eqref{K_1} and~\eqref{K_2}.

\subsubsection{Parameter integrals and matrix expansion}
\label{sec:matrix}
The expanded exponent in~\eqref{Delta_F} is compact yet contains all information for arbitrary multiplicity of low-energy photon couplings. We now show how to perform the parameter integrals in a way complimentary to~\cite{Ahmadiniaz:2023jwd} -- they used a matrix decomposition for a specific configuration of (anti-)parallel fields; we however, keep our homogeneous field arbitrary. And, at the same time demonstrate an equivalent expansion resulting in a recursive evaluation. To maintain a compact final form, we wish to keep the field dependence explicit and under a trace. 

To begin, we will need to employ the following identities for dual field strength $\widetilde{F}^{\mu\nu}=\frac{1}{2} \epsilon^{\mu\nu\alpha \beta} F_{\alpha \beta}$, and likewise for $\widetilde{f}$, (these incidentally hold for arbitrary field in (3+1)-dimensions):
\begin{equation}\label{eq:Fidentity}
    Ff-\widetilde{f}\widetilde{F}=-I_{f F}\,,
    \quad\widetilde{F}f+\widetilde{f}F=2I_{f\widetilde{F}}\,, 
\end{equation}
and their transposes, and the following corollaries:
\begin{align}
\label{eq:F1identity}
    &[F^{2},f]=[\widetilde{F}^{2},f]
    =F\widetilde{f}\widetilde{F}-\widetilde{F}\widetilde{f}F\,,\\
    &\{F^{2}+\widetilde{F}^{2},f\}
    =4I_{f\widetilde{F}}\widetilde{F}-2I_{f F}F\,,
    \label{eq:F2identity}
\end{align}
where the Lorentz invariants read $I_{fF}=\frac{1}{2}f_{\mu\nu}F^{\mu\nu}$ and $I_{f\widetilde{F}}=-\frac{1}{4}f^{\mu\nu}\widetilde{F}_{\mu\nu}$.

Let us show how the above identities aid in the evaluation of the parameter integrals in~\eqref{Delta_F}. First we write for an $f$ insertion
\begin{equation}\label{f_insertion}
f=\frac{1}{2} \bigl( \{f,F^2 +\widetilde{F}^2\}+[f,F^2 +\widetilde{F}^2]  \bigr) \cdot \frac{1}{F^2 +\widetilde{F}^2}\,,
\end{equation}
where we can now see that the anticommutator part, according to~\eqref{eq:F2identity}, commutes with all other $F$ and $\widetilde{F}$ dependent elements under the trace of~\eqref{Delta_F}. Then for the anticommutator part of (\ref{f_insertion}) we can simplify one parameter integral according to
\begin{align}
	&\int_{0}^{T}d\tau' \Gd_{\tau\tau'} \cdot \{f,F^2 +\widetilde{F}^2\} \cdot \Gd_{\tau'\tau''}\notag\\
	&=2\int_{0}^{T}d\tau' \Gd_{\tau\tau'} \cdot \Gd_{\tau'\tau''} \cdot [2I_{f\widetilde{F}}\widetilde{F}-I_{f F}F]\,.
	\label{eqAComm}
\end{align}
After performing many such operations (for anti-commutator insertions) one will be left with a number of folded integrals involving $\Gd$, which we elaborate on below. 

The other set of integrals we will encounter after applying~\eqref{f_insertion} involve the commutator. Fortunately, one may show that the following identity holds for such cases:
\begin{equation}
	e^{2}\int_{0}^{T}\!d\tau' \dot{\mathcal{G}}_{\tau\tau'} \cdot [f,F^2 \tp \widetilde{F}^2] \cdot \dot{\mathcal{G}}_{\tau'\tau''} =
	-2[\dot{\mathcal{G}}_{\tau\tau''},\{eF,f\}]\,,
	\label{eqComm}
\end{equation}
in effect reducing the rank of the Green functions from two to one. The above can be proven by first applying the defining differential equation, and its cousin following from the time translational symmetry of the worldline action, i.e. $(\partial_\tau^2 -2eF\partial_\tau)\G_{\tau\tau'}=2\delta(\tau-\tau')-2/T$ and $\G_{\tau\tau'}(\overleftarrow{\partial}_{\tau'}^2 +2eF\overleftarrow{\partial}_{\tau'})=2\delta(\tau-\tau')-2/T$ and the periodic BC properties of $\mathcal{G}$, to find
\begin{equation}
	\int_{0}^{T}\!d\tau \, \G_{\tau'\tau} \cdot \overleftarrow{\partial}_\tau 2[f,eF] \cdot \dot{\G}_{\tau\tau''}=2[\dot{\G}_{\tau'\tau''}\,,f]\,.
\end{equation}
Then using the above,~\eqref{eq:F1identity}, and the fact that $[f,F^2]=F[f,F]+[f,F]F$, one arrives at~\eqref{eqComm}.

In \cite{Ahmadiniaz:2023jwd}, the background field was chosen to consist of anti-parallel electric and magnetic fields to facilitate evaluation of the parameter integrals entering $\mathcal{I}^{(n)}$ leading to a closed form expression. This, however, complicates the Lorentz structure of the result and required various projection operators to arrive at closed form results for the parameter integrals. We take an alternative route here, which is closer to worldline calculations in vacuum and simplifies the Lorentz structure (at the cost of complicating the parameter integrals where they do not appear under a trace). We claim that application of the above manipulations allow all parameter integrals to be reduced to ones over ``closed cycles'' of $\dot{\mathcal{G}}$:
\begin{equation}
	\mathcal{J}^{(n)} = \prod_{i = 1}^{n} \int_{0}^{T}d\tau_{i}\, \dot{\mathcal{G}}_{12}\cdot  \dot{\mathcal{G}}_{23}\cdot \ldots \cdot \dot{\mathcal{G}}_{n1}
\end{equation}
that, crucially, are factorised from the remaining Lorentz tensors (again, this holds under the trace). This is reminiscent of the worldline calculation of the \textit{finite-}energy amplitudes \textit{in vacuum}, but holds here only in the low energy limit. Before showing this, let us provide a formula for $n>1$ such parameter integrals. Since
\begin{equation}
	\frac{1}{[\hat{\partial}_\tau-2eF]^n}=\frac{1}{(n-1)!}\Bigl(\frac{\partial_{e}}{2F}\Bigr)^{n-1} \frac{1}{\hat{\partial}_\tau-2eF}\,,
\end{equation}
and using that
\begin{equation}
	\dot{\G}_{\tau \tau'}=\frac{1}{Z}-\frac{\e^{-Z\cdot \dot{G}_{\tau\tau'}}}{\sinh(Z)}=\frac{1}{Z}+\sum_{j=0}^{\infty}2\e^{[2j+1-\dot{G}_{\tau\tau'}]\cdot Z}\,,
\end{equation}
one may readily find that
\begin{align}
	&\langle \tau | \frac{1}{[\hat{\partial}_\tau-2eF]^n} |\tau' \rangle =\frac{-1}{T}\Bigl(\frac{-1}{2eF}\Bigr)^{n}\\
	&+\sum_{j=0}^{\infty}\frac{1}{(n-1)!}\Bigl(\frac{T}{2}[2j+1-\dot{G}_{\tau\tau'}]\Bigr)^{n-1}\e^{[2j+1-\dot{G}_{\tau\tau'}]\cdot eFT}\,.\notag
\end{align}
Although a closed form solution for this series seems difficult, we can make progress for the functional trace, 
\begin{equation}
	\mathcal{J}^{(n)} = \int_{0}^{T} d\tau\, \langle \tau | \frac{2^{n}}{[\hat{\partial}_\tau-2eF]^n} |\tau \rangle\,.
	\label{Jn}
\end{equation}
Since only even powers of $\dot{G}(\tau, \tau)$ survive, we have $\e^{\dot{G}(\tau, \tau)eFT} = \cosh(eFT)$, $\mathcal{J}^{(1)} = T\big[\frac{1}{eFT} - \textrm{coth}(eFT)\big]$ so that
\begin{equation}
	\mathcal{J}^{(n)} = \frac{T}{(n-1)!} \Bigl(\frac{\partial_{e}}{F}\Bigr)^{n-1} \Big[\frac{1}{eFT} - \coth(eFT)\Big]\,.
	\label{JnDeriv}
\end{equation}
We provide a closed form expression for this derivative in Appendix \ref{AppTraces}. 

It remains, then, to prove the claim that the parameter integrals in $\mathcal{I}^{(n)}$ can indeed be reduced to a product of the cycle integrals of $\mathcal{J}^{(n)}$ and some (potentially complicated) function of the various field strength tensors. To do this we proceed by induction on $n$. For $n = 1$ there is nothing to do. For the base case $n = 2$, equations (\ref{eqAComm}) and (\ref{eqComm}) exhibit the fact that the dependence on photon field strength can be separated from a one- or two-cycle of $\dot{\mathcal{G}}$(s):
\begin{align}
\hspace{-1em}	\mathcal{I}^{(2)} = &-\tr\Big(\int_{0}^{T} d\tau_{1} \int_{0}^{T} d\tau_{2}\, \frac{ \dot{\mathcal{G}}_{12} \cdot \dot{\mathcal{G}}_{21}}{F^{2} + \tilde{F}^{2}} \cdot \big[2 I_{f\tilde{F}} \tilde{F} - I_{f F}F\big] \cdot f\Big) \nonumber \\
\hspace{-1em}	&+ \tr\Big(\int_{0}^{T}d\tau_{1} \, \frac{1}{F^{2} + \tilde{F}^{2}} \cdot \big[ \dot{\mathcal{G}}_{11}, \big\{ eF, f\} \big] \cdot \frac{f}{e^{2}} \Big)
\end{align}
A little further rearrangement finds
 \begin{align}
 \mathcal{I}^{(2)}=&-\frac{1}{2}\int_{0}^{T}\! d\tau_{1}d\tau_{2}\,\mathrm{tr}\Bigl[\dot{\mathcal{G}}_{12} \cdot \dot{\mathcal{G}}_{21} \cdot \frac{\{f,F^{2}+\widetilde{F}^{2}\}}{F^{2}+\widetilde{F}^{2}}\cdot f\Bigr]\\&+T\int_{0}^{T}\! d\tau_{1}\, \mathrm{tr}\Bigl[\frac{\dot{\mathcal{G}}_{11}}{(Z^{2}+\widetilde{Z}^{2})^{2}} \cdot [\{Z,f\},(Z^{2}+\widetilde{Z}^{2})\cdot f]\Bigr]\notag\,,
 \end{align}
 where we remark that $[(F^{2}+\widetilde{F}^{2})^{2}]_{\mu\nu}=(I_{FF}^{2}+4I_{F\tilde{F}}^{2})\eta_{\mu\nu}$ commutes with $f$. For later reference, we now pause to apply~\eqref{JnDeriv} or~\eqref{finalJsol} to arrive at the following compact form for the resulting parameter integrals
 \begin{align}
\label{I2_compact}
 \mathcal{I}^{(2)}=&-\frac{T^{2}}{2}\mathrm{tr}\Bigl[\frac{\sinh^{-2}(Z)-Z^{-2}}{Z^{2}+\widetilde{Z}^{2}}\cdot \{f,Z^{2}+\widetilde{Z}^{2}\}\cdot f\Bigr]\notag\\&-2T^{2}\mathrm{tr}\Bigl[\frac{Z^{-1}-\coth Z}{(Z^{2}+\widetilde{Z}^{2})^{2}} \cdot [Z^{2},f] \cdot \{Z,f\}\Bigr]\,.
 \end{align}
Supposing, then, that the factorisation of cycle integrals holds at order $n$, we consider $\mathcal{I}^{(n+1)}$ and apply the results (\ref{eqAComm}) and (\ref{eqComm}) to, say, the variable $\tau_{1}$:
\begin{align}
	\hspace{-1.25em}\mathcal{I}^{(n+1)} &= \prod_{i=1}^{n+1} \int_{0}^{T}d\tau_{i}\, \tr \Big( \frac{2 I_{f \tilde{F}} \tilde{F} - I_{f F}F}{F^{2} + \tilde{F}^{2}} \cdot  \dot{\mathcal{G}}_{12} \cdot f \cdot \ldots \cdot f \cdot \dot{\mathcal{G}}_{n+1, 1}  \Big)\nonumber\\
	\hspace{-1.5em}&- \frac{1}{e^{2}}\prod_{i=2}^{n+1} \int_{0}^{T}d\tau_{i}\, \tr\Big( \big[ \dot{\mathcal{G}}_{n+1, 2}, \big\{eF, f \big\} \big] \cdot  \frac{1}{F^{2} + \tilde{F}^{2}}  \nonumber \\
	& \hspace{11em}\cdot f \cdot \dot{\mathcal{G}}_{23} \cdot f \cdot \ldots \cdot f  \Big)\,.
\end{align}
Now, the first line on the RHS has already factorised into some Lorentz tensors multiplied by a cycle with only $n$ photon field strength tensors so can be reduced to fully factorised form by the inductive hypothesis. For the second line on the RHS we expand the commutator and use cyclicity of the trace to obtain two contributions under the trace:
\begin{align}
	&\big\{eF, f\big\} \cdot \frac{1}{F^{2} + \tilde{F}^{2}} \cdot f \cdot \dot{\mathcal{G}}_{23} \cdot f \cdot \ldots \cdot f \cdot \dot{\mathcal{G}}_{n+1, 2} \nonumber \\
	-& \big\{eF,  f\big\} \cdot \frac{1}{F^{2} + \tilde{F}^{2}} \cdot \dot{\mathcal{G}}_{n+1, 2} \cdot f \cdot \dot{\mathcal{G}_{23}}\cdot \ldots \cdot f\,.
\end{align}
In both cases, the parameter integrals are again of order $n$ and so can be reduced by assumption. Hence the result is proved and, in principle, the problem of calculating the low energy amplitudes is solved by this procedure and the evaluation of the $\mathcal{J}^{(n)}$ in Appendix \ref{AppTraces}. It should be noted that the same approach also proves the factorisation into cycles for spinor QED, presented below.

Let us now present an entirely different expansion, which in turn illustrates the evaluated form of the parameter integrals. Since the scattering photon fields and background field superpose we can avoid a functional expansion by instead performing a power series expansion about $f$ in~\eqref{Det_unexp}:
\begin{align}\label{taylor}
   K[F\tp f] &= \e^{f\cdot \partial_F}K[F]\,,\;\\
   K_N[F]&=\tfrac{1}{N!}(f\!\cdot\!\partial_F)^N K[F]\Big|_{\text{lin}N}\,,
\end{align}
where here $f\cdot \partial F \equiv f_{\mu\nu} \frac{\partial}{\partial F_{\mu\nu}}$. Above, however, we determined that all $N$ corrections can be encapsulated in an exponential term, $\mathcal{I}^{(n)}$. Thus, to make the connection to our previous findings it is convenient to re-express this expansion using di Bruno's formula such that
\begin{equation}
	K[F\tp f]=K[F]\e^{-\frac{1}{2}\sum_{n=1}^\infty \frac{(f\cdot \partial_F)^n}{n!} \tr \ln[\frac{\sinh (Z)}{Z}] }\,,
\end{equation}
where now we can identify
\begin{equation}\label{eq:Del_n}
	\mathcal{I}^{(n)}=-\frac{1}{(n-1)!} \Bigl(\frac{f\cdot \partial_F}{-ie}\Bigr)^n \tr \ln\Bigl[\frac{\sinh (Z)}{Z}\Bigr]\,.
\end{equation}
Although this can be evaluated recursively, a straightforward power series expansion is inhibited because\\ 
$[f\cdot\partial_Fg(F),F]\neq 0$. Order by order, however, it is possible, and it is particularly simple for $\mathcal{I}^{(1)}$; this is because of the cyclicity of the trace with only one insertion of $f$. First, let us write
\begin{equation}\label{eq:Del_1}
	\mathcal{I}^{(1)}=\frac{i}{e}\mathrm{tr}\Bigl[\frac{\sinh(Z)}{Z}f \!\cdot\!\partial_F \frac{Z}{\sinh(Z)}\Bigr]\,.
\end{equation}
To analyse the matrix derivative under the trace let's consider a suitable power expansion of the matrix function $g(F(x))=\sum_{n=0}^{\infty}a_{n}F(x)^{n}$, alongside an arbitrary function $h(F(x))$. Using cyclicity it is straightforward to show that
\begin{equation}
	\mathrm{tr}[h(F(x))\partial_{x}g(F(x))]=\mathrm{tr}[h(F(x))\partial_{x}F(x) g'(F(x))]\,,
\end{equation}
where $g'(F(x)) = \partial_{q} g(q) |_{q\to F}$ with $q$ a scalar weighted function or otherwise commutable matrix. Let's show how the matrix derivative acts for our purposes with a few simple examples: $f\cdot \partial_F F=f$, $f\cdot \partial_F \widetilde{F}=\widetilde{f}$, $f\cdot \partial_F I_{fF}=I_{ff}$, etc.
 Then we find the first order correction to be
\begin{equation}\label{eq:Delta_1}
	\mathcal{I}^{(1)} =-iT\,\mathrm{tr}[(\coth(Z)-Z^{-1})f]\,.
\end{equation}
This agrees with (\ref{K_1}) (we recall $J^{(1)}$ given above) and confirms the known result~\cite{Ahmadiniaz:2023jwd, MishaLow} when converted to Minkowski spacetime.

Determining the higher orders of $\mathcal{I}^{(n)}$ can also be achieved, however to apply the same principles as employed to find $\mathcal{I}^{(1)}$ we must first reduce the $f$ under the trace to linear order.  We accomplish this by employing the identities shown above in~\eqref{eq:F1identity}-\eqref{eq:F2identity}. Let us show the procedure for $\mathcal{I}^{(2)}$. We insert $(F^{2}+\widetilde{F}^{2})/(F^{2}+\widetilde{F}^{2})$ under the trace in~\eqref{eq:Delta_1}, and apply cyclicity of the trace to form the anti-commutator on the LHS of~\eqref{eq:F2identity}. This then commutes with $F$, so we may apply the same steps leading to $\mathcal{I}^{(1)}$ in~\eqref{eq:Delta_1}, but for $\mathcal{I}^{(2)}$. This approach extends to higher orders. Let us show the explicit form of $\mathcal{I}^{(2)}$ as 
\begin{align}\label{eq:Delta_2}
	\mathcal{I}^{(2)}=&-\frac{T}{e}\mathrm{tr}\frac{1}{F^{2}+\widetilde{F}^{2}}\Bigl\{(Z^{-1}-\coth(Z))\\
	&\times\Bigl[2I_{f\widetilde{f}}\widetilde{F}-I_{ff}F+2I_{f\widetilde{F}}\widetilde{f}-I_{fF}f\notag\\
	&-\frac{(2I_{f\widetilde{F}}\widetilde{F}-I_{fF}F)(2Ff+2\widetilde{F}\widetilde{f})}{F^{2}+\widetilde{F}^{2}}\Bigr]\notag\\
	&+eT(-Z^{-2}+\sinh^{-2}(Z))(2I_{f\widetilde{F}}\widetilde{F}-I_{fF}F)f\Bigr\}\notag\,.
\end{align}
Higher orders increase in complexity for our case of a generic constant field, and showing their form for the coming cases is not illuminating. However, the recursion relation is readily applicable. One need only insert again $(F^{2}+\widetilde{F}^{2})/(F^{2}+\widetilde{F}^{2})$ under the trace, and use~\eqref{eq:F2identity}. Then since all the elements under the trace commute, we can straightforwardly take the $f\cdot\partial_F$ derivative. 

To make the connection to~\eqref{I2_compact} as presented from the parameter integral approach, we can use the above~\eqref{eq:Delta_2}, rearranging terms; however, in practice it is simplest to keep all Lorentz structures under the trace. To show the comparison, let us rework $\mathcal{I}^{(2)}$, but instead of using the Lorentz invariants in~\eqref{eq:F2identity} we use the anti-commutation relation directly. To begin we write $\mathcal{I}^{(2)}$ using the above principles as
\begin{align}
	\mathcal{I}^{(2)}=&-\frac{T^{2}}{2}f\cdot\partial_{Z}\mathrm{tr}\Big[\frac{Z^{-1}-\coth Z}{Z^{2}+\widetilde{Z}^{2}}\{f,Z^{2}+\widetilde{Z}^{2}\}\Big]\\\notag=&-\frac{T^{2}}{2}\mathrm{tr}\Bigl[\frac{\sinh^{-2}(Z)-Z^{-2}}{Z^{2}+\widetilde{Z}^{2}}\{f,Z^{2}+\widetilde{Z}^{2}\}f\\ \notag&+\frac{Z^{-1}-\coth Z}{(Z^{2}+\widetilde{Z}^{2})^{2}}\Bigl(-\{f,Z^{2}+\widetilde{Z}^{2}\}f\cdot\partial_{Z}(Z^{2}+\widetilde{Z}^{2})\\&+(Z^{2}+\widetilde{Z}^{2})\{f,f\cdot\partial_{Z}(Z^{2}+\widetilde{Z}^{2})\}\Bigr)\Bigr]\,.
\end{align}
Then we make use of yet another useful identity, $\{F,f\}=\{\widetilde{F},\widetilde{f}\}$, found from~\eqref{eq:Fidentity}, and consequently $f\cdot\partial_{Z}(Z^{2}+\widetilde{Z}^{2})=2\{F,f\}$ to find that $\mathcal{I}^{(2)}$ does indeed reproduce~\eqref{I2_compact}. In the form of~\eqref{I2_compact} one may more readily make the connection to the expression found in~\cite{Ahmadiniaz:2023jwd, MishaLow}, once continued to Euclidean spacetime.

\subsection{Propagator}
We now address the scalar propagator dressed by low energy photons. The Master Formula we will obtain is of particular interest for its connection to tree-level scattering observables at arbitrary multiplicity. As before we need the propagator evaluated for homogeneous fields, that we split into a superposition of two fields. The exact propagator in a constant background is well-known~\cite{Schwinger}; let us present it in propertime form, after integrating over trajectories in~\eqref{eqDN}\footnote{Here we have chosen to use Fock-Schwinger gauge ``instantaneously'' centred at the reference point $\hat{x} = x$, the initial point of a given worldline as in \cite{Ahmad_2017} -- see the Appendix for a more thorough discussion.}, and solution as
\begin{align}
	D^{x'x}[A\tp a]&=-i\!\int^\infty_0\!\frac{dT}{(4\pi T)^2}\e^{-im^2T}K^{x'x}[F\tp f]\,, 
	\label{D_homo}\\
	K^{x'x}[F\tp f]&=K[F\tp f]\e^{-i\frac{x^{2}_-}{4T}+ix_- \cdot \frac{(Z\tp z)^2}{T^{4}} \cdot \oDelo(F\tp f)\cdot x_-} \notag\\
	&=K[F\tp f]\e^{-\frac{i}{4T}x_- \cdot (Z\tp z) \cdot \coth(Z\tp z) \cdot x_-}\,,
	\label{D_homo_expand}
\end{align} 
where $x_-\coloneqq x'-x$. In the worldline formalism the exponent comes from completing the square in the worldline action, requiring the Green function in an arbitrary homogeneous field background on the Hilbert space of functions obeying \textit{Dirichlet boundary conditions}, denoted DBC (note the normalisation),
\begin{equation}\label{Del_1}
	\Del(\tau,\tau'|F\tp f)=\langle \tau | \frac{1}{\hat{\partial}_\tau^2-2e(F\tp f)\hat{\partial}_\tau} |\tau'\rangle\,,
\end{equation}
c.f.,~\eqref{G_1}. We further adopt the usual notation of $\dDel(\tau,\tau')=\partial_\tau \Del(\tau,\tau')$ and $\oDel(\tau')=\int^T_0d\tau\, \Del(\tau,\tau')$ (and likewise for the right-hand side), keeping implicit where appropriate the propertime arguments. With Minkowski space metric it is convenient to express $\Del{}_{ij} := \Delta(\tau_{i}, \tau_{j})$ in terms of the string-inspired Green function, viz. 
\begin{equation}\label{Green_Del}
	2 \Del{}_{ij} = \G_{ij} - \G_{i0} - \G_{0j} + \G_{00}\,,
\end{equation}
where zeroes here indicate that e.g. $\G(\tau,0)=\G(\tau,\tau'\!\to\! 0)$. As before we must decompose $f$ into a sum over external low-energy photon field strengths, $f_{i}$; here now in addition to a determinant as we just treated with the effective action, we must also deal with the expansion of the exponent involving the the Green function.

Let us first treat the determinant. This is similar to~\eqref{Det_unexp}, but is evaluated on the space of functions with DBCs. This is known, however, to take the same value as the determinant under PBCs. For considering~\eqref{Delta_F}, observe that the replacement of $\dot{\G}_{ij}\to\dot{\G}_{ij}-\dot{\G}_{i0}$ is possible under the trace, since $\int^T_0\!d\tau_i\, \dot{\G}_{i0}=0$. Hence one may replace $\dot{\G}_{ij}$ with $2 \dDel_{ij}$ in~\eqref{Delta_F} correctly reproducing the determinant but under Dirichlet BCs. Thus $K[F\tp f]$ in~\eqref{D_homo_expand} is indeed given by~\eqref{ExpDeltaF}. 

For the expansion of the additional exponential depending on the endpoints of propagation,  we rewrite the operator in~\eqref{Del_1} as
\begin{equation}
	\frac{1}{\hat{\partial}_\tau^2-2e(F\tp f)\hat{\partial}_\tau}=\frac{1}{1-2e\frac{1}{\hat{\partial}_\tau^2-2eF\hat{\partial}_\tau} f\hat{\partial}_\tau }\frac{1}{\hat{\partial}_\tau^2-2eF\hat{\partial}_\tau}\,,
\end{equation}
then perform a geometric series expansion in $f$ to find
\begin{align}
    &\Del(\tau,\tau'|F\tp f)=\Del(\tau,\tau')+\sum_{n=1}^\infty(-ie)^n\Del{}^{(n)}(\tau,\tau')\,,\\
	&\Del{}^{(n)}(\tau,\tau')=(2i)^n\prod_{i=1}^n 
    \int^T_0\!d\tau_i \,\Del{}_{\tau 1} \tcdot f \tcdot \dDel_{12} \tcdot f \tcdot \!... \,\tcdot \dDel_{n\tau'}\,, \label{Del_expanded}
\end{align}
where we have abbreviated $\Del(\tau,\tau'|F)\eqqcolon\Del(\tau,\tau')$. 

Gathering both the expanded determinant and exponent carrying the dependence on propagator endpoints we can write for the worldline kernel
\begin{align}
	\hspace{-0.5em}&K^{x'x}[F\tp f]=K^{x'x}[F]\e^{\sum_{n=1}^\infty \frac{(-ie)^n}{2n}\mathcal{I}^{(n)} }\nonumber \\
		\hspace{-0.5em}&\e^{\frac{i}{T^4}x_- \cdot    \bigl[ 2z \tcdot \oDelo \tcdot Z  +z \tcdot \oDelo \tcdot z  +\sum_{n=1}^\infty (-ie)^n (Z\tp z) \tcdot \oDel{}^{(n)}{}^\circ \tcdot (Z\tp z)\bigr] \cdot x_- }\,.
	\label{Kxx_expanded}
\end{align}
Then from an operational standpoint to compute $D^{x'x}_N[A]$ and hence $K^{x'x}_N[F]=K^{x'x}[F+f]|_{\text{lin}N}$ one need only truncate the sum in the exponent to $N$, i.e., $\sum_{n=1}^\infty\to \sum_{n=1}^N$. Then take all combinations of $N$ powers of $f$, recalling that $z=efT$. Let us show this for the first order correction to the kernel as
\begin{align}\label{K_1xx}
	K^{x'x}_1[F]&=-\frac{ie}{2}K^{x'x}[F]\,\mathcal{I}^{x'x(1)}\Big|_{\text{lin}1}\,,\\
	\mathcal{I}^{x'x(1)}&=\mathcal{I}^{(1)}-\frac{2}{T^4}x_-\tcdot [
	 2Tf \tcdot \oDelo \tcdot Z -iZ \tcdot \oDel{}^{(1)}{}^\circ \tcdot Z]\tcdot x_-\,.
	 \label{K_1xx2}
\end{align}
And the second order correction becomes
\begin{align}
	K^{x'x}_2[F]&=\frac{(-ie)^2}{4}K^{x'x}[F]\,\Bigl[\frac{1}{2}(\mathcal{I}^{x'x(1)})^2+\mathcal{I}^{x'x(2)}\Bigr]\Big|_{\text{lin}2}\,,\\
	\mathcal{I}^{x'x(2)}&=\mathcal{I}^{(2)}-\frac{4}{T^4}x_-\tcdot \big[
	 iT^2 f \tcdot \oDelo{} \tcdot f  \nonumber\\
	 &+2Tf \tcdot \oDel{}^{(1)}{}^\circ \tcdot Z -iZ \tcdot \oDel{}^{(2)}{}^\circ \tcdot Z \big]\tcdot x_-\,.\label{Ixx2}
\end{align}
For completeness, in Appendix \ref{AppExp} we provide explicit expressions for the multi-linear expansion of (\ref{Kxx_expanded}) for arbitrary photon multiplicity, $N$. In the following subsection we will verify these results using the path integral approach.

\subsubsection{Worldline photon vertices}
Here we show how the additional endpoint contributions arise from the perspective of the photon vertex operators for open lines. We begin by writing the following expression for the scalar propagator in coordinate space from~\eqref{eqDN}
\begin{align}\label{eqDz2}
	&\mathcal{D}_{N}^{x^\prime x}[A]= (-ie)^N\int_{0}^{\infty}dT\,\e^{-im^2T}\e^{-i\frac{x_{-}^{2}}{4T}} \\
	&\int_{\mathrm{DBC}}\hspace{-0.75em}\mathcal{D}q(\tau)\,\e^{-i\int_{0}^{T}d\tau\,\left(\frac{\dot{q}^2}{4}+\frac{e}{2}q\cdot F \cdot \dot{q} - \frac{e}{T}\,x_{-} \cdot F \cdot q\right)} \prod_{i = 1}^{N}  V[f_i] \, ,\notag
\end{align}
where we have decomposed the arbitrary trajectory $x(\tau)$ into a straight-line and a fluctuation part $x(\tau) = x + x_-\frac{\tau}{T} + q(\tau)$ with DBC on $q(\tau)$ and used the Fock-Schwinger gauge,~\eqref{FS}, with reference point $\hat{x}=x$ being one of the worldline endpoints. After expanding about the straight line, the scalar vertex operators,~\eqref{lowEvertex_line}, are expressed in terms of the fluctuation as
\begin{equation}
	V[f_i]=\int_{0}^{T}d\tau_i\;\frac{1}{2}\Big(q_i\cdot f_i\cdot\dot{q}_i-\frac{2}{T}x_{-}\cdot f_i\cdot q_i\Big)\, ,\label{eqVtxOp}
\end{equation}
where we used the short notation $q_{i}=q(\tau_{i})$. This can be recast as a worldline expectation value
\begin{align}
	\mathcal{D}_{N}^{x^{\prime}x}[A] &= -i\int_{0}^{\infty}   \frac{dT}{(4\pi T)^2}  \,\e^{-im^2T}K^{x'x}_N[F]\,,\\
	K^{x'x}_N[F]&=(-ie)^NK[F]\e^{-i\frac{x_{-}^{2}}{4T}}\notag\\
	&\times \Big\langle \e^{\frac{ie}{T}\int_{0}^{T}d\tau\,x_{-}\cdot F\cdot q}  \prod_{i=1}^{N}V[f_i]\Big\rangle_{\mathrm{DBC}}\,, \label{eqDzVtxOp}
\end{align}
where Wick contraction of fields inside the expectation value are determined according to $\langle q^{\mu}(\tau)q^{\nu}(\tau^{\prime}) \rangle_{\mathrm{DBC}} = 2i\Del^{\mu\nu}(\tau,\tau^{\prime})$, since the function on the RHS inverts the kinetic operator in the worldline path integral. 

To corroborate that we obtain the same result as with the previous method, let us consider the one photon amplitude (corresponding to the lowest order); to do that, we use \eqref{eqDzVtxOp} with $N=1$. The first order correction becomes $ K^{x'x}_1[F]= -\frac{ie}{2}K^{x'x}[F]\mathcal{I}_1^{x'x(1)}$ where
\begin{align}
	&K^{x'x}[F] = K[F]\e^{-i \frac{x_{-}^{2}}{4T}}\Big\langle \e^{\frac{ie}{T}\int_{0}^{T}d\tau\,x_{-}\cdot F\cdot q}\Big\rangle_{\mathrm{DBC}}\,,\\
	&\mathcal{I}_1^{x'x(1)}=\int_{0}^{T}d\tau_1\,\Big[ 2i \tr(\dDel_{11} \tcdot f_{1})
\label{eqVtxOpExpan}\\
	& -\frac{4}{T^4}\,x_{-} \tcdot \bigl( Tf_{1} \tcdot \Delo_{1} \tcdot Z
	\nonumber +Z \tcdot \oDel_{1} \tcdot f_{1} \tcdot \dDelo_{1} \tcdot Z \bigr)\tcdot x_{-} \Big]\,,
\end{align}
which is in agreement with~\eqref{K_1xx2} (the trace matches the first order contribution from expanding the functional determinant and the second line corresponds with the expansion of the exponential carrying the dependence on the endpoints). 

To proceed with higher orders we introduce a new notation: we denote an ordered cycle by
\begin{equation}
	\mathcal{I}_{i_1i_2...i_n}^{x'x(n)}=\mathcal{I}^{x'x(n)}(f_{i_1},f_{i_2},...,f_{i_n})\,,
\end{equation}
where each instance of $f$ in $\mathcal{I}_1^{x'x(1)}$ is replaced with the specific $f_i$ in the order given. Then the $N=2$ case can be expressed as
\begin{equation}
\hspace{-0.5em}K^{x'x}_2[F]=\frac{(-ie)^2}{4}K^{x'x}[F][\mathcal{I}^{x'x(1)}_1\mathcal{I}^{x'x(1)}_2+\mathcal{I}^{x'x(2)}_{12}+\mathcal{I}^{x'x(2)}_{21}]\,
\label{eqK2xx}
\end{equation}
with, explicitly,
\begin{align}
	&\mathcal{I}^{x'x(2)}_{ij}\!=\!-4\int^T_0\!d\tau d\tau' \, \Bigl\{
	\mathrm{tr}[\dDel{}_{\tau \tau'} \tcdot f_i \tcdot \dDel{}_{\tau\tau'} \tcdot f_j]\notag\\
	&+\frac{i}{T^2}x_-\tcdot f_i \tcdot \oDelo \tcdot f_j \tcdot x_-+ \frac{4i}{T^3}x_-\tcdot f_i \tcdot \Del{}_{\tau \tau'} \tcdot f_j \tcdot \dDelo{}_{\tau'} \tcdot Z \tcdot x_-\notag\\
	&+\frac{4i}{T^4}x_-\tcdot Z \tcdot \oDel{}_\tau \tcdot f_i \tcdot \dDel{}_{\tau \tau'} \tcdot f_j \tcdot \dDelo{}_{\tau'} \tcdot Z \tcdot x_-
	\Bigr\}\,,
\end{align}
which is fully in agreement with~\eqref{Ixx2} (as before the terms involving traces correspond to the expansion of the functional determinant whereas those with endpoint dependence match with the expansion of the exponential there; the first term in (\ref{eqK2xx}) contains cross terms from these expansions).

\subsubsection{Matrix expansion}
As performed in the previous section for the effective action, we may also achieve a matrix expansion of the propagator via a Taylor series in the superposition of scattering and background fields . The expansion takes the same form as on the loop in~\eqref{taylor}, but with $K[F+f]\to K^{x'x}[F+f]$, where we write
\begin{equation}
	\hspace{-1em}K^{x'x}[F\tp f]=K^{x'x}[F]\e^{\sum_{n=1}^\infty\big[\frac{(-ie)^n}{2n}\mathcal{I}^{(n)} -\frac{i}{4T}\frac{f\cdot \partial_F}{n!} x_-\cdot Z\coth(Z)\cdot x_-\big]}
\end{equation}
with $\mathcal{I}^{(n)}$ given by~\eqref{eq:Del_n}. However, in addition to the determinant factor we must also expand the exponential factor depending on the endpoints. We again Lorentz decompose the exponent by writing
\begin{align}
	x_-\cdot Z\coth (Z)\cdot x_- =x_-^2 \frac{1}{2}\mathrm{tr}&\Bigl[\frac{\widetilde{Z}^{2}}{Z^2+\widetilde{Z}^2}\coth(Z)Z\Bigr]\notag\\
	+x_-\cdot Z^2 \cdot x_-\frac{1}{2}\mathrm{tr}&\Bigl[\frac{1}{Z^2+\widetilde{Z}^{2}}\coth(Z)Z\Bigr]\,,
	\label{matpropxid}
\end{align}
where we have made use of projection operators with respect to $Z$, first finding the magnitude of the eigenvalues of $Z$, extracting the Lorentz scalars involving $x_{-}$ and then re-assembling the eigenvalues into their matrix form; see~\cite{Copinger:2022gfz} for details on formulae and projection operators. 

To determine the contribution for one photon, we factorise $K^{x'x}_1[F]$ as in~\eqref{K_1xx}, arriving straightforwardly at
\begin{align}
	&\mathcal{I}^{x'x(1)}=\mathcal{I}^{(1)}+\frac{1}{2eT}f\cdot\partial_F x_- \cdot Z\coth(Z)\cdot x_-\\
	&=\mathcal{I}^{(1)}+\frac{x_-\tcdot Zf\tcdot x_-}{2}\mathrm{tr}\Bigl[\frac{1}{Z^{2}+\widetilde{Z}^{2}}\coth(Z)Z\Bigr]\\
	&+\frac{x_-^{2}}{4}\mathrm{tr}\Bigl[\frac{\widetilde{Z}^{2}}{(Z^{2}+\widetilde{Z}^{2})^{2}}\coth(Z)Z\Bigl(2\frac{Z^{2}}{\widetilde{Z}}\widetilde{f}+\frac{\widetilde{Z}^{2}-Z^{2}}{Z}f\notag\\
	&-\frac{Z(Z^{2}+\widetilde{Z}^{2})}{\sinh(2Z)}f\Bigr)\Bigr]+\frac{x_-\tcdot Z^{2}\tcdot x_-}{4}\mathrm{tr}\Bigl[\frac{\coth(Z)Z}{(Z^{2}+\widetilde{Z}^{2})^{2}}\notag\\
	&\times\Bigl(-2\widetilde{Z}\widetilde{f}+\frac{\widetilde{Z}^{2}-Z^{2}}{Z}f-\frac{2(Z^{2}+\widetilde{Z}^{2})}{\sinh(2Z)}Zf\Bigr)\Bigr]\,,\notag
\end{align}
where $\mathcal{I}^{(1)}$ is given in~\eqref{eq:Delta_1}. Here too, for the propagator, higher order expressions are conceptually simple to produce, but result in larger expressions; therefore we confine our attention to just the first order case here. The various traces above can be easily taken with application of projection operators as in ~\cite{Copinger:2022gfz}. To arrive at the above from~\eqref{eqVtxOpExpan}, one must complete the parameter integrals, then use projection operators to rewrite the remaining expression in terms of $x_-^2$ and $x_-\cdot Z^2 \cdot x_-$ as done in~\eqref{matpropxid}. This completes our discussion of the scalar propagator in position space -- we will return to the momentum space propagator and its connection to scattering amplitudes later after turning to the more physical theory of spinor matter.

%%%%%%%%%%%%%%%%%%%%%%%%%%%%%%%%%%%%%%%%%%%
\section{Master Formulae in spinor SFQED}
\label{secSpinor}
For the case of spin 1/2 fermions, we should add the contributions from the spin degrees of freedom factors. In the worldline formalism these are represented by ``spin factors'' \cite{Feyn2, Strass1}, described in more detail below. For the effective action this leads to a proper time representation
\begin{align}
    \Gamma_{\text{sp}}[A + a] &= \frac{i}{2}\mathcal{V}_4 \int_{0}^{\infty} \!\frac{dT\,\e^{-im^{2}T}}{(4\pi)^2 T^3} K_\text{sp}[F+f]\,,\\
K_\text{sp}[F+f]&= \det{\!}^{\!-\frac{1}{2}} \Big[ \frac{\tanh(Z\tp z)}{Z\tp z}\Big]\,,
\end{align}
with an additional factor of $\det^{\frac{1}{2}} [ \cosh(Z + z) ]$ entering the matrix determinant from the spin factor:
\begin{align}
     \tr \mathcal{P} \e^{-i\frac{e}{2} \int_{0}^{T}d\tau\, \sigma^{\mu\nu}(F_{\mu\nu}+f_{\mu\nu}) } &= \det{}^{\frac{1}{2}}[ \cosh(Z+z)]\\
    &= \e^{\frac{1}{2} \Tr \ln [ \hat{\partial}_{\tau} + 2e(F+f)]}\,.
\end{align}
We have again written the determinant in its operator form and used $\sigma^{\mu\nu} = \frac{i}{2}[\gamma^{\mu}, \gamma^{\nu}]$. As before we will need a Green function with which we can perform a suitable expansion. As spin degrees of freedom obey anti-periodic boundary conditions on the worldline, this is  $\GF(\tau, \tau' |F) \eqqcolon \GF(\tau,\tau')=2\langle \tau | [\hat{\partial}_{\tau} +2eF ]^{-1} | \tau' \rangle$, satisfying $\GF{}_{0\tau'}=-\GF{}_{T\tau'}$ and $\GF{}_{\tau 0}=-\GF{}_{\tau T}$. Let us go ahead and provide its solution as~\cite{Copinger:2023ctz}
\begin{align}
	\GF(\tau,\tau')&= \mathrm{sgn}(\tau - \tau')\frac{\e^{Z \dot{G}(\tau, \tau')}}{\cosh(Z)}\\
	&=\e^{-\frac{2Z}{T}(\tau-\tau')}\bigl[\mathrm{sgn}(\tau-\tau')+\tanh( Z)\big]\,.
\end{align}
With the anti-periodic Green function we again expand the logarithm in the weak field to obtain
\begin{align}
  	  \frac{\Det^{\frac{1}{2}} \big[ \hat{\partial}_{\tau} + 2e(F+f)\big]}{\Det^{\frac{1}{2}} \big[ \hat{\partial}_{\tau} + 2e F\big]}  
	&= \e^{\frac{1}{2} \Tr \ln \bigl[1 +  2e (\hat{\partial}_{\tau} + 2e F)^{-1} f\bigr]} \nonumber \\
 	   &=\e^{-\sum_{n=1}^\infty\frac{(-ie)^n}{2n}\mathcal{I}^{(n)}_F}\,,
    \label{eqDetSpin}
\end{align}
where
\begin{equation}
	\mathcal{I}^{(n)}_F=\!(-i)^n\prod_{i=1}^n 
    \int^T_0\!d\tau_i \,\tr (\GF{}_{12}\cdot f \cdot  \GF{}_{23} \cdot f \cdot ... \,\cdot \GF{}_{n1} \cdot f )\,.
\end{equation}
Written in the form above we can see that this respects the loop replacement rule transforming between scalar and spinor matter already at the level of the exponent (see \cite{ChrisRev, UsRep}), since we now acquire
\begin{equation}
	K_{\text{sp}}[F+f]=K_{\text{sp}}[F]\e^{\sum_{n=1}^\infty\frac{(-ie)^n}{2n}[\mathcal{I}^{(n)} -\mathcal{I}^{(n)}_F]  }\,.
\end{equation}
So, to determine the spinor extension to the complex scalar case one may simply replace $\mathcal{I}^{(n)}\to\mathcal{I}^{(n)}-\mathcal{I}^{(n)}_F$. This replacement rule usually holds in the worldline or string-inspired setting only after expanding the exponent to multi-linear order in the field strength polarisation vectors (and carrying out an integration by parts). It holds here at the level of the unexpanded exponent due to the projection onto the low energy limit of the scattering photons. In this sense the so-called ``Q-representation'' of the amplitudes (see \cite{ChrisRev}) is the natural one for this expansion. 

For the open line, we must also add the boundary information corresponding to the spin degrees of freedom. Indeed, the propagator can be compactly written in first quantised form as \cite{ReducibleSpinor, fppaper3}:
\begin{equation}
    \mathcal{S}^{x'x}[A +a] = [-m - i\Ds_{x'}]\mathcal{K}^{x'x}[A+a]\,,
    \label{eqSK}
\end{equation}
where here $D_{\mu} = \partial_{\mu} + ie(A + a)_{\mu}$ and we have introduced the propagator spacetime ``kernel'' for a second-order formalism of the fermion degrees of freedom (see \cite{fppaper1, fppaper2}) which has propertime representation
\begin{align}
	\mathcal{K}^{x'x}[A\tp a]=&-i\!\int^\infty_0\!\frac{dT\,\e^{-im^2T}}{(4\pi T)^2}\symb^{-1}\bigl\{ K^{x'x}_{\text{sp}}[F\tp f]\bigr\} \label{K_homo}\\
	K^{x'x}_{\text{sp}}[F\tp f]=&K_{\text{sp}}[F\tp f]\e^{-\frac{i}{4T}x_-\cdot(Z\tp z) \cdot \coth(Z\tp z)\cdot x_-}\notag\\
	& \e^{- \eta \cdot \tanh(Z+z) \cdot \eta} \,. \label{eqEta}
\end{align} 
In the above we have introduced the inverse ``symbol map'' which maps the Grassmann variables, $\eta^{\mu}$ to antisymmetrised products of $\gamma$ matrices, which produces the Dirac matrix structure of the spinor propagator:
\begin{align}
    \symb\Big\{\gamma^{[\mu_{1}}\ldots \gamma^{\mu_{n}]} \Big\} = (-i\sqrt{2})^{n}\eta^{\mu_{1}}\ldots\, \eta^{\mu_{n}}\,,
\end{align}
where antisymmetrisation includes the combinatorial factor $\frac{1}{n!}$. See~\cite{Copinger:2023ctz} for further details including explicit formulae for their operation in (3+1)-dimensions. The factor inside the inverse symbol map is generated in the worldline formalism from an operator inversion:
\begin{equation}
    \e^{-\eta \cdot \tanh(Z\tp z) \cdot \eta}  = \e^{-\eta \cdot (Z\tp z) \cdot \eta+ \frac{1}{T^{2}} \eta \cdot (Z\tp z) \cdot \oGFo(Z\tp z) \cdot (Z\tp z) \cdot \eta}\,.
    \label{eqETanE}
\end{equation}
Again, our strategy is to expand the exponent in powers of the dynamical photon field strength tensors. Doing this effects the replacement
\begin{align}
    \hspace{-1em}&\GF(\tau,\tau'|F\tp f)=\GF(\tau,\tau')+\sum_{n=1}^\infty(-ie)^n  \G_F^{(n)} (\tau,\tau')\,,\\
	\hspace{-1em}&\G_F^{(n)}(\tau,\tau')=(-i)^n\prod_{i=1}^n 
    \int^T_0\!d\tau_i \,\GF{}_{\tau 1} \cdot f \cdot \GF{}_{12}\cdot f \cdot ... \cdot\GF{}_{n\tau'}\,.
\end{align}
Using the above we can determine the expansion of the propertime kernel as
\begin{align}
	&K^{x'x}_{\text{sp}}[F\tp f]=K^{x'x}_{\text{sp}}[F]\,\e^{\sum_{n=1}^\infty  \frac{(-ie)^n}{2n}(\mathcal{I}^{(n)}-\mathcal{I}^{(n)}_F ) } \notag \\
	&\e^{\frac{i}{T^4}x_-\cdot    \bigl[ 2z \cdot \oDelo \cdot Z  +z \cdot \oDelo \cdot z  +\sum_{n=1}^\infty (-ie)^n (Z\tp z) \cdot \oDel{}^{(n)}{}^\circ \cdot (Z\tp z)\bigr] \cdot x_- }\notag\\
	&\e^{-\eta \cdot z \cdot \eta+\frac{1}{T^{2}}  \eta \cdot \bigl[ 2z \cdot \oG{}^{\circ}_F \cdot Z+z \cdot  \oG{}^{\circ}_F \cdot z +  \sum_{n=1}^\infty (-ie)^n (Z\tp z) \cdot \oG{}^{(n) \circ}_F \cdot (Z\tp z) \bigr]\cdot \eta}\,,
	\label{eqKSp}
\end{align}
now ready for an expansion onto multi-linear order.

Remarkably we see from (\ref{Kxx_expanded}), (\ref{eqKSp}) that a replacement rule persists even on the line, in the low-energy photon limit. We see that in addition to the loop replacement rule that $\mathcal{I}^{(n)}\to\mathcal{I}^{(n)}-\mathcal{I}^{(n)}_F$, we have the following accompanying rule on the line that incorporates the contribution from the end points as well:
\begin{align}
	x_-&...  \oDel{}^{(n)}{}^\circ... x_- 
	\to x_-...  \oDel{}^{(n)}{}^\circ... x_-
	-iT^{2}  \eta ...  \oG{}^{(n) \circ}_F ... \eta\,,
	\label{eqRep}
\end{align}
(for the $n = 0$ case replace $\oDel{}^{(n)}{}^\circ\to \oDelo$ and $\oG{}^{(n) \circ}_F \to \oG{}^{\circ}_F$ in (\ref{eqRep})) alongside introducing the additional $-\eta\cdot z\cdot \eta$ factor into the exponent.

As before we can readily determine the first few corrections. The first order correction is
\begin{align}\
	K^{x'x}_{\text{sp}\,1}[F]&=-\frac{ie}{2}K^{x'x}_{\text{sp}}[F]\,\bigl[\mathcal{I}^{x'x(1)}-\mathcal{I}^{x'x(1)}_F\bigr]\Big|_{\text{lin}1}\,,\label{K_sp_1def}\\
	\mathcal{I}^{x'x(1)}_F&=\mathcal{I}^{(1)}_F    +2iT\eta \tcdot f \tcdot \eta\notag\\
	&\;\;\;-\frac{2}{T^{2}}  \eta \tcdot \bigl[ 2iTf \tcdot \oG{}^{\circ}_F \tcdot Z+ Z \tcdot \oG{}^{(1) \circ}_F \tcdot Z \bigr]\tcdot \eta\,.
	\label{K_sp1}
\end{align}
And the second order correction becomes
\begin{align}
	&K^{x'x}_{\text{sp}\,2}[F]=\frac{(-ie)^2}{4}K^{x'x}_{\text{sp}}[F]\\
	&\bigl[\frac{1}{2}(\mathcal{I}^{x'x(1)}-\mathcal{I}^{x'x(1)}_F)^2+\mathcal{I}^{x'x(2)}-\mathcal{I}^{x'x(2)}_F\bigr]\Big|_{\text{lin}2}\,,\notag\\
	&\mathcal{I}^{x'x(2)}_F=\mathcal{I}^{(2)}_F \label{I_F_coor_2}\\&+
	\frac{4}{T^{2}}  \eta \tcdot \bigl[T^2 f \tcdot \oG{}^{\circ}_F \tcdot f -2i Tf \tcdot \oG{}^{(1) \circ}_F \tcdot Z
	- Z \tcdot \oG{}^{(2) \circ}_F \tcdot Z \bigr]\tcdot \eta \notag\,.
\end{align}
Again, in Appendix \ref{AppExp} we give the full expansion of (\ref{eqETanE}) suitable for projecting onto the multi-linear contribution at arbitrary multiplicity.

To convert these expressions into an expansion of the spinor propagator proper, we return to (\ref{eqSK}) which now contains two types of contributions. In analogy to \cite{fppaper1, fppaper2, Copinger:2023ctz}, we refer to the contributions in which all $N$ of the scattering photons' field strength tensors are taken from $\mathcal{K}$ as the ``leading contribution'' and the contributions in which one of the field strength tensors is taken from the leading covariant derivative as the ``subleading contributions,'' which form the distinct parts of the decomposition
\begin{equation}
    \mathcal{S}_{N}^{x'x} =- [m + i (\slashed{\partial}_{x'} + ie\slashed{A}(x'))]\mathcal{K}_{N}^{x'x} + e\sum_{i = 1}^{N} \slashed{a}_{i}(x') \mathcal{K}_{N-1}^{x'x}\,,
\end{equation}
where in the $\mathcal{K}_{N-1}$ we omit photon $i$. In vacuum \cite{fppaper1} and in a plane wave background \cite{Copinger:2023ctz}, the subleading terms do not contribute to scattering amplitudes, as they do not survive the LSZ reduction which extracts them from $\mathcal{S}_{N}$, once matter legs go on-shell. In the homogeneous fields considered here, LSZ is troublesome as the fields do not vanish (the gauge potential does not become constant) asymptotically. Rather than performing LSZ, one may truncate the external lines of the coordinate propagator with asymptotic in and out wavefunctions in a constant field to arrive at an amplitude as is done in the standard approach to strong field QED. Therefore we halt at the above form in the coordinate representation (but see the momentum space formulae below).

\subsection{Worldline photon vertices}
For completeness we also show how the contribution from the spin factor can be obtained from low-energy photon vertex operators in the worldline approach, which are now augmented by the presence of the spin degrees of freedom. We do this for the open line, since we can always obtain the corresponding integrand for the one-loop amplitudes by fixing $x_{-} = 0$ (and change the measure on the $T$ integral).

In analogy with the scalar case, for the spinor case we start from the path integral expanded about the straight line
\begin{align}\label{eqKz2}
	&\mathcal{K}_{N}^{x^\prime x}[A]= (-ie)^N\int_{0}^{\infty}dT\,\e^{-im^2T}
	\e^{-i\frac{x_{-}^{2}}{4T}} 	\notag\\
	&\int_{\mathrm{DBC}}\hspace{-0.75em}\mathcal{D}q(\tau)\,\e^{-i\int_{0}^{T}d\tau\,\bigl(\frac{\dot{q}^2}{4}+\frac{e}{2}q\cdot F \cdot \dot{q} - \frac{e}{T}\,x_{-} \cdot F \cdot q\bigr)}\notag\\
	&2^{-\frac{D}{2}}\,\symb^{-1}\Bigl\{ \!\int_{ABC}\!\mathcal{D}\psi(\tau)\e^{iS_{\text{sp}}[F]}\,\prod_{i=1}^{N}V_{\eta}^{x'x}[f_i] \Bigr\},
\end{align}
where $ABC$ stands for anti-periodic boundary conditions. The corresponding action in the path integral over Grassmann variables producing the Feynman spin factor reads $S_{\text{sp}}[F]=i\int_{0}^{T}d\tau\bigl(\frac{1}{2}\psi\cdot\dot{\psi}+e\psi_\eta \cdot F \cdot \psi_\eta \bigr)$, and for brevity we write $\psi_\eta\coloneqq\psi+\eta$. In the above the low-energy spinor vertex operator is obtained by expanding the spinor vertex operator given in \cite{fppaper3, Copinger:2023ctz, Ahmadiniaz:2017rrk} to linear order in the photon momentum (in fact the spin-coupling to the line already starts at this order). Explicitly this leads to
\begin{equation}
	V_\eta[f_i] \coloneqq V[f_i]-i\int_{0}^{T}\!d\tau_i\,\psi_\eta(\tau_i)\cdot f_i\cdot\psi_\eta(\tau_i)\,.
\end{equation}
Similarly to the scalar case, the spinor propagator kernel can be written as a worldline expectation value as
\begin{align}
	\mathcal{K}_{N}^{x^{\prime}x}[A] &= -i\int_{0}^{\infty}   \frac{dT\e^{-im^2T}}{(4\pi T)^2}  \symb^{-1} K^{x'x}_{\text{sp}\,N}[F]\,,\\
	K^{x'x}_{\text{sp}\,N}[F]&=(-ie)^NK[F]\e^{-i\frac{x_{-}^{2}}{4T}}\notag\\
	&\times\Big\langle \e^{\frac{ie}{T}\int_{0}^{T}d\tau\,x_{-}\cdot F\cdot q}  \prod_{i=1}^{N}V_\eta[f_i]\Big\rangle_{\mathrm{DBC},\, \mathrm{ABC}}\,, \label{eqKzVtxOp}
\end{align}
where we write for the anti-periodic Grassmann expectation value
\begin{equation}
	\langle \mathcal{O} \rangle_{\mathrm{ABC}}= \frac{\int_{ABC}\!\mathcal{D}\psi(\tau)\,\mathcal{O}\e^{iS_{\text{sp}}[F]}}{\int_{ABC}\!\mathcal{D}\psi(\tau)\,\e^{iS_{\text{sp}}[F]}}\,,
\end{equation}
with the basic rule for Wick contractions of Grassmann fields being $\langle \psi^{\mu}(\tau)\psi^{\nu}(\tau^{\prime}) \rangle = \frac{1}{2}\GF^{\mu\nu}(\tau,\tau^{\prime})$.

We again focus on the one photon amplitude in the spinor case, choosing $N=1$ in the expectation value showed above, but this time we notice that, since scalar and spinor variables do not mix, it is sufficient to focus only on Wick contractions for the product of vertex operators with Grassmann variables (the scalar part has been already worked in the previous section). Hence, it is sufficient to work with the spinor part: $\langle \int_{0}^{T}d\tau_1\,\psi_\eta(\tau_1)\cdot f_1\cdot \psi_\eta(\tau_1)\rangle_{\mathrm{A}} $. Computing the expectation value we arrive at the following expression for the first order correction as $K^{x'x}_{\text{sp}\,1}[F]=-\frac{ie}{2}K^{x'x}_{\text{sp}}[F][\mathcal{I}^{x'x(1)}_1-\mathcal{I}^{x'x(1)}_{F1}]$ with
\begin{equation}
	\mathcal{I}^{x'x(1)}_{F1}=-i\int_{0}^{T}\!d\tau_1\Big[\tr(\GF{}_{11} \cdot f_{1})-2 \eta \cdot \bar{\mathcal{G}}^T_1 \cdot f_{1} \cdot \bar{\mathcal{G}}_1 \cdot \eta \Big]\label{eqSpinN1}\,,
\end{equation}
where we have written $\bar{\mathcal{G}}_i\coloneqq g- T^{-1}\GFo{}_{i} \cdot Z$ and $\bar{\mathcal{G}}^T_i=g - T^{-1}Z \cdot \oGF{}_{i}$ with $g$ the flat Minkowski metric. And thus confirming the first order expression found in~\eqref{K_sp1}, along with $\mathcal{I}^{x'x(1)}$ already confirmed in~\eqref{eqVtxOpExpan}.

Confirmation of the $N=2$ case proceeds similarly. Namely we find upon gathering the various Wick contractions that the following structure emerges for the worldline kernel
\begin{align}
	&K^{x'x}_{\text{sp}\,2}[F]=\frac{(-ie)^2}{4}K^{x'x}_{\text{sp}}[F]\bigl[\mathcal{I}^{x'x(2)}_{12}+\mathcal{I}^{x'x(2)}_{21}-\mathcal{I}^{x'x(2)}_{F12} \notag\\
	&-\mathcal{I}^{x'x(2)}_{F21}+(\mathcal{I}^{x'x(1)}_1-\mathcal{I}^{x'x(1)}_{F1})(\mathcal{I}^{x'x(1)}_2-\mathcal{I}^{x'x(1)}_{F2})\bigr]\,,
\label{K_line_2}
\end{align}
where now the only new component to evaluate reads
\begin{align}
    \mathcal{I}^{x'x(2)}_{F12}&=\int^T_0\!d\tau_1d\tau_2 \Bigl[-\mathrm{tr}(\GF{}_{12} \cdot f_2 \cdot \GF{}_{21} \cdot f_1)\notag\\
    &+4\eta \cdot \bar{\mathcal{G}}^{T}_1 \cdot f_1 \cdot \GF{}_{12} \cdot f_2 \cdot \bar{\mathcal{G}}_2 \cdot \eta\Bigr]\,,
\end{align}
in agreement with~\eqref{I_F_coor_2}. As anticipated, we see that with the worldline path integral approach the replacement rule from scalar to spinor low-energy $N$-photon coupling insertion is manifest:  for $N=2$ above this is $\mathcal{I}^{x'x(2)}_{12}\to \mathcal{I}^{x'x(2)}_{12}-\mathcal{I}^{x'x(2)}_{F12}$ and likewise for $1\leftrightharpoons 2$, and in $\mathcal{I}^{x'x(1)}_1\to \mathcal{I}^{x'x(1)}_1-\mathcal{I}^{x'x(1)}_{F1}$ and likewise for $1\to 2$, going from scalar in~\eqref{eqK2xx} to spinor, above in~\eqref{K_line_2}.

\subsection{Parameter integrals and matrix expansion}
The formula provided in Section \ref{sec:matrix} can be used here too so as to reduce the evaluation of $\mathcal{I}_{F}^{(n)}$ to cycle integrals of the form (but see \cite{Ahmadiniaz:2023jwd} for a tensor decomposition of $\mathcal{I}_{F}^{(n)}$)
\begin{equation}
	\mathcal{J}_{F}^{(n)} = \prod_{i=1}^{n} \int_{0}^{T}d\tau_{i} \, \mathcal{G}_{F12} \cdot \mathcal{G}_{F23} \cdot \ldots \cdot \mathcal{G}_{Fn1}\,,
\end{equation}
which is proved by induction with only necessary changes in the Green function. Evaluation of this integral is again achieved by derivation:
\begin{align}
	\mathcal{J}_{F}^{(n)} &= \int_{0}^{T}d\tau\, \big\langle \tau \big| \frac{2^{n}}{(\hat{\partial}_{\tau} + 2eF)^{n}} \big| \tau \big\rangle\\
	&= \frac{T}{(n-1)!}\Bigl(\frac{\partial_{e}}{F}\Bigr)^{n-1} \tanh(eFT)\,,
	\label{eqJF}
\end{align}
with $\mathcal{J}_{F}^{(1)} = T\tanh(eFT)$. An explicit formula for the above derivative is given in Appendix \ref{AppTraces}.

In the same way as was done for the complex scalar field case we may also perform a matrix expansion. Let us write the expanded worldline kernel, then, in the following form
\begin{align}
	&K^{x'x}_{\text{sp}}[F\tp f]=K^{x'x}_{\text{sp}}[F]\e^{\sum_{n=1}^\infty\frac{(-ie)^n}{2n}[\mathcal{I}^{(n)}-\mathcal{I}^{(n)}_F]}\\
&\e^{\sum_{n=1}^\infty[ -\frac{i}{4T}\frac{f\cdot \partial_F}{n!} x_-\cdot Z \cdot \coth(Z)\cdot x_- - \frac{f\cdot \partial_F}{n!} \eta \cdot \tanh(Z) \cdot \eta]}\notag\,,
\end{align}
which immediately identifies
\begin{equation}
	\mathcal{I}^{(n)}_F=-\frac{1}{(n-1)!} \Bigl(\frac{f\cdot \partial_F}{-ie}\Bigr)^n \tr \ln[\cosh (Z)]\,.
\end{equation}
To take the matrix derivative for the $\eta$ dependent portion we must first decompose the term into its available Lorentz invariants. Using projection operators we determine that
\begin{align}
	&\eta\cdot \tanh(Z)\cdot \eta = \eta\cdot Z\cdot \eta \frac{1}{2}\mathrm{tr}\Bigl[\frac{\widetilde{Z}^{2}}{Z^{2}+\widetilde{Z}^{2}} \cdot \frac{\tanh(Z)}{Z}\Bigr]\notag\\
	&+(\eta\cdot Z^3\cdot \eta)\frac{1}{2}\mathrm{tr}\Bigl[\frac{1}{Z^{2}+\widetilde{Z}^{2}}\cdot \frac{\tanh(Z)}{Z}\Bigr]\,.
\end{align}

Let us show the first order correction from the above, where $K_{\text{sp}\,1}^{x'x}[F]$ is given by~\eqref{K_sp_1def}. First let us evaluate $\mathcal{I}^{(1)}_F$ in compact form as
\begin{equation}
	\mathcal{I}^{(1)}_F=-iT\mathrm{tr}[\tanh(Z) \cdot f]\,.
\end{equation}
Then we may evaluate entire first order contribution as
\begin{align}
	\hspace{-0.75em}&\mathcal{I}^{x'x(1)}_F=\mathcal{I}^{(1)}_F+\frac{2i}{e}f\cdot\partial_F \eta\cdot\tanh(Z)\cdot \eta\\
	\hspace{-0.75em}&=\mathcal{I}^{(1)}_F  +i\eta\tcdot Z\tcdot \eta \,\mathrm{tr}\Bigl[\frac{T\widetilde{Z}^{2}\tanh(Z)}{(Z^{2}+\widetilde{Z}^{2})^{2}Z}\Bigl(2\frac{Z^{2}}{\widetilde{Z}}\widetilde{f}-\frac{3Z^{2}+\widetilde{Z}^{2}}{Z}f\notag\\
	\hspace{-0.75em}&+\frac{2(Z^{2}+\widetilde{Z}^{2})}{\sinh(2Z)}f\Bigr)\Bigr]+i\eta\tcdot (2Z^2f\tp ZfZ)\tcdot \eta\mathrm{tr}\Bigl[\frac{T\tanh(Z)}{(Z^{2}+\widetilde{Z}^{2})Z}\Bigr]\notag\\
	\hspace{-1em}&-i(\eta\tcdot Z^{3}\tcdot \eta)\mathrm{tr}\Bigl[\frac{T}{(Z^{2}+\widetilde{Z}^{2})^{2}}\frac{\tanh(Z)}{Z}\Bigl(2\widetilde{Z}\widetilde{f}+\frac{3Z^{2}+\widetilde{Z}^{2}}{Z}f\notag\\
	\hspace{-0.75em}&-\frac{2(Z^{2}+\widetilde{Z}^{2})}{\sinh(2Z)}f\Bigr)\Bigr]+i\eta\tcdot f\tcdot \eta \mathrm{tr}\,\Bigl[\frac{T\widetilde{Z}^{2}\tanh(Z)}{(Z^{2}+\widetilde{Z}^{2})Z}\Bigr]
	\,,
\end{align}
which also gives an explicit form for the parameter integration. Higher orders may be found through recursively applying the derivative operator, taking the steps outlined in  Sec.~\ref{sec:matrix}. And also, similar to the scalar propagator case, to arrive at the above expression from~\eqref{eqSpinN1}, one should compute the parameter integrals using the formulae presented above, and then use the projection operators to express the result terms of $\eta\cdot Z\cdot \eta$ and $\eta\cdot Z^3\cdot \eta$.

%%%%%%%%%%%%%%%%%%%%%%%%%%%%%%%%%%%%%
\section{Momentum space formulae}
\label{secmom}
For the purpose of calculating scattering amplitudes with external matter legs, it is of course more convenient to work with the correlation functions in momentum space. In the worldline formalism this is usually achieved by first computing the path integral with vertex operator insertions to obtain a Master Formula in position space, which is then Fourier transformed to momentum space (see, for example \cite{UsRep, ChrisRev, fppaper1, fppaper2, Copinger:2023ctz, Ahmad_2017}). Since the scattering photons' field, $a(x)$, appears already in the path integral exponent, for plane wave states at arbitrary energies the transition to momentum space is not immediate. 

In the present case, however, the projection onto the low-energy limit of the photon scattering amplitudes allows this ordering to be reversed, so that we can \textit{first} carry out the Fourier transform, thereby obtaining path integral representations of the dressed propagator directly in momentum space. 

Let's begin with the first approach -- we take the Fourier transform of the coordinate space propagator directly, \textit{viz.} $\mathcal{D}^{p'p}[A+a]=\int d^4x'd^4x\, \e^{i(p'\cdot x'-p\cdot x)}\mathcal{D}^{x'x}[A+a]$. As before we need only examine the case of superposed homogeneous fields and then later implement the plane wave expansion of the scattering photons and subsequent projection onto the multi-linear contribution. Starting with \eqref{eqDz2} without vertices ($N = 0$) we compute the Fourier transform before integrating out the fluctuations. After changing variables to $x_{+} = \frac{1}{2}(x' + x)$ and $x_{-} = x'-x$ and computing the Gaussian integrals we find (using $\hat{\delta}_{p'p}=2\pi\delta(p'-p)$)
\begin{align}\label{Dpp_1}
	&\mathcal{D}^{p^\prime p}[A\tp a] =-i \hat{\delta}^4_{p'p} \!\int_{0}^{\infty}\hspace{-0.5em}dT\,\e^{-im^2T}K^{pp}[F\tp f]\,,\\
	&K^{pp}[F\tp f] = (4\pi T)^2 \int_{\mathrm{DBC}}\hspace{-1em}\mathcal{D}q(\tau)\e^{-i\int_{0}^{T}\!d\tau[\frac{\dot{q}^2}{4}+\frac{e}{2}q\cdot(F\tp f)\cdot\dot{q}]}\notag\\
	\times &\e^{ip^2T+i\int_{0}^{T}\!d\tau\bigl[2e p\cdot (F\tp f)\cdot q-\frac{e^2}{T}\int_{0}^{T}\!d\tau^\prime\,q(\tau)\cdot (F\tp f)^2\cdot q(\tau^\prime)\bigr]}\,.
	\label{eqDpp}
\end{align}
Unusually for worldline calculations, this path integral appears with a new non-local kinetic term -- note that we are still integrating over trajectories with homogeneous DBC in coordinate space with the momenta appearing as parameters under the integral. Of course we may also perform the Fourier transform after having integrated over trajectories (i.e. from (\ref{D_homo})) leading to the known result~\cite{Ahmad_2017}
\begin{align}
	K^{pp}[F\tp f]&= \e^{ip\cdot \frac{T\tanh(Z\tp z)}{Z\tp z}\cdot p} \det{}^{\!-1/2}[\cosh(Z+z)]\,.
	\label{eqDpp_soln}
\end{align}
We shall show that (\ref{eqDpp}) reproduces this in the following section and develop a matrix expansion of (\ref{eqDpp_soln}) along similar lines to our above work in position space. At this point we remind the reader that the momentum space propagator can take on different functional forms, connected to gauge choices in the gauge covariant coordinate space propagator (the holonomy factor interacts non-trivially with the Fourier integrals). We explore this tangential point in the Appendix~\ref{appFS}.

For the spinor propagator ``kernel'' in momentum space, i.e. $\mathcal{K}^{p^\prime p}[A\tp a]=\int d^4x\int d^4x^\prime\,\e^{i(p'\cdot x'-p\cdot x)}\,\mathcal{K}^{x^\prime x}[A\tp a]$,  we will have by analogy with the scalar case (since the spinor degrees of freedom do not couple to the initial and final spacetime points):
\begin{align}
	&\mathcal{K}^{p^\prime p}[A\tp a] =-i \hat{\delta}^4_{p'p} \!\int_{0}^{\infty}\hspace{-0.5em}dT\,\e^{-im^2T}\symb^{-1}\{K^{pp}_{\text{sp}}[F\tp f]\}\,, \label{eqKpp}\\
	&K_{\text{sp}}^{pp}[F\tp f] \coloneqq (4\pi T)^2 \int_{\mathrm{DBC}}\hspace{-1em}\mathcal{D}q(\tau)\e^{-i\int_{0}^{T}\!d\tau[\frac{\dot{q}^2}{4}+\frac{e}{2}q\cdot(F\tp f)\cdot\dot{q}]}\notag\\
	&\e^{ip^2T+i\int_{0}^{T}\!d\tau\bigl[2e p\cdot (F\tp f)\cdot q-\frac{e^2}{T}\int_{0}^{T}\!d\tau^\prime\,q(\tau)\cdot (F\tp f)^2\cdot q(\tau^\prime)\bigr]}\notag\\
&\times 2^{-\frac{D}{2}}\,\int_{A}\!\mathcal{D}\psi(\tau)\e^{iS_{\text{sp}}[F]}\,.
	\label{eqKppSp}
\end{align}
Therefore in this section we confine our attention to the scalar case, noting that the resulting formulae  for spinors follow with minimal modification.

To compute this path integral we need to study the non-diagonal operator $\hat{\chi}$ associated to the non-local kinetic term, defined with matrix elements
\begin{align}
	\mathcal{\chi}(\tau, \tau' | F) &\coloneqq \big\langle \tau \big| \hat{\partial}_{\tau}^{2} - 2eF\hat{\partial}_{\tau} - \frac{4e^{2}}{T} \hat{F}^{2} \big| \tau' \big\rangle \\
	&\coloneqq \big(\partial_{\tau}^{2} - 2eF\partial_{\tau}\big)\delta(\tau - \tau') - \frac{4e^{2}}{T} F^{2}\,.
\end{align}
The inverse of this operator, $\hat{\Xi} = \hat{\chi}^{-1}$, satisfies the ``position-space'' Green equation (there are no zero modes)
\begin{equation}
	\big(\partial_{\tau}^{2} - 2eF\partial_{\tau}\big)\Xi(\tau, \tau') - \frac{4e^{2}}{T} F^{2}  {}^{\circ}\Xi(\tau') = \delta(\tau - \tau')\,,
	\label{eqXiDE}
\end{equation}
where again we compactly write for the background field case $\Xi(\tau, \tau'|F)\eqqcolon \Xi(\tau,\tau')$, and borrow the derivative and integral notation introduced for $\Del$ below (\ref{Del_1}). The Green equation is solved by (see Appendix \ref{AppOps})
\begin{align}
	\Xi(\tau, \tau')&= \Del{}(\tau, \tau') + \frac{4}{T^{3}} \Delo{}(\tau) \cdot \frac{Z^{2}}{ 1 -\frac{4}{T^{3}} \oDelo \cdot Z^{2}} \cdot \oDel(\tau')\\
	&= \Del(\tau, \tau') + \frac{4}{T^{3}} \Delo(\tau) \cdot Z \cdot \tanh(Z) \cdot \oDel(\tau')\,.
	\label{eqXi}
\end{align}
We also require the path integral normalisation obtained from the functional determinant (also in Appendix \ref{AppOps})
\begin{align}
	 \widetilde{K}[F] &= \Det{}^{\!-\frac{1}{2}}\Big[  \hat{\partial}_{\tau}^{2} - 2eF\hat{\partial}_{\tau} - \frac{4e^{2}}{T} \hat{\rm F}^{2} \Big]\\
	 &= \frac{-i}{(4\pi T)^2}\det{}^{\!-\frac{1}{2}}(\cosh(Z) ) \,.
	\label{eqDetXi}
\end{align}

With these results in hand, we now employ the path integral representation of the momentum space propagator to develop expansions in $f$ of the background-dressed propagators analogous to those in sections \ref{secScalar} and \ref{secSpinor}. Let us compute the Gaussian path integral in~\eqref{eqDpp} to find
\begin{equation}
	K^{pp}[F\tp f]\eqqcolon\widetilde{K}[F\tp f]\e^{ip^2T+i\frac{4}{T^2} p\cdot (Z\tp z) \cdot \oXi{}^{\circ} \cdot (Z\tp z) \cdot p }\,,
	\label{eqKppF}
\end{equation}
where here $\Xi(\tau, \tau') = \Xi(\tau, \tau' | Z\tp z)$. This already rederives (\ref{eqDpp_soln}) once we use the formulae for integrals of $\Xi$ given in Appendix \ref{AppOps}.

As before we may now formally exponentiate the functional determinant factor
\begin{align}
	\widetilde{K}[F\tp f]&=\widetilde{K}[F]\e^{-\frac{1}{2} \Tr \ln \big\{ 1 -\hat{\Xi} [2e f\hat{\partial}_{\tau} + \frac{4e^2}{T} (\hat{F} \hat{f} + \hat{f}  \hat{F} + \hat{f}^{2}) ]\big\} }\,,
\end{align}
where at this stage $\hat{\Xi} = \hat{\Xi}(F)$ depends only on the background field. The presence of terms both linear and quadratic in $F$ make a closed formula for the expansion in the low energy field strength tensors more challenging now; however we can provide results order by order.

Let us first expand the functional determinant. By expanding the logarithm in the weak field and introducing suitable resolutions of the identity we can produce the contribution corresponding to insertions of scattering photons. For example, at $\mathcal{O}(f)$ in the exponent we get the following (reusing the notation $\Xi_{\tau \tau'} = \Xi(\tau, \tau')$)
\begin{equation}\label{mom_det1}
	\widetilde{K}[F\tp f]=\widetilde{K}[F]\e^{e\,\mathrm{tr}[  \int^T_0\!d\tau (f \cdot \dXi{}_{\tau\tau}) + \frac{4}{T^2} f \cdot Z \cdot \oXio]+\mathcal{O}(f^2)}\,.
\end{equation}
Next we expand the momentum dependent exponent of (\ref{eqKppF}). This involves a geometric series expansion in $f$ for the Green function $\hat{\Xi}(F + f)$ and a simple binomial expansion for the polynomial operators sandwiching the various $\Xi(F)$. We again restrict attention to the first order correction. For the Green function this is
\begin{align}
	\Xi(\tau,\tau'|F\tp f) &= \Xi_{\tau\tau'} +2e\int^T_0\!d\sigma\, \Xi_{\tau\sigma} \cdot f \cdot \dXi_{\sigma\tau'}\notag\\
		&+\frac{4e}{T^2} \Xio{}_{\tau} \cdot [f \cdot Z+Z \cdot f ] \cdot\oXi_{\tau'}+\mathcal{O}(f^2)\,.
\end{align}
Then using the above along with the expansion for the determinant to $\mathcal{O}(f)$ in~\eqref{mom_det1} we can readily find the $N=1$ contribution as
\begin{align}\label{Kpp1}
	K^{pp}_1[F]&=eK^{pp}[F]\Bigl\{\mathrm{tr}\Bigl[  \frac{4}{T^2} f \cdot Z \cdot \oXio+ \int^T_0\!d\tau f \cdot \dXi{}_{\tau\tau} \Bigr] \notag\\
	&+\frac{8i}{T^2}p\tcdot \Bigl[ Tf \tcdot Z \tcdot \oXio +  \int^T_0\!d\sigma Z \tcdot \oXi_{\sigma} \tcdot f \tcdot \dXio{}_{\sigma} \tcdot Z  \notag\\
&+\frac{4}{T^2}Z \tcdot \oXio \tcdot f \tcdot Z \tcdot \oXio \tcdot Z \Bigr]\tcdot p 
	\Bigr\}\,.
\end{align}
We shall verify this result by a worldline calculation below (one can, of course, compare this with a direct Fourier transform of the position space result, (\ref{K_1xx}), where agreement is found after substituting the explicit form of $\Xi$ from (\ref{eqXi}) and using its integral properties in (\eqref{eqXiprop}).

\subsection{Momentum space vertex operator}
We are now well-situated to derive the vertex operator for low-energy photon scattering amplitudes already in momentum space.  In fact, to the best of the authors' knowledge, this is the first time that a momentum space vertex operator has been derived on the worldline \textit{using Dirichlet boundary conditions} (the obstacle to doing this for finite external photon momenta is that the worldline action ceases to be quadratic in the path integral variable); a momentum space vertex operator has been considered for a path integral with Neumann boundary conditions in~\cite{Bonezzi:2025iza}. Let us remark that the low-energy limit applies only to the photons and not to the matter momenta, where we can take $p$ to be large.

To derive the vertex operator let us start from $K^{pp}[F+f]$  before taking the path integral,~\eqref{eqDpp}. To derive the photon vertex operators we wish to consider $f = \sum_{i = 1}^{N}f_{i}$ and take the part multi-linear in the $f_{i}$. To achieve this we first linearise the term in the action that is quadratic in $f$, we do this by introducing an auxiliary
field, $X(\tau)$, as in \cite{Yukawa} so we can rewrite\footnote{We observe that, in this case, a valid integral representation could be achieved by using an auxiliary \textit{variable}, $X$ (say). However, we ascribe to the worldline philosophy that all degrees of freedom should be represented by path integrals on equal terms.}
\begin{equation}
	\hspace{-1em}\e^{-i\frac{e^2}{T}\int_{0}^{T}\!d\tau \int_{0}^{T}d\tau' \, q(\tau) \cdot f^2\cdot q(\tau')} = \int \mathcal{D}X(\tau)\,\e^{-i\int_{0}^{T} \!d\tau(\frac{X^2}{4}+\frac{e}{T}X \cdot f \cdot Q)},
\end{equation}
where $Q\coloneqq\int^T_0\!d\tau\, q(\tau)$, the Wick contractions for the auxiliary field are defined by $\langle X(\tau)X(\tau^\prime)\rangle=2i\delta(\tau-\tau^\prime)$ and its free path integral is normalised to unity. If we use the above equation in the momentum space worldline kernel, it is rewritten in terms of an expectation value of new, momentum space vertex operators
\begin{align}
	K_N^{pp}[F\tp f] &=(-ie)^N \widetilde{K}[F]\e^{ip^2T}\\
	&\Bigl \langle \e^{2ie\int_{0}^{T}\!d\tau  p\cdot F\cdot q} \, \prod_{i=1}^N V^p[f_i]
	\Bigr\rangle_{\text{DBC, X}}\,,\notag
\end{align}
where the expectation values are taken with respect to both $q(\tau)$ and $X(\tau)$. Here the low-energy vertices read
\begin{align}
	\nonumber V^{p}[f_i]:=& \int_{0}^{T} \! d\tau_i\; \Big(\frac{1}{2}q_i\cdot f_i\cdot\dot{q}_i - 2\,p\cdot f_i\cdot q_i \\
	&+ \frac{1}{T}X\cdot f_i\cdot Q + \frac{2}{T^2}q_i\cdot f_i  \cdot Z \cdot Q \Big).
\end{align}
Of course, the zero background field limit, $F \to 0$, yields a new worldline path integral representation of the open line, low-energy scattering amplitudes in vacuum.

Considering the basic Wick contraction of the fields in momentum space $\langle q^{\mu}(\tau)q^{\nu}(\tau^{\prime}) \rangle = 2i\Xi^{\mu\nu}(\tau,\tau^{\prime})$, at order $N = 1$ the expectation value yields
\begin{align}\label{K1spin}
	\hspace{-0.5em}&K^{pp}_1[F]=-ieK^{pp}[F]\int_{0}^{T}\!d\tau_1\,\Big[ i \tr(\dXi_{11} \cdot f_{1})\\
	\hspace{-0.5em}& -\frac{8}{T^2}\,p\cdot \bigl( Tf_{1} \cdot \Xio_{1} \cdot Z
	\nonumber +Z \cdot \oXi_{1} \cdot f_{1} \cdot \dXio_{1} \cdot Z \bigr)\cdot p \notag\\
	\hspace{-0.5em}&+\frac{4}{T^{2}}\int d\tau_{2}\Bigl(i\mathrm{tr}(Z \tcdot f_1 \tcdot \Xi_{12})-\frac{8}{T^{2}}p\tcdot Z \tcdot \oXi{}_1 \tcdot f_1Z \tcdot \Xio{}_2 \tcdot Z \tcdot p\Bigr)\Big]\notag\,,
\end{align}
in agreement with~\eqref{Kpp1} and, therefore, the Fourier transform of \eqref{eqVtxOpExpan}, which would be the usual way of arriving at momentum space formulae in the worldline formalism.

For the spinor case, referring to (\ref{eqKppSp}), it is clear that we only need to append the spin degrees of freedom contractions inside the symbol map. These were, however, already worked out in (\ref{eqSpinN1}), so it remains to take the linear part of either the scalar or the spinor contributions. At higher order, (e.g. $N = 2$) these contributions can mix: there will be a second order contribution from the spin part, a second order contribution from the scalar part, and the product of the first order contributions from each part. The formalism presented here produces all such contributions via a worldline path integral with vertex operator insertions.

\subsection{Matrix expansion}
The matrix expansion technique may be applied to the momentum space propagator here without difficulty. Let us see how this expansion looks like for the worldline kernel (again here in this subsection we restrict our attention to the case of scalars). Let us perform the Taylor expansion of the exponent in~\eqref{eqDpp_soln} to find
\begin{equation}
\hspace{-0.5em}	K^{pp}[F\tp f]=K^{pp}[F]\e^{\sum_{n=1}^\infty \!\frac{(f\cdot \partial_F)^n}{n!}\big\{ ip\cdot \frac{T\tanh(Z)}{Z}\cdot p -\frac{1}{2}\mathrm{tr}\ln [\cosh(Z)] \big\}}\,.
\end{equation}
The expansion of the functional determinant factor is essentially the same as for the spin contribution to the effective action, so we do not require any new results here. However to do the matrix derivative of the momentum dependent term we will require its Lorentz invariant decomposition for factors with $p$. This can, of course, be found using projection operators, as in Sec.~\ref{sec:matrix}, as
\begin{align}
	&p\cdot \frac{\tanh(Z)}{Z}\cdot p = p^{2}\frac{1}{2}\mathrm{tr}\Bigl[\frac{\widetilde{Z}^{2}}{Z^{2}+\widetilde{Z}^{2}}\frac{\tanh(Z)}{Z}\Bigr]\notag\\
	&+(p\cdot Z^{2}\cdot p)\frac{1}{2}\mathrm{tr}\Bigl[\frac{1}{Z^{2}+\widetilde{Z}^{2}}\frac{\tanh(Z)}{Z}\Bigr]\,.
\end{align}

So, for the $N = 1$ insertion, performing the matrix derivatives one may readily find an alternative representation of (\ref{K1spin})
\begin{align}
	&K^{pp}_1[F]=\frac{-ieT}{2}K^{pp}[F]\Bigl\{-i\mathrm{tr}\big[\tanh(Z) \cdot f\big]\\
	&-p^{2}\mathrm{tr}\Bigl[\frac{T\widetilde{Z}^{2}}{(Z^{2}+\widetilde{Z}^{2})^{2}}\frac{\tanh(Z)}{Z}\Bigl(2\frac{Z^{2}}{\widetilde{Z}}\widetilde{f}-\frac{3Z^{2}+\widetilde{Z}^{2}}{Z}f\notag\\
	&+\frac{2(Z^{2}+\widetilde{Z}^{2})}{\sinh(2Z)}f\Bigr)\Bigr]-2\big(p\cdot Z \cdot f\cdot p\big)\mathrm{tr}\Bigl[\frac{T}{Z^{2}+\widetilde{Z}^{2}}\frac{\tanh(Z)}{Z}\Bigr]\notag\\
	&+(p\cdot Z^{2}\cdot p)\mathrm{tr}\Bigl[\frac{T}{(Z^{2}+\widetilde{Z}^{2})^{2}}\frac{\tanh(Z)}{Z}\Bigl(2\widetilde{Z}\widetilde{f}+\frac{3Z^{2}+\widetilde{Z}^{2}}{Z}f\notag\\
	&-\frac{2(Z^{2}+\widetilde{Z}^{2})}{\sinh(2Z)}f\Bigr)\Bigr]
	\Bigr\}\,,
	\end{align}
with higher orders obtainable by performing the steps as outlined in Sec.~\ref{sec:matrix}. The advantage of this approach is that the parameter integrals do not appear. Lastly, as before one may arrive at the above expression from~\eqref{K1spin} by computing the parameter integrals, and using projection operators to rewrite the result in terms of $p^2$, $p\cdot Z\cdot f\cdot p$, and $p\cdot Z^2 \cdot p$.

Finally, let us remark that the formulae derived for open lines here correspond to photon-dressed correlation functions. To convert these into scattering amplitudes is not immediate due to the difficulty of defining asymptotic states for an LSZ reduction in a constant background. To compare to existing literature in the standard formalism one could amputate with \textit{free} propagators and the known form of the (Volkov) wavefunctions for external matter in a crossed constant field.

\pagebreak
%%%%%%%%%%%%%%%%%%%%%%%%%%%%%%%%%%%%%%%%%%%
\section{Conclusions}
In this article, we have derived three different ways to calculate the effective action and propagator in a homogeneous background electromagnetic field additionally dressed by scattering photons in the low energy limit. These are associated to one-loop (respectively tree-level) photon scattering amplitudes in the constant background. We did this for both scalar and spinor matter in SFQED, and in both coordinate space and momentum space on the line. We showed that in the low-energy photon limit both the one-loop effective action and propagator can be related to their counterparts dressed by a superposition of a constant background field and constant field strengths produced by the scattering photons. The study of the scattering amplitudes is thus reduced to a multi-linear expansion of the dressed effective action or propagator in the scattering photons.

For each case we developed a functional expansion based on series representations of matrix elements associated to first quantised operators that arise in the worldline approach to QFT. This approach recovered existing results for the low energy one-loop amplitudes for scalars and spinors \cite{Ahmadiniaz:2023jwd} and extends them to arbitrary background configurations. For the propagator, we derived new compact Master Formulae that include all possible contributions from the expansion, which includes propagation endpoint dependence. En route we also discover a replacement rule from scalar to spinor quantities that extended \textit{to the open line} (that holds in the low-energy limit). To treat the momentum space propagator case, we developed a novel approach using a worldline path integral over trajectories (with Dirichlet boundary conditions) for the momentum space correlation function. 

We also corroborated the expansions (up to order $N=2$) through a direct computation of the effective action and propagator using the worldline formalism, with scattering photons represented by vertex operator insertions under the path integrals. Last, we presented a different but equivalent way to evaluate the expansion recursively using standard calculus, deriving formulae for a Taylor expansion of the path integral determinant, obviating the difficulty of evaluating parameter integrals for an arbitrary background field. Of particular significance we show that it is possible to evaluate the worldline parameter integrals in an arbitrary homogeneous field by reducing them to familiar cycle integrals; this required new identities and Lorentz invariants obtained in terms of the field strength tensors and their duals. 

There are several avenues for future work. The closed-form evaluation of the parameter integrals for  arbitrary multiplicity scattering off matter lines would be a naturally mathematical line of inquiry generalising existing studies of worldline integrals in vacuum \cite{Ahmadiniaz:2024rvi, Ahmadiniaz:2022yam}. Closed form expressions could provide access to important information on the behaviour of particle propagation and cascades in a strong homogeneous background field. Selecting a specific configuration for the electromagnetic field (parallel electric and magnetic fields or constant cross fields) relative to on-shell scattering photons could help simplify the tensor structure of the results presented here, and shed further light on the problem of scattering in this background \cite{MishaLow}. A merit of our low-energy approach is that higher-multiplicity (e.g. $\geq N=4$) tree-level scattering quantities are readily available -- as opposed to the difficulty of carrying out such calculations in the Furry picture at generic photon energy. And it would be interesting to compare low-energy scattering with high-powered laser observations in a phenomenological setting, where vacuum birefringence, photon splitting and general light-by-light interactions are studied. Last, the evaluation of higher loop contributions to the propagator, while technically more challenging, is of high importance in SFQED in the context of the famous Ritus-Narozhny conjecture \cite{RN1, RN2}. It may be hoped that the known efficiency of the worldline formalism will allow easier access to higher loop calculations and their asymptotic behaviour.

\begin{acknowledgments}
\noindent JPE gratefully acknowledges discussion with Christian Schubert who made many insightful comments on questions of gauge dependence and light-by-light scattering. JPE and PC are supported by EPSRC Standard Grant EP/X02413X/1. IA is funded by a University of Plymouth doctoral scholarship.
\end{acknowledgments}

\bigskip
\bigskip

%%%%%%%%%%%%%%%%%%%%%%%%%%%%%%%%%%%%%%%%%%%
\appendix
\section{Exponential expansion to multi-linear order}
\label{AppExp}
In this appendix we record the explicit expressions for the expansion of the propagator with respect to the field strength tensors of the scattering photons. We provide these expansions at the level of the \textit{exponent}, with the understanding that a final projection onto the multi-linear sector needs to be implemented. 

Starting from the worldline kernel, (\ref{Kxx_expanded}), we specialise the external field strength tensor to $f = \sum_{k = 1}^{N} f_{k}$ and retain only those contributions that could enter the multi-linear sector. Since we have already provided the expansion of the determinant factor we present here the corresponding expansion for the exponent carrying the endpoint dependence:
\begin{widetext}
	\begin{align}
		&\e^{-i\frac{x_{-}}{4T} \cdot (Z+z) \cdot \coth(Z+z) \cdot x_{-}} \rightarrow \e^{-i\frac{x_{-}}{4T} \cdot Z \cdot \coth(Z) \cdot x_{-} -i\frac{e}{2}  \sum\limits_{\{k\}}x_{-} \cdot f_{k} \cdot (\coth(Z) - \frac{1}{Z}) \cdot x_{-} -i\frac{e^{2}T}{4} \sum\limits_{\{\ell, k \}} x_{-} \cdot f_{k} \cdot \big(\frac{\coth(Z)}{Z} - \frac{1}{Z^{2}}\big) \cdot f_{\ell} \cdot x_{-} }  \nonumber \\
		&\e^{\frac{i}{T^{4}} \sum\limits_{n = 1}^{\infty} \!\sum\limits_{\{i_{1} \ldots i_{n}\}}\! (2e)^{n} \!\prod\limits_{i = 1}^{n} \int\limits_{0}^{T}d\tau_{i}\, x_{-} \cdot Z \cdot \oDel{}_{1} \cdot f_{i_{1}} \cdot \dDel{}_{12} \cdot f_{i_{2}} \cdot \ldots \cdot \dDelo{}_{n } \cdot Z \cdot x_{-}  + \frac{2ie}{T^{3}} \sum\limits_{n = 1}^{\infty}\! \sum\limits_{\{k, i_{1}\, \ldots i_{n} \}} \!(2e)^{n}  \prod\limits_{i = 1}^{n} \int\limits_{0}^{T}d\tau_{i}\, x_{-} \cdot f_{k} \cdot \oDel{}_{1} \cdot f_{i_{1}} \cdot \dDel{}_{12} \cdot f_{i_{2}} \cdot \ldots \cdot \dDelo{}_{n} \cdot Z \cdot x_{-} }\nonumber \\
		&\e^{\frac{ie^{2}}{T^{2}}  \sum\limits_{n = 1}^{\infty} \sum\limits_{\{k, i_{1}\, \ldots i_{n}, \ell \}} (2e)^{n} \prod\limits_{i = 1}^{n} \int\limits_{0}^{T}d\tau_{i}\, x_{-} \cdot f_{k} \cdot \oDel{}_{1} \cdot  f_{i_{1}} \cdot \dDel{}_{12} \cdot f_{i_{2}} \cdot \ldots \cdot \dDelo{}_{n} \cdot f_{\ell} \cdot x_{-} } \Big|_{\text{lin}N}\,,
	\end{align}
where all sums go over all orderings of assigning unequal field strength tensors within the products. 

Similarly, for the spin degrees of freedom, (\ref{eqETanE}) is expanded as
	\begin{align}
		&\e^{-\eta \cdot \tanh(Z+z) \cdot \eta} \rightarrow  \e^{- \eta \cdot \tanh(Z) \cdot \eta- eT\sum\limits_{\{k\}}\eta \cdot \bigl[ f_k +2  f_k \cdot \big(\frac{\tanh(Z)}{Z} - 1 \big) \bigr]\cdot \eta- e^{2}T^{2}\sum\limits_{\{\ell , k\}}\eta \cdot f_k \cdot \big(\frac{\tanh(Z)}{Z^{2}} - \frac{1}{Z}\big) \cdot f_l \cdot \eta}\nonumber \\
		& \e^{\frac{1}{T^2}  \sum\limits_{n=1}^{\infty} \!\sum\limits_{\{ i_{1}, \ldots i_{n} \}}\!(-e)^n \prod\limits_{i = 1}^{n} \int\limits_{0}^{T} d\tau_{i} \,\eta \cdot Z \cdot \oGF{}_{1} \cdot f_{i_1} \cdot \ldots \cdot f_{i_n} \cdot \GFo{}_{n} \cdot Z \cdot \eta + \frac{2e}{T} \sum\limits_{n = 1}^{\infty}\sum\limits_{\{k, i_{1}, \ldots i_{n} \}}(-e)^n \prod\limits_{i = 1}^{n} \int\limits_{0}^{T}d\tau_{i} \eta \cdot f_k \cdot \oGF{}_{1} \cdot f_{i_1} \cdot \ldots \cdot f_{i_n} \cdot \GFo{}_{n} \cdot Z \cdot \eta} \nonumber \\
		&  \e^{e^{2} \sum\limits_{n = 1}^{\infty}  \sum\limits_{\{k, i_{1}, \ldots i_{n},\ell \}}(-e)^n  \prod\limits_{i = 1}^{n} \int\limits_{0}^{T}d\tau_{i} \eta \cdot f_k \cdot \oGF{}_{1} \cdot f_{i_1} \cdot \ldots \cdot f_{i_n} \cdot \GFo{}_{n} \cdot f_{\ell} \cdot \eta} \Big|_{\textrm{lin}N}\,. \label{eqETanExpan}
	\end{align}
\end{widetext}

\section{Momentum space propagator and Fock-Schwinger gauges}
\label{appFS}
The use of Fock-Schwinger gauge is particularly convenient in the worldline formalism, both in constant fields \cite{Ahmad_2017,Dittrich:2000wz,  Reuter:1996zm} and in inhomogeneous fields \cite{Schmidt:1993rk, Fliegner_1998, Auer:2003rq}. In the constant field case, the advantage is that the gauge potential,
$A_{\mu}(x) = -\frac{1}{2}F_{\mu\nu} (x - \hat{x})^{\nu}$
is linear in coordinates, so that the action remains quadratic and the path integral is Gaussian. Usually the reference point, $\hat{x}$, where $A_{\mu}$ vanishes, is taken to be the loop centre of mass (respectively initial point of the line) when calculating the effective action (respectively matter propagator). This may appear strange, since the loop centre of mass is integrated over, and likewise for the propagator endpoints when transforming to momentum space. Hence both the effective action and momentum space propagator so constructed receive contributions from path integrals corresponding to \textit{different} choices of reference point, and hence different gauges. This is not problematic, since the effective action (propagator) are gauge (co-)variant, and so physical observables are not affected.

For completeness, let us present here the matter propagator (we restrict to scalar QED in $D = 4$) instead calculated with a \textit{fixed} reference point, $\hat{x}$. In configuration space, this is easily related to our results of the main text by the gauge holonomy
$\mathcal{D}^{x'x}(\hat{x}) = \e^{ie \Phi(x',x)}\mathcal{D}^{x'x}(x)$,
where $ \Phi(x',x) = \frac{1}{2} x_{-} \cdot F \cdot (x - \hat{x})$.
Note, then, that in contrast to shifting between two \textit{constant} reference points (e.g. $\hat{x} \rightarrow \hat{x}^{\prime})$, this holonomy factor is quadratic in the endpoints. Hence, a linear shift in the reference gauge already introduces a non-linear variation in the momentum space propagator:
\begin{align}
	&\mathcal{D}^{p'p}(\hat{x})=-i\e^{ip_-\cdot \hat{x}}\int_{0}^{\infty}\!dT\,(4\pi T)^{2}\e^{-im^{2}T}K^{p_+p_+}[Z]\notag\\
	&\mathrm{det}^{1/2}\Bigl[\frac{\coth(Z)}{Z}\Bigr] \e^{-ip_+\cdot\frac{T\tanh(Z)}{Z}\cdot p_++ip_-\cdot\frac{T\coth(Z)}{Z}\cdot p_-+2i  p_+\cdot\frac{T}{Z}\cdot p_- }\,,
	\label{eqppDxhat}
\end{align}
for $p_-=p'-p$ and $p_+=\frac 1 2 (p'+p)$. Let us note that the choice of a fixed reference point, $\hat{x}$, also breaks the translation invariance of the propagator, so we do not obtain a momentum conserving $\delta$-function (except for in the zero field limit). This should be restored for the calculation of physical scattering amplitudes in this reference gauge.

To recover the more familiar worldline expression for the propagator one may take the 2-way convolution of the momentum space propagator with (the inverse) Fourier transformed gauge holonomy factor. First, the Fourier transform of the multiplicative gauge holonomy is
\begin{equation}
	\widetilde{\e^{-ie\Phi}}(q', q) = (2\pi)^{4} \det\Big(\frac{2T}{Z}\Big) \, \e^{i q_- \cdot \hat{x} - 2iT q_+ \cdot Z^{-1} \cdot q_-}\,. 
\end{equation}
Then, taking the 2-way convolution with the propagator~\eqref{eqppDxhat} we obtain
\begin{equation}
	\int \frac{d^{4}q}{(2\pi)^{4}} \int \frac{d^{4}q'}{(2\pi)^{4}} \,\mathcal{D}^{q'q}(\hat{x}) \widetilde{\e^{-ie\Phi}}(p'-q',p- q) = \mathcal{D}^{q'q}\,,
\end{equation}
with $\mathcal{D}^{q'q}$ given by~\eqref{Dpp_1}, the usual worldline propagator. In the above a momentum conserving $\delta$-function arose from integrating over the sum of momenta, $q_{+}$, which couples to the transfer of momenta $p_-$, while the integral over $q_{-}$ gives only a normalising matrix determinant.

\section{Functional traces}
\label{AppTraces}
The derivatives in (\ref{JnDeriv}) can be calculated by appealing to the derivative polynomials of the cotangent function \cite{wintucky1971formulas, CVIJOVIC20093002, Derivs}:
\begin{align}
	\partial_{Z}^{n} \coth(Z) &= C_{n}(\coth(Z))\\
	C_{n}(\coth(Z)) &= (-2)^{n}(\coth(Z)+1)\omega_{n}\Big( \frac{\coth(Z)-1}{2} \Big) \\
	\omega_{n}(z) &= \sum_{k=0}^{n} \Big\{ \begin{matrix}n\\k\end{matrix} \Big\} k! \,z^{k}\,,
\end{align}
where $\Big\{ \begin{matrix}n\\k\end{matrix} \Big\}$ are the Stirling numbers of the second kind. Note that these polynomials satisfy the recursion relation
\begin{equation}
	C_{n}(z) = (1 - z^{2})C_{n-1}'(x)\,
\end{equation}
with $C_{0}(z) \equiv z$. With this we can write (with $Z = eFT$ as above)
\begin{equation}\label{finalJsol}
	\mathcal{J}^{(n)} =  T^{n} \Big[   \frac{(-1)^{n-1}}{z^{n}} - \frac{C_{n-1}(\coth(Z))}{(n-1)!}\Big]\,.
\end{equation}
For the anti-periodic Green function we will also need the derivative polynomials for the hyperbolic tangent. These turn out to be the same as for the hyperbolic cotangent, since both satisfy the same recursion relation, so 
\begin{equation}
	\partial_{Z}^{n} \tanh(Z) = C_{n}(\tanh(Z))\,.
\end{equation}
Hence we can write for (\ref{eqJF})
\begin{equation}\label{finalJFsol}
	\mathcal{J}_{F}^{(n)} =  \frac{T^{n}}{(n-1)!} (-1)^{n-1} C_{n-1}(\tanh(Z))\,.
\end{equation}

\section{Operators, determinants and inverses}
\label{AppOps}
Here we derive the inverse and determinant corresponding to the kinetic operators that appear in the worldline actions. First we consider the better-known diagonal operator relevant in position space, $\hat{\partial}_{\tau}^{2} -2eF \hat{\partial}_{\tau}$. Its Green function for string inspired boundary conditions (periodic in the space orthogonal to the constant zero mode) in Minkowski spacetime is calculated by expanding about the vacuum part (see \cite{Reuter:1996zm, ChrisRev}),
\begin{align}
	\mathcal{G}(\tau- \tau') &= 2\langle \tau | (1 - 2eF \hat{\partial}_{\tau}^{-1})^{-1}\hat{\partial}_{\tau}^{-2} |\tau' \rangle \\
	&= 2\sum_{n = 0}^{\infty} (2eF)^{n}  \langle \tau | \hat{\partial}_{\tau}^{-(2 + n)} | \tau'\rangle \,.
\end{align}
Using \cite{Reuter:1996zm, ChrisRev} that $\langle \tau | \hat{\partial}_{\tau}^{-n} | \tau' \rangle = -\frac{T^{n}}{n!} \sigma^{n}(\tau - \tau')B_{n} (|\tau - \tau'|)$ for Bernoulli polynomials $B_{n}$, we get
\begin{align}
	\mathcal{G}_{\tau\tau'} &= -2\sum_{n = 0}^{\infty} \frac{ (2eFT )^{n} } {(n+2)!} \sigma^{n}_{\tau\tau'} B_{n+2} (|\tau - \tau'|)\\
	&= \frac{T^{2}}{2Z^{2}} \Big(1-\frac{Z}{\sinh(Z)} \e^{-Z \dot{G}_{\tau \tau'}} - Z \dot{G}_{\tau \tau'} \Big)\,,
\end{align}
where the generating function for the Bernoulli polynomials was employed. The Green function for DBC follows from the identity~\eqref{Green_Del}. One of the reasons for deriving this here is that we can use this Green function for an alternative derivation of the path integral normalisation obtained as the functional determinant:
\begin{align}
	&\ln \frac{\Det\big[\hat{\partial}_{\tau}^{2} -2eF \hat{\partial}_{\tau}\big]}{\Det\big[\hat{\partial}_{\tau}^{2}\big]}= \Tr \ln \big[\hat{\partial}_{\tau}^{2} -2eF \hat{\partial}_{\tau} \big] - \Tr \ln \big[\hat{\partial}_{\tau}^{2} \big]  \notag\\
	&=  -2e \int_{0}^{F} dF' \, \Tr \big[\hat{\partial}_{\tau}\big(\hat{\partial}_{\tau}^{2} -2eF' \hat{\partial}_{\tau}\big)^{-1} \big]\\
	&= -e \tr\int_{0}^{F} dF'\, \int_{0}^{T}d\tau\, \dot{G}(\tau, \tau | F')\\
	&= \tr \ln \Big[ \frac{\sinh(Z)}{Z} \Big] = \ln \det \Big[ \frac{\sinh(Z)}{Z} \Big]  \,.
\end{align}
This result holds also for Dirichlet boundary conditions. 

Now we turn to the non-local operator, $\hat{\chi}$. We solve (\ref{eqXiDE}) by resumming a geometric series obtained by expanding about the diagonal part of $\hat{\chi}$: 
\begin{align}
	&\Xi(\tau, \tau') =  \langle \tau | \Big[\hat{\partial}_{\tau}^{2} - 2eF\hat{\partial}_{\tau} - \frac{4 e^{2}}{T} \hat{ F}^{2}\Big]^{-1} | \tau' \rangle  \\
	&= \langle \tau | \frac{1}{1 - \frac{4 e^{2}}{T} \frac{1}{\hat{\partial}_{\tau}^{2} -2eF \hat{\partial}_{\tau}}\hat{F}^{2} }\frac{1}{\hat{\partial}_{\tau}^{2} -2eF \hat{\partial}_{\tau} } |\tau' \rangle \\
	&= \Del{}_{\tau \tau'} + \frac{4}{T^3}  \Delo{}_{\tau} \tcdot Z^{2} \tcdot \oDel{}_{\tau'} 
	+ \Big(\frac{4}{T^3} \Big)^{2} \Delo_{\tau} \tcdot Z^{2} \tcdot \oDelo \tcdot Z^{2} \tcdot \oDel_{\tau'} + \ldots \notag\\
	&= \Del{}_{\tau \tau'} + \frac{4}{T^{3}} \Delo{}_\tau \cdot \frac{Z^{2}}{ 1 -\frac{4}{T^{3}} \oDelo \cdot Z^{2}} \cdot \oDel{}_{\tau'}
	\label{Del_Z2}\\
	&= \Del(\tau, \tau') + \frac{4}{T^{3}} \cdot \Delo(\tau) \cdot Z \cdot \tanh(Z) \cdot \oDel(\tau')\,,
\end{align}
where we have used the fact that $\oDelo = \frac{T^{3}}{4Z}(\frac{1}{Z}-\coth(Z))$. It is easy to verify this function satisfies the associated Green equation and Dirichlet boundary conditions.

Some useful integrals of this Green function are derived from the corresponding results for $\Del$ and $\mathcal{G}$:
\begin{align}
	\nonumber \dXi(\tau,\tau^\prime) &= \dDel(\tau,\tau^\prime) + \frac{4}{T^3}\dDelo(\tau) \cdot Z \cdot \tanh(Z) \cdot \oDel(\tau^\prime)\\
	\nonumber \dXid(\tau,\tau^\prime) &= \dDeld(\tau,\tau^\prime) + \frac{4}{T^3}\dDelo(\tau) \cdot Z \cdot \tanh(Z) \cdot \oDeld(\tau^\prime)\\
	\nonumber \oXi(\tau^\prime) &= \Big( 1 + \frac{4}{T^3}\oDelo \cdot Z \cdot \tanh(Z) \Big) \cdot \oDel(\tau^\prime) \\
	&= \nonumber \frac{\tanh(Z)}{Z} \cdot \oDel(\tau^\prime)\,, \\
	\nonumber \oXid(\tau^\prime) &= \frac{\tanh(Z)}{Z} \cdot \oDeld(\tau^\prime)  \\
	\nonumber \oXio &= \frac{\tanh(Z)}{Z} \cdot \oDelo \\
	&= \frac{T^{3}}{4Z^{2} } \Big( \frac{\tanh(Z)}{Z}-1\Big)\,.
	\label{eqXiprop}
\end{align}
\medskip

\noindent It also has the following Lorentz properties 
\begin{align}
\Xi^{\mathrm T}(\tau, \tau') &= \Xi(\tau', \tau)\\
\dXi{}^{\mathrm T}(\tau, \tau') &= \Xid(\tau', \tau)\\
\dXid{}^{\mathrm T}(\tau, \tau') &= \dXid(\tau', \tau)\\
\oXi^{\mathrm T}(\tau') &= \Xio(\tau')\\
\oXio{}\,^{\mathrm T} &= \oXio\,
\end{align}
(the corresponding identities for $\Delta$ and $\mathcal{G}$ were given in \cite{Ahmad_2017}).

To calculate the path integral normalisation we adapt the method presented for the diagonal operator by writing
\begin{align}
	&\ln \frac{\Det\big[\hat{\partial}_{\tau}^{2} -2eF \hat{\partial}_{\tau} - \frac{4e^{2}}{T} \hat{F}^{2}\big]}{\Det\big[\hat{\partial}_{\tau}^{2} - 2eF \hat{\partial}_{\tau}\big]} = \nonumber \\
	 &\int_{0}^{1} d\lambda \, \Tr\big[ -\frac{4e^{2}}{T}\hat{F}^{2} \big( \hat{\partial}_{\tau}^{2} -2eF \hat{\partial}_{\tau} - \frac{4\lambda e^{2}}{T} \hat{F}^{2} \big)^{-1}\big] \\
	 &=- \frac{4e^{2}}{T} \tr \int_{0}^{1} \!d\lambda  \int_{0}^{T}\!d\tau \int_{0}^{T}\!d\tau'\, [\Xi_{\tau\tau'}|_{\hat{F}^2\to \lambda\hat{F}^2}]F^{2} \,.
\end{align}
To find $\Xi$ with the shifted argument for $\hat{F}$ one need only replace $\hat{F}^2\to \lambda\hat{F}^2$ in~\eqref{Del_Z2} (but leave the $\Del$ unchanged). Evaluating the parameter integrals leads to
\begin{equation}
	-\frac{4}{T^{3}} \tr \int_{0}^{1}d\lambda\,  \Big(\oDelo + \frac{4}{T^{3}}  \frac{\lambda Z^{2}}{1 -\frac{4}{T^{3}} \oDelo \lambda \cdot Z^{2}} \cdot (\oDelo)^2\Big) \cdot Z^{2}\,.
\end{equation}
So that upon integrating over $\lambda$ we find
\begin{equation}
	\tr \ln\Big[\frac{Z}{\tanh(Z)}\Big] = \ln \det\Big[\frac{Z}{\tanh(Z)} \Big]\,.
\end{equation}
Hence 
\begin{equation}
\frac{\Det[\hat{\partial}_{\tau}^{2} -2eF \hat{\partial}_{\tau} - \frac{4e^{2}}{T} \hat{F}^{2}]}{\Det[\hat{\partial}_{\tau}^{2} ]} = \det [\cosh(Z)]
\end{equation}
as required.

\bibliography{bibLow}

%apsrev4-2.bst 2019-01-14 (MD) hand-edited version of apsrev4-1.bst
%Control: key (0)
%Control: author (8) initials jnrlst
%Control: editor formatted (1) identically to author
%Control: production of article title (0) allowed
%Control: page (0) single
%Control: year (1) truncated
%Control: production of eprint (0) enabled
\begin{thebibliography}{143}%
\makeatletter
\providecommand \@ifxundefined [1]{%
 \@ifx{#1\undefined}
}%
\providecommand \@ifnum [1]{%
 \ifnum #1\expandafter \@firstoftwo
 \else \expandafter \@secondoftwo
 \fi
}%
\providecommand \@ifx [1]{%
 \ifx #1\expandafter \@firstoftwo
 \else \expandafter \@secondoftwo
 \fi
}%
\providecommand \natexlab [1]{#1}%
\providecommand \enquote  [1]{``#1''}%
\providecommand \bibnamefont  [1]{#1}%
\providecommand \bibfnamefont [1]{#1}%
\providecommand \citenamefont [1]{#1}%
\providecommand \href@noop [0]{\@secondoftwo}%
\providecommand \href [0]{\begingroup \@sanitize@url \@href}%
\providecommand \@href[1]{\@@startlink{#1}\@@href}%
\providecommand \@@href[1]{\endgroup#1\@@endlink}%
\providecommand \@sanitize@url [0]{\catcode `\\12\catcode `\$12\catcode
  `\&12\catcode `\#12\catcode `\^12\catcode `\_12\catcode `\%12\relax}%
\providecommand \@@startlink[1]{}%
\providecommand \@@endlink[0]{}%
\providecommand \url  [0]{\begingroup\@sanitize@url \@url }%
\providecommand \@url [1]{\endgroup\@href {#1}{\urlprefix }}%
\providecommand \urlprefix  [0]{URL }%
\providecommand \Eprint [0]{\href }%
\providecommand \doibase [0]{https://doi.org/}%
\providecommand \selectlanguage [0]{\@gobble}%
\providecommand \bibinfo  [0]{\@secondoftwo}%
\providecommand \bibfield  [0]{\@secondoftwo}%
\providecommand \translation [1]{[#1]}%
\providecommand \BibitemOpen [0]{}%
\providecommand \bibitemStop [0]{}%
\providecommand \bibitemNoStop [0]{.\EOS\space}%
\providecommand \EOS [0]{\spacefactor3000\relax}%
\providecommand \BibitemShut  [1]{\csname bibitem#1\endcsname}%
\let\auto@bib@innerbib\@empty
%</preamble>
\bibitem [{\citenamefont {Laporta}(2017)}]{Laporta4}%
  \BibitemOpen
  \bibfield  {author} {\bibinfo {author} {\bibfnamefont {S.}~\bibnamefont
  {Laporta}},\ }\bibfield  {title} {\bibinfo {title} {High-precision
  calculation of the 4-loop contribution to the electron g-2 in qed},\ }\href
  {https://doi.org/doi:10.1016/j.physletb.2017.06.056} {\bibfield  {journal}
  {\bibinfo  {journal} {Phys. Lett. B}\ }\textbf {\bibinfo {volume}
  {\textbf{772}}},\ \bibinfo {pages} {232} (\bibinfo {year} {(2017)})},\
  \Eprint {https://arxiv.org/abs/hep-ph/1704.06996} {arXiv:hep-ph/1704.06996}
  \BibitemShut {NoStop}%
\bibitem [{\citenamefont {Volkov}(2024)}]{Volkov5}%
  \BibitemOpen
  \bibfield  {author} {\bibinfo {author} {\bibfnamefont {S.}~\bibnamefont
  {Volkov}},\ }\bibfield  {title} {\bibinfo {title} {Calculation of the total
  10th order qed contribution to the electron magnetic moment},\ }\href
  {https://doi.org/doi:10.1103/PhysRevD.110.036001} {\bibfield  {journal}
  {\bibinfo  {journal} {Phys. Rev. D}\ }\textbf {\bibinfo {volume}
  {\textbf{110}}},\ \bibinfo {pages} {036001} (\bibinfo {year} {(2024)})},\
  \Eprint {https://arxiv.org/abs/hep-ph/2404.00649} {arXiv:hep-ph/2404.00649}
  \BibitemShut {NoStop}%
\bibitem [{\citenamefont {T.~Aoyama}\ and\ \citenamefont
  {Nio}(2015)}]{Kinoshita5}%
  \BibitemOpen
  \bibfield  {author} {\bibinfo {author} {\bibfnamefont {T.~K.}\ \bibnamefont
  {T.~Aoyama}, \bibfnamefont {M.~Hayakawa}}\ and\ \bibinfo {author}
  {\bibfnamefont {M.}~\bibnamefont {Nio}},\ }\bibfield  {title} {\bibinfo
  {title} {Tenth-order electron anomalous magnetic moment --- contribution of
  diagrams without closed lepton loops},\ }\href
  {https://doi.org/doi:10.1103/PhysRevD.91.033006} {\bibfield  {journal}
  {\bibinfo  {journal} {Phys. Rev. D}\ }\textbf {\bibinfo {volume}
  {\textbf{91}}},\ \bibinfo {pages} {033006} (\bibinfo {year} {(2015)})},\
  \Eprint {https://arxiv.org/abs/hep-ph/1412.8284} {hep-ph/1412.8284}
  \BibitemShut {NoStop}%
\bibitem [{\citenamefont {Euler}\ and\ \citenamefont
  {Kockel}(1935)}]{Euler:1935zz}%
  \BibitemOpen
  \bibfield  {author} {\bibinfo {author} {\bibfnamefont {H.}~\bibnamefont
  {Euler}}\ and\ \bibinfo {author} {\bibfnamefont {B.}~\bibnamefont {Kockel}},\
  }\bibfield  {title} {\bibinfo {title} {{The scattering of light by light in
  Dirac\textquoteright{}s theory}},\ }\href
  {https://doi.org/10.1007/BF01493898} {\bibfield  {journal} {\bibinfo
  {journal} {Naturwiss.}\ }\textbf {\bibinfo {volume} {23}},\ \bibinfo {pages}
  {246} (\bibinfo {year} {1935})}\BibitemShut {NoStop}%
\bibitem [{\citenamefont {Heisenberg}\ and\ \citenamefont
  {Euler}(1936)}]{Heisenberg:1936nmg}%
  \BibitemOpen
  \bibfield  {author} {\bibinfo {author} {\bibfnamefont {W.}~\bibnamefont
  {Heisenberg}}\ and\ \bibinfo {author} {\bibfnamefont {H.}~\bibnamefont
  {Euler}},\ }\bibfield  {title} {\bibinfo {title} {{Consequences of Dirac's
  theory of positrons}},\ }\href {https://doi.org/10.1007/BF01343663}
  {\bibfield  {journal} {\bibinfo  {journal} {Z. Phys.}\ }\textbf {\bibinfo
  {volume} {98}},\ \bibinfo {pages} {714} (\bibinfo {year} {1936})},\ \Eprint
  {https://arxiv.org/abs/physics/0605038} {arXiv:physics/0605038} \BibitemShut
  {NoStop}%
\bibitem [{\citenamefont {Akhieser}\ \emph {et~al.}(1936)\citenamefont
  {Akhieser}, \citenamefont {Landau},\ and\ \citenamefont
  {Pomeranchook}}]{Akhieser}%
  \BibitemOpen
  \bibfield  {author} {\bibinfo {author} {\bibfnamefont {A.}~\bibnamefont
  {Akhieser}}, \bibinfo {author} {\bibfnamefont {L.}~\bibnamefont {Landau}},\
  and\ \bibinfo {author} {\bibfnamefont {I.}~\bibnamefont {Pomeranchook}},\
  }\bibfield  {title} {\bibinfo {title} {Scattering of light by light},\
  }\href@noop {} {\bibfield  {journal} {\bibinfo  {journal} {Nature}\ }\textbf
  {\bibinfo {volume} {138}},\ \bibinfo {pages} {206} (\bibinfo {year}
  {1936})}\BibitemShut {NoStop}%
\bibitem [{\citenamefont {Karplus}\ and\ \citenamefont
  {Neuman}(1950)}]{Karplus:1950zza}%
  \BibitemOpen
  \bibfield  {author} {\bibinfo {author} {\bibfnamefont {R.}~\bibnamefont
  {Karplus}}\ and\ \bibinfo {author} {\bibfnamefont {M.}~\bibnamefont
  {Neuman}},\ }\bibfield  {title} {\bibinfo {title} {{Non-Linear Interactions
  between Electromagnetic Fields}},\ }\href
  {https://doi.org/10.1103/PhysRev.80.380} {\bibfield  {journal} {\bibinfo
  {journal} {Phys. Rev.}\ }\textbf {\bibinfo {volume} {80}},\ \bibinfo {pages}
  {380} (\bibinfo {year} {1950})}\BibitemShut {NoStop}%
\bibitem [{\citenamefont {Karplus}\ and\ \citenamefont
  {Neuman}(1951)}]{Karplus:1950zz}%
  \BibitemOpen
  \bibfield  {author} {\bibinfo {author} {\bibfnamefont {R.}~\bibnamefont
  {Karplus}}\ and\ \bibinfo {author} {\bibfnamefont {M.}~\bibnamefont
  {Neuman}},\ }\bibfield  {title} {\bibinfo {title} {{The scattering of light
  by light}},\ }\href {https://doi.org/10.1103/PhysRev.83.776} {\bibfield
  {journal} {\bibinfo  {journal} {Phys. Rev.}\ }\textbf {\bibinfo {volume}
  {83}},\ \bibinfo {pages} {776} (\bibinfo {year} {1951})}\BibitemShut
  {NoStop}%
\bibitem [{\citenamefont {De~Tollis}(1964)}]{DeTollis:1964una}%
  \BibitemOpen
  \bibfield  {author} {\bibinfo {author} {\bibfnamefont {B.}~\bibnamefont
  {De~Tollis}},\ }\bibfield  {title} {\bibinfo {title} {{Dispersive approach to
  photon-photon scattering}},\ }\href {https://doi.org/10.1007/BF02735895}
  {\bibfield  {journal} {\bibinfo  {journal} {Nuovo Cim.}\ }\textbf {\bibinfo
  {volume} {32}},\ \bibinfo {pages} {757} (\bibinfo {year} {1964})}\BibitemShut
  {NoStop}%
\bibitem [{\citenamefont {De~Tollis}(1965)}]{DeTollis:1965vna}%
  \BibitemOpen
  \bibfield  {author} {\bibinfo {author} {\bibfnamefont {B.}~\bibnamefont
  {De~Tollis}},\ }\bibfield  {title} {\bibinfo {title} {{The scattering of
  photons by photons}},\ }\href {https://doi.org/10.1007/BF02735534} {\bibfield
   {journal} {\bibinfo  {journal} {Nuovo Cim.}\ }\textbf {\bibinfo {volume}
  {35}},\ \bibinfo {pages} {1182} (\bibinfo {year} {1965})}\BibitemShut
  {NoStop}%
\bibitem [{\citenamefont {Scharnhorst}(2017)}]{Scharnhorst:2017wzh}%
  \BibitemOpen
  \bibfield  {author} {\bibinfo {author} {\bibfnamefont {K.}~\bibnamefont
  {Scharnhorst}},\ }\bibfield  {title} {\bibinfo {title} {{Photon-photon
  scattering and related phenomena. Experimental and theoretical approaches:
  The early period}},\ }\href@noop {} {\  (\bibinfo {year} {2017})},\ \Eprint
  {https://arxiv.org/abs/1711.05194} {arXiv:1711.05194 [physics.hist-ph]}
  \BibitemShut {NoStop}%
\bibitem [{\citenamefont {Abramowicz}\ \emph {et~al.}(2023)\citenamefont
  {Abramowicz} \emph {et~al.}}]{LUXE:2023crk}%
  \BibitemOpen
  \bibfield  {author} {\bibinfo {author} {\bibfnamefont {H.}~\bibnamefont
  {Abramowicz}} \emph {et~al.} (\bibinfo {collaboration} {LUXE}),\ }\bibfield
  {title} {\bibinfo {title} {{Technical Design Report for the LUXE
  Experiment}}\ }\href {https://doi.org/10.1140/epjs/s11734-024-01164-9}
  {10.1140/epjs/s11734-024-01164-9} (\bibinfo {year} {2023}),\ \Eprint
  {https://arxiv.org/abs/2308.00515} {arXiv:2308.00515 [hep-ex]} \BibitemShut
  {NoStop}%
\bibitem [{\citenamefont {{Christine Clarke et al.}}(2022)}]{Clarke:2022rbd}%
  \BibitemOpen
  \bibfield  {author} {\bibinfo {author} {\bibnamefont {{Christine Clarke et
  al.}}},\ }\bibfield  {title} {\bibinfo {title} {{Facet-ii}},\ }\href
  {https://doi.org/10.18429/JACoW-LINAC2022-WE1AA03} {\bibfield  {journal}
  {\bibinfo  {journal} {JACoW}\ }\textbf {\bibinfo {volume} {LINAC2022}},\
  \bibinfo {pages} {631} (\bibinfo {year} {2022})},\ \bibinfo {note}
  {doi:10.18429/JACoW-LINAC2022-WE1AA03}\BibitemShut {NoStop}%
\bibitem [{\citenamefont {Yakimenko}\ \emph {et~al.}(2019)\citenamefont
  {Yakimenko}, \citenamefont {Alsberg}, \citenamefont {Bong}, \citenamefont
  {Bouchard}, \citenamefont {Clarke}, \citenamefont {Emma}, \citenamefont
  {Green}, \citenamefont {Hast}, \citenamefont {Hogan}, \citenamefont
  {Seabury}, \citenamefont {Lipkowitz}, \citenamefont {O'Shea}, \citenamefont
  {Storey}, \citenamefont {White},\ and\ \citenamefont
  {Yocky}}]{PhysRevAccelBeams.22.101301}%
  \BibitemOpen
  \bibfield  {author} {\bibinfo {author} {\bibfnamefont {V.}~\bibnamefont
  {Yakimenko}}, \bibinfo {author} {\bibfnamefont {L.}~\bibnamefont {Alsberg}},
  \bibinfo {author} {\bibfnamefont {E.}~\bibnamefont {Bong}}, \bibinfo {author}
  {\bibfnamefont {G.}~\bibnamefont {Bouchard}}, \bibinfo {author}
  {\bibfnamefont {C.}~\bibnamefont {Clarke}}, \bibinfo {author} {\bibfnamefont
  {C.}~\bibnamefont {Emma}}, \bibinfo {author} {\bibfnamefont {S.}~\bibnamefont
  {Green}}, \bibinfo {author} {\bibfnamefont {C.}~\bibnamefont {Hast}},
  \bibinfo {author} {\bibfnamefont {M.~J.}\ \bibnamefont {Hogan}}, \bibinfo
  {author} {\bibfnamefont {J.}~\bibnamefont {Seabury}}, \bibinfo {author}
  {\bibfnamefont {N.}~\bibnamefont {Lipkowitz}}, \bibinfo {author}
  {\bibfnamefont {B.}~\bibnamefont {O'Shea}}, \bibinfo {author} {\bibfnamefont
  {D.}~\bibnamefont {Storey}}, \bibinfo {author} {\bibfnamefont
  {G.}~\bibnamefont {White}},\ and\ \bibinfo {author} {\bibfnamefont
  {G.}~\bibnamefont {Yocky}},\ }\bibfield  {title} {\bibinfo {title} {Facet-ii
  facility for advanced accelerator experimental tests},\ }\href
  {https://doi.org/10.1103/PhysRevAccelBeams.22.101301} {\bibfield  {journal}
  {\bibinfo  {journal} {Phys. Rev. Accel. Beams}\ }\textbf {\bibinfo {volume}
  {22}},\ \bibinfo {pages} {101301} (\bibinfo {year} {2019})}\BibitemShut
  {NoStop}%
\bibitem [{\citenamefont {Ahmadiniaz}\ \emph {et~al.}(2025)\citenamefont
  {Ahmadiniaz} \emph {et~al.}}]{Ahmadiniaz:2024xob}%
  \BibitemOpen
  \bibfield  {author} {\bibinfo {author} {\bibfnamefont {N.}~\bibnamefont
  {Ahmadiniaz}} \emph {et~al.},\ }\bibfield  {title} {\bibinfo {title}
  {{Towards a vacuum birefringence experiment at the Helmholtz International
  Beamline for Extreme Fields (Letter of Intent of the BIREF@HIBEF
  Collaboration)}},\ }\href {https://doi.org/10.1017/hpl.2024.70} {\bibfield
  {journal} {\bibinfo  {journal} {High Power Laser Sci. Eng.}\ }\textbf
  {\bibinfo {volume} {13}},\ \bibinfo {pages} {e7} (\bibinfo {year} {2025})},\
  \Eprint {https://arxiv.org/abs/2405.18063} {arXiv:2405.18063
  [physics.ins-det]} \BibitemShut {NoStop}%
\bibitem [{\citenamefont {Ahmadiniaz}\ \emph
  {et~al.}(2023{\natexlab{a}})\citenamefont {Ahmadiniaz}, \citenamefont
  {Cowan}, \citenamefont {Grenzer}, \citenamefont {Franchino-Vi\~nas},
  \citenamefont {Garcia}, \citenamefont {\v{S}m\'\i{}d}, \citenamefont
  {Toncian}, \citenamefont {Trejo},\ and\ \citenamefont
  {Schützhold}}]{Ahmadiniaz:2022nrv}%
  \BibitemOpen
  \bibfield  {author} {\bibinfo {author} {\bibfnamefont {N.}~\bibnamefont
  {Ahmadiniaz}}, \bibinfo {author} {\bibfnamefont {T.~E.}\ \bibnamefont
  {Cowan}}, \bibinfo {author} {\bibfnamefont {J.}~\bibnamefont {Grenzer}},
  \bibinfo {author} {\bibfnamefont {S.}~\bibnamefont {Franchino-Vi\~nas}},
  \bibinfo {author} {\bibfnamefont {A.~L.}\ \bibnamefont {Garcia}}, \bibinfo
  {author} {\bibfnamefont {M.}~\bibnamefont {\v{S}m\'\i{}d}}, \bibinfo {author}
  {\bibfnamefont {T.}~\bibnamefont {Toncian}}, \bibinfo {author} {\bibfnamefont
  {M.~A.}\ \bibnamefont {Trejo}},\ and\ \bibinfo {author} {\bibfnamefont
  {R.}~\bibnamefont {Schützhold}},\ }\bibfield  {title} {\bibinfo {title}
  {{Detection schemes for quantum vacuum diffraction and birefringence}},\
  }\href {https://doi.org/10.1103/PhysRevD.108.076005} {\bibfield  {journal}
  {\bibinfo  {journal} {Phys. Rev. D}\ }\textbf {\bibinfo {volume} {108}},\
  \bibinfo {pages} {076005} (\bibinfo {year} {2023}{\natexlab{a}})},\ \Eprint
  {https://arxiv.org/abs/2208.14215} {arXiv:2208.14215 [physics.optics]}
  \BibitemShut {NoStop}%
\bibitem [{\citenamefont {Sch\"utze}\ \emph {et~al.}(2024)\citenamefont
  {Sch\"utze}, \citenamefont {Doyle}, \citenamefont {Schreiber}, \citenamefont
  {Zepf},\ and\ \citenamefont {Karbstein}}]{Schutze:2024kzu}%
  \BibitemOpen
  \bibfield  {author} {\bibinfo {author} {\bibfnamefont {F.}~\bibnamefont
  {Sch\"utze}}, \bibinfo {author} {\bibfnamefont {L.}~\bibnamefont {Doyle}},
  \bibinfo {author} {\bibfnamefont {J.}~\bibnamefont {Schreiber}}, \bibinfo
  {author} {\bibfnamefont {M.}~\bibnamefont {Zepf}},\ and\ \bibinfo {author}
  {\bibfnamefont {F.}~\bibnamefont {Karbstein}},\ }\bibfield  {title} {\bibinfo
  {title} {{Dark-field setup for the measurement of light-by-light scattering
  with high-intensity lasers}},\ }\href
  {https://doi.org/10.1103/PhysRevD.109.096009} {\bibfield  {journal} {\bibinfo
   {journal} {Phys. Rev. D}\ }\textbf {\bibinfo {volume} {109}},\ \bibinfo
  {pages} {096009} (\bibinfo {year} {2024})},\ \Eprint
  {https://arxiv.org/abs/2403.06762} {arXiv:2403.06762 [physics.optics]}
  \BibitemShut {NoStop}%
\bibitem [{\citenamefont {Sirunyan}\ \emph {et~al.}(2019)\citenamefont
  {Sirunyan} \emph {et~al.}}]{CMS}%
  \BibitemOpen
  \bibfield  {author} {\bibinfo {author} {\bibfnamefont {A.~M.}\ \bibnamefont
  {Sirunyan}} \emph {et~al.} (\bibinfo {collaboration} {CMS}),\ }\bibfield
  {title} {\bibinfo {title} {{Evidence for light-by-light scattering and
  searches for axion-like particles in ultraperipheral PbPb collisions at
  $\sqrt{s_\mathrm{NN}} =$ 5.02 TeV}},\ }\href
  {https://doi.org/10.1016/j.physletb.2019.134826} {\bibfield  {journal}
  {\bibinfo  {journal} {Phys. Lett. B}\ }\textbf {\bibinfo {volume} {797}},\
  \bibinfo {pages} {134826} (\bibinfo {year} {2019})},\ \Eprint
  {https://arxiv.org/abs/1810.04602} {arXiv:1810.04602 [hep-ex]} \BibitemShut
  {NoStop}%
\bibitem [{\citenamefont {Aad}\ \emph {et~al.}(2019)\citenamefont {Aad},
  \citenamefont {Abbott} \emph {et~al.}}]{ATLAS}%
  \BibitemOpen
  \bibfield  {author} {\bibinfo {author} {\bibfnamefont {G.}~\bibnamefont
  {Aad}}, \bibinfo {author} {\bibnamefont {Abbott}}, \emph {et~al.} (\bibinfo
  {collaboration} {ATLAS}),\ }\bibfield  {title} {\bibinfo {title} {Observation
  of light-by-light scattering in ultraperipheral $\mathrm{Pb}+\mathrm{Pb}$
  collisions with the atlas detector},\ }\href
  {https://doi.org/10.1103/PhysRevLett.123.052001} {\bibfield  {journal}
  {\bibinfo  {journal} {Phys. Rev. Lett.}\ }\textbf {\bibinfo {volume} {123}},\
  \bibinfo {pages} {052001} (\bibinfo {year} {2019})},\ \Eprint
  {https://arxiv.org/abs/1904.03536} {arXiv:1904.03536 [hep-ex]} \BibitemShut
  {NoStop}%
\bibitem [{\citenamefont {Yamaji}(2016)}]{Sacla}%
  \BibitemOpen
  \bibfield  {author} {\bibinfo {author} {\bibfnamefont {T.~e.~a.}\
  \bibnamefont {Yamaji}},\ }\bibfield  {title} {\bibinfo {title} {An experiment
  of x-ray photon–photon elastic scattering with a laue-case beam collider},\
  }\href@noop {} {\bibfield  {journal} {\bibinfo  {journal} {Phys. Lett. B}\
  }\textbf {\bibinfo {volume} {763}},\ \bibinfo {pages} {454} (\bibinfo {year}
  {2016})}\BibitemShut {NoStop}%
\bibitem [{\citenamefont {Yoon}\ \emph {et~al.}(2021)\citenamefont {Yoon},
  \citenamefont {Kim}, \citenamefont {Choi}, \citenamefont {Sung},
  \citenamefont {Lee}, \citenamefont {Lee},\ and\ \citenamefont
  {Nam}}]{Corels}%
  \BibitemOpen
  \bibfield  {author} {\bibinfo {author} {\bibfnamefont {J.~W.}\ \bibnamefont
  {Yoon}}, \bibinfo {author} {\bibfnamefont {Y.~G.}\ \bibnamefont {Kim}},
  \bibinfo {author} {\bibfnamefont {I.~W.}\ \bibnamefont {Choi}}, \bibinfo
  {author} {\bibfnamefont {J.~H.}\ \bibnamefont {Sung}}, \bibinfo {author}
  {\bibfnamefont {H.~W.}\ \bibnamefont {Lee}}, \bibinfo {author} {\bibfnamefont
  {S.~K.}\ \bibnamefont {Lee}},\ and\ \bibinfo {author} {\bibfnamefont {C.~H.}\
  \bibnamefont {Nam}},\ }\bibfield  {title} {\bibinfo {title} {Realization of
  laser intensity over 1023 w/cm${}^{2}$},\ }\href@noop {} {\bibfield
  {journal} {\bibinfo  {journal} {Optica}\ }\textbf {\bibinfo {volume} {8}},\
  \bibinfo {pages} {630} (\bibinfo {year} {2021})}\BibitemShut {NoStop}%
\bibitem [{ELI()}]{ELI}%
  \BibitemOpen
  \href@noop {} {}\bibinfo {note}
  {\href{https://eli-laser.eu/}{https://eli-laser.eu/}}\BibitemShut {NoStop}%
\bibitem [{\citenamefont {Yukie}\ and\ \citenamefont {Lianghong}(2021)}]{SEL}%
  \BibitemOpen
  \bibfield  {author} {\bibinfo {author} {\bibfnamefont {P.}~\bibnamefont
  {Yukie}}\ and\ \bibinfo {author} {\bibfnamefont {Y.}~\bibnamefont
  {Lianghong}},\ }\bibfield  {title} {\bibinfo {title} {Overview and status of
  station of extreme light toward 100 pw},\ }\href@noop {} {\bibfield
  {journal} {\bibinfo  {journal} {Reza Kenkyu}\ }\textbf {\bibinfo {volume}
  {49}},\ \bibinfo {pages} {93} (\bibinfo {year} {2021})}\BibitemShut {NoStop}%
\bibitem [{\citenamefont {d'Enterria}\ and\ \citenamefont
  {da~Silveira}(2013)}]{dEnterria:2013zqi}%
  \BibitemOpen
  \bibfield  {author} {\bibinfo {author} {\bibfnamefont {D.}~\bibnamefont
  {d'Enterria}}\ and\ \bibinfo {author} {\bibfnamefont {G.~G.}\ \bibnamefont
  {da~Silveira}},\ }\bibfield  {title} {\bibinfo {title} {{Observing
  light-by-light scattering at the Large Hadron Collider}},\ }\href
  {https://doi.org/10.1103/PhysRevLett.111.080405} {\bibfield  {journal}
  {\bibinfo  {journal} {Phys. Rev. Lett.}\ }\textbf {\bibinfo {volume} {111}},\
  \bibinfo {pages} {080405} (\bibinfo {year} {2013})},\ \bibinfo {note}
  {[Erratum: Phys.Rev.Lett. 116, 129901 (2016)]},\ \Eprint
  {https://arxiv.org/abs/1305.7142} {arXiv:1305.7142 [hep-ph]} \BibitemShut
  {NoStop}%
\bibitem [{\citenamefont {Adam}(2021)}]{PhysRevLett.127.052302}%
  \BibitemOpen
  \bibfield  {author} {\bibinfo {author} {\bibfnamefont {J.~e.~a.}\
  \bibnamefont {Adam}} (\bibinfo {collaboration} {STAR Collaboration}),\
  }\bibfield  {title} {\bibinfo {title} {Measurement of
  ${e}^{+}{e}^{\ensuremath{-}}$ momentum and angular distributions from
  linearly polarized photon collisions},\ }\href
  {https://doi.org/10.1103/PhysRevLett.127.052302} {\bibfield  {journal}
  {\bibinfo  {journal} {Phys. Rev. Lett.}\ }\textbf {\bibinfo {volume} {127}},\
  \bibinfo {pages} {052302} (\bibinfo {year} {2021})},\ \Eprint
  {https://arxiv.org/abs/1910.12400} {arXiv:1910.12400 [nucl-ex]} \BibitemShut
  {NoStop}%
\bibitem [{\citenamefont {Sauter}(1931)}]{Sauter}%
  \BibitemOpen
  \bibfield  {author} {\bibinfo {author} {\bibfnamefont {F.}~\bibnamefont
  {Sauter}},\ }\bibfield  {title} {\bibinfo {title} {On the behavior of an
  electron in a homogeneous electric field in dirac’s relativistic theory},\
  }\href@noop {} {\bibfield  {journal} {\bibinfo  {journal} {Z. Phys}\ }\textbf
  {\bibinfo {volume} {69}},\ \bibinfo {pages} {742} (\bibinfo {year}
  {1931})}\BibitemShut {NoStop}%
\bibitem [{\citenamefont {Schwinger}(1951)}]{Schwinger}%
  \BibitemOpen
  \bibfield  {author} {\bibinfo {author} {\bibfnamefont {J.}~\bibnamefont
  {Schwinger}},\ }\bibfield  {title} {\bibinfo {title} {On gauge invariance and
  vacuum polarization},\ }\href {https://doi.org/10.1103/PhysRev.82.664}
  {\bibfield  {journal} {\bibinfo  {journal} {Phys. Rev.}\ }\textbf {\bibinfo
  {volume} {82}},\ \bibinfo {pages} {664} (\bibinfo {year} {1951})}\BibitemShut
  {NoStop}%
\bibitem [{\citenamefont {V.Weisskopf}(1936)}]{Weisskopf}%
  \BibitemOpen
  \bibfield  {author} {\bibinfo {author} {\bibnamefont {V.Weisskopf}},\
  }\bibfield  {title} {\bibinfo {title} {Über die elektrodynamik des vakuums
  auf grund der quantentheorie des elektrons},\ }\href@noop {} {\bibfield
  {journal} {\bibinfo  {journal} {Kong. Dans. Vid. Selsk. Math-fys. Medd}\
  }\textbf {\bibinfo {volume} {XIV}} (\bibinfo {year} {(1936)})},\ \bibinfo
  {note} {reprinted in Quantum Electrodynamics, J. Schwinger (Ed.) (Dover, New
  York, 1958).}\BibitemShut {Stop}%
\bibitem [{\citenamefont {Dittrich}\ and\ \citenamefont
  {Gies}(2000)}]{Dittrich:2000zu}%
  \BibitemOpen
  \bibfield  {author} {\bibinfo {author} {\bibfnamefont {W.}~\bibnamefont
  {Dittrich}}\ and\ \bibinfo {author} {\bibfnamefont {H.}~\bibnamefont
  {Gies}},\ }\href {https://doi.org/10.1007/3-540-45585-X} {\emph {\bibinfo
  {title} {{Probing the quantum vacuum. Perturbative effective action approach
  in quantum electrodynamics and its application}}}},\ Vol.\ \bibinfo {volume}
  {166}\ (\bibinfo  {publisher} {Springer Berlin, Heidelberg},\ \bibinfo {year}
  {2000})\BibitemShut {NoStop}%
\bibitem [{\citenamefont {Fradkin}\ \emph {et~al.}(1991)\citenamefont
  {Fradkin}, \citenamefont {Gitman},\ and\ \citenamefont
  {Shvartsman}}]{Fradkin:1991zq}%
  \BibitemOpen
  \bibfield  {author} {\bibinfo {author} {\bibfnamefont {E.~S.}\ \bibnamefont
  {Fradkin}}, \bibinfo {author} {\bibfnamefont {D.~M.}\ \bibnamefont
  {Gitman}},\ and\ \bibinfo {author} {\bibfnamefont {S.~M.}\ \bibnamefont
  {Shvartsman}},\ }\href@noop {} {\emph {\bibinfo {title} {{Quantum
  electrodynamics with unstable vacuum}}}}\ (\bibinfo  {publisher}
  {Springer-Verlag},\ \bibinfo {year} {1991})\BibitemShut {NoStop}%
\bibitem [{\citenamefont {Karbstein}\ \emph {et~al.}(2022)\citenamefont
  {Karbstein}, \citenamefont {Ullmann}, \citenamefont {Mosman},\ and\
  \citenamefont {Zepf}}]{PhysRevLett.129.061802}%
  \BibitemOpen
  \bibfield  {author} {\bibinfo {author} {\bibfnamefont {F.}~\bibnamefont
  {Karbstein}}, \bibinfo {author} {\bibfnamefont {D.}~\bibnamefont {Ullmann}},
  \bibinfo {author} {\bibfnamefont {E.~A.}\ \bibnamefont {Mosman}},\ and\
  \bibinfo {author} {\bibfnamefont {M.}~\bibnamefont {Zepf}},\ }\bibfield
  {title} {\bibinfo {title} {Direct accessibility of the fundamental constants
  governing light-by-light scattering},\ }\href
  {https://doi.org/10.1103/PhysRevLett.129.061802} {\bibfield  {journal}
  {\bibinfo  {journal} {Phys. Rev. Lett.}\ }\textbf {\bibinfo {volume} {129}},\
  \bibinfo {pages} {061802} (\bibinfo {year} {2022})},\ \Eprint
  {https://arxiv.org/abs/2207.09866} {arXiv:2207.09866 [hep-ph]} \BibitemShut
  {NoStop}%
\bibitem [{\citenamefont {Macleod}\ \emph {et~al.}(2023)\citenamefont
  {Macleod}, \citenamefont {Edwards}, \citenamefont {Heinzl}, \citenamefont
  {King},\ and\ \citenamefont {Bulanov}}]{Macleod:2023asi}%
  \BibitemOpen
  \bibfield  {author} {\bibinfo {author} {\bibfnamefont {A.~J.}\ \bibnamefont
  {Macleod}}, \bibinfo {author} {\bibfnamefont {J.~P.}\ \bibnamefont
  {Edwards}}, \bibinfo {author} {\bibfnamefont {T.}~\bibnamefont {Heinzl}},
  \bibinfo {author} {\bibfnamefont {B.}~\bibnamefont {King}},\ and\ \bibinfo
  {author} {\bibfnamefont {S.~V.}\ \bibnamefont {Bulanov}},\ }\bibfield
  {title} {\bibinfo {title} {{Strong-field vacuum polarisation with high energy
  lasers}},\ }\href {https://doi.org/10.1088/1367-2630/acf1c0} {\bibfield
  {journal} {\bibinfo  {journal} {New J. Phys.}\ }\textbf {\bibinfo {volume}
  {25}},\ \bibinfo {pages} {093002} (\bibinfo {year} {2023})},\ \Eprint
  {https://arxiv.org/abs/2304.02114} {arXiv:2304.02114 [hep-ph]} \BibitemShut
  {NoStop}%
\bibitem [{\citenamefont {Ahmadiniaz}\ \emph
  {et~al.}(2022{\natexlab{a}})\citenamefont {Ahmadiniaz}, \citenamefont
  {Cowan}, \citenamefont {Ding}, \citenamefont {Lopez}, \citenamefont
  {Sauerbrey}, \citenamefont {Shaisultanov},\ and\ \citenamefont
  {Schützhold}}]{Ahmadiniaz:2022mcy}%
  \BibitemOpen
  \bibfield  {author} {\bibinfo {author} {\bibfnamefont {N.}~\bibnamefont
  {Ahmadiniaz}}, \bibinfo {author} {\bibfnamefont {T.~E.}\ \bibnamefont
  {Cowan}}, \bibinfo {author} {\bibfnamefont {M.}~\bibnamefont {Ding}},
  \bibinfo {author} {\bibfnamefont {M.~A.~L.}\ \bibnamefont {Lopez}}, \bibinfo
  {author} {\bibfnamefont {R.}~\bibnamefont {Sauerbrey}}, \bibinfo {author}
  {\bibfnamefont {R.}~\bibnamefont {Shaisultanov}},\ and\ \bibinfo {author}
  {\bibfnamefont {R.}~\bibnamefont {Schützhold}},\ }\bibfield  {title}
  {\bibinfo {title} {{Field-assisted birefringent Compton scattering}},\
  }\href@noop {} {\  (\bibinfo {year} {2022}{\natexlab{a}})},\ \Eprint
  {https://arxiv.org/abs/2212.03350} {arXiv:2212.03350 [hep-ph]} \BibitemShut
  {NoStop}%
\bibitem [{\citenamefont {Karbstein}(2022)}]{Karbstein:2021otv}%
  \BibitemOpen
  \bibfield  {author} {\bibinfo {author} {\bibfnamefont {F.}~\bibnamefont
  {Karbstein}},\ }\bibfield  {title} {\bibinfo {title} {{Vacuum Birefringence
  at the Gamma Factory}},\ }\href {https://doi.org/10.1002/andp.202100137}
  {\bibfield  {journal} {\bibinfo  {journal} {Annalen Phys.}\ }\textbf
  {\bibinfo {volume} {534}},\ \bibinfo {pages} {2100137} (\bibinfo {year}
  {2022})},\ \Eprint {https://arxiv.org/abs/2106.06359} {arXiv:2106.06359
  [hep-ph]} \BibitemShut {NoStop}%
\bibitem [{\citenamefont {Karbstein}(2018)}]{Karbstein:2018omb}%
  \BibitemOpen
  \bibfield  {author} {\bibinfo {author} {\bibfnamefont {F.}~\bibnamefont
  {Karbstein}},\ }\bibfield  {title} {\bibinfo {title} {{Vacuum birefringence
  in the head-on collision of x-ray free-electron laser and optical
  high-intensity laser pulses}},\ }\href
  {https://doi.org/10.1103/PhysRevD.98.056010} {\bibfield  {journal} {\bibinfo
  {journal} {Phys. Rev. D}\ }\textbf {\bibinfo {volume} {98}},\ \bibinfo
  {pages} {056010} (\bibinfo {year} {2018})},\ \Eprint
  {https://arxiv.org/abs/1807.03302} {arXiv:1807.03302 [quant-ph]} \BibitemShut
  {NoStop}%
\bibitem [{\citenamefont {Schubert}(2024)}]{Schubert:2024heu}%
  \BibitemOpen
  \bibfield  {author} {\bibinfo {author} {\bibfnamefont {C.}~\bibnamefont
  {Schubert}},\ }\bibfield  {title} {\bibinfo {title} {{Photon-photon
  scattering}},\ }\href {https://doi.org/10.1051/epjconf/202430103003}
  {\bibfield  {journal} {\bibinfo  {journal} {EPJ Web Conf.}\ }\textbf
  {\bibinfo {volume} {301}},\ \bibinfo {pages} {03003} (\bibinfo {year}
  {2024})}\BibitemShut {NoStop}%
\bibitem [{\citenamefont {Heinzl}\ \emph {et~al.}(2006)\citenamefont {Heinzl},
  \citenamefont {Liesfeld}, \citenamefont {Amthor}, \citenamefont {Schwoerer},
  \citenamefont {Sauerbrey},\ and\ \citenamefont {Wipf}}]{HEINZL2006318}%
  \BibitemOpen
  \bibfield  {author} {\bibinfo {author} {\bibfnamefont {T.}~\bibnamefont
  {Heinzl}}, \bibinfo {author} {\bibfnamefont {B.}~\bibnamefont {Liesfeld}},
  \bibinfo {author} {\bibfnamefont {K.-U.}\ \bibnamefont {Amthor}}, \bibinfo
  {author} {\bibfnamefont {H.}~\bibnamefont {Schwoerer}}, \bibinfo {author}
  {\bibfnamefont {R.}~\bibnamefont {Sauerbrey}},\ and\ \bibinfo {author}
  {\bibfnamefont {A.}~\bibnamefont {Wipf}},\ }\bibfield  {title} {\bibinfo
  {title} {On the observation of vacuum birefringence},\ }\href
  {https://doi.org/https://doi.org/10.1016/j.optcom.2006.06.053} {\bibfield
  {journal} {\bibinfo  {journal} {Optics Communications}\ }\textbf {\bibinfo
  {volume} {267}},\ \bibinfo {pages} {318} (\bibinfo {year} {2006})},\ \Eprint
  {https://arxiv.org/abs/hep-ph/0601076} {arXiv:hep-ph/0601076} \BibitemShut
  {NoStop}%
\bibitem [{\citenamefont {Costantini}\ \emph {et~al.}(1971)\citenamefont
  {Costantini}, \citenamefont {De~Tollis},\ and\ \citenamefont
  {Pistoni}}]{Costantini:1971cj}%
  \BibitemOpen
  \bibfield  {author} {\bibinfo {author} {\bibfnamefont {V.}~\bibnamefont
  {Costantini}}, \bibinfo {author} {\bibfnamefont {B.}~\bibnamefont
  {De~Tollis}},\ and\ \bibinfo {author} {\bibfnamefont {G.}~\bibnamefont
  {Pistoni}},\ }\bibfield  {title} {\bibinfo {title} {{Nonlinear effects in
  quantum electrodynamics}},\ }\href {https://doi.org/10.1007/BF02736745}
  {\bibfield  {journal} {\bibinfo  {journal} {Nuovo Cim. A}\ }\textbf {\bibinfo
  {volume} {2}},\ \bibinfo {pages} {733} (\bibinfo {year} {1971})}\BibitemShut
  {NoStop}%
\bibitem [{\citenamefont {Mignani}\ \emph {et~al.}(2016)\citenamefont
  {Mignani}, \citenamefont {Testa}, \citenamefont {González~Caniulef},
  \citenamefont {Taverna}, \citenamefont {Turolla}, \citenamefont {Zane},\ and\
  \citenamefont {Wu}}]{10.1093mnrasstw2798}%
  \BibitemOpen
  \bibfield  {author} {\bibinfo {author} {\bibfnamefont {R.~P.}\ \bibnamefont
  {Mignani}}, \bibinfo {author} {\bibfnamefont {V.}~\bibnamefont {Testa}},
  \bibinfo {author} {\bibfnamefont {D.}~\bibnamefont {González~Caniulef}},
  \bibinfo {author} {\bibfnamefont {R.}~\bibnamefont {Taverna}}, \bibinfo
  {author} {\bibfnamefont {R.}~\bibnamefont {Turolla}}, \bibinfo {author}
  {\bibfnamefont {S.}~\bibnamefont {Zane}},\ and\ \bibinfo {author}
  {\bibfnamefont {K.}~\bibnamefont {Wu}},\ }\bibfield  {title} {\bibinfo
  {title} {{Evidence for vacuum birefringence from the first
  optical-polarimetry measurement of the isolated neutron star RX
  J1856.5-3754}},\ }\href {https://doi.org/10.1093/mnras/stw2798} {\bibfield
  {journal} {\bibinfo  {journal} {Monthly Notices of the Royal Astronomical
  Society}\ }\textbf {\bibinfo {volume} {465}},\ \bibinfo {pages} {492}
  (\bibinfo {year} {2016})},\ \Eprint
  {https://arxiv.org/abs/https://academic.oup.com/mnras/article-pdf/465/1/492/8593962/stw2798.pdf}
  {https://academic.oup.com/mnras/article-pdf/465/1/492/8593962/stw2798.pdf}
  \BibitemShut {NoStop}%
\bibitem [{\citenamefont {Kim}\ and\ \citenamefont
  {Kim}(2024)}]{10.10635.0215939}%
  \BibitemOpen
  \bibfield  {author} {\bibinfo {author} {\bibfnamefont {C.~M.}\ \bibnamefont
  {Kim}}\ and\ \bibinfo {author} {\bibfnamefont {S.~P.}\ \bibnamefont {Kim}},\
  }\bibfield  {title} {\bibinfo {title} {{Magnetars as laboratories for strong
  field QED}},\ }\href {https://doi.org/10.1063/5.0215939} {\bibfield
  {journal} {\bibinfo  {journal} {AIP Conf. Proc.}\ }\textbf {\bibinfo {volume}
  {2874}},\ \bibinfo {pages} {020013} (\bibinfo {year} {2024})},\ \Eprint
  {https://arxiv.org/abs/2112.02460} {arXiv:2112.02460 [astro-ph.HE]}
  \BibitemShut {NoStop}%
\bibitem [{\citenamefont {{Kim}}(2019)}]{2019arXiv190513439K}%
  \BibitemOpen
  \bibfield  {author} {\bibinfo {author} {\bibfnamefont {S.~P.}\ \bibnamefont
  {{Kim}}},\ }\bibfield  {title} {\bibinfo {title} {{Astrophysics in Strong
  Electromagnetic Fields and Laboratory Astrophysics}},\ }\href
  {https://doi.org/10.48550/arXiv.1905.13439} {\bibfield  {journal} {\bibinfo
  {journal} {arXiv e-prints}\ ,\ \bibinfo {eid} {arXiv:1905.13439}} (\bibinfo
  {year} {2019})},\ \Eprint {https://arxiv.org/abs/1905.13439}
  {arXiv:1905.13439 [gr-qc]} \BibitemShut {NoStop}%
\bibitem [{\citenamefont {{Abishev}}\ \emph {et~al.}(2018)\citenamefont
  {{Abishev}}, \citenamefont {{Toktarbay}}, \citenamefont {{Beissen}},
  \citenamefont {{Belissarova}}, \citenamefont {{Khassanov}}, \citenamefont
  {{Kudussov}},\ and\ \citenamefont {{Abylayeva}}}]{2018MNRAS.481...36A}%
  \BibitemOpen
  \bibfield  {author} {\bibinfo {author} {\bibfnamefont {M.~E.}\ \bibnamefont
  {{Abishev}}}, \bibinfo {author} {\bibfnamefont {S.}~\bibnamefont
  {{Toktarbay}}}, \bibinfo {author} {\bibfnamefont {N.~A.}\ \bibnamefont
  {{Beissen}}}, \bibinfo {author} {\bibfnamefont {F.~B.}\ \bibnamefont
  {{Belissarova}}}, \bibinfo {author} {\bibfnamefont {M.~K.}\ \bibnamefont
  {{Khassanov}}}, \bibinfo {author} {\bibfnamefont {A.~S.}\ \bibnamefont
  {{Kudussov}}},\ and\ \bibinfo {author} {\bibfnamefont {A.~Z.}\ \bibnamefont
  {{Abylayeva}}},\ }\bibfield  {title} {\bibinfo {title} {{Effects of
  non-linear electrodynamics of vacuum in the magnetic quadrupole field of a
  pulsar}},\ }\href {https://doi.org/10.1093/mnras/sty2272} {\bibfield
  {journal} {\bibinfo  {journal} {Monthly Notices of the Royal Astronomical
  Society}\ }\textbf {\bibinfo {volume} {481}},\ \bibinfo {pages} {36}
  (\bibinfo {year} {2018})}\BibitemShut {NoStop}%
\bibitem [{\citenamefont {{Heyl}}\ and\ \citenamefont
  {{Caiazzo}}(2018)}]{2018Galax...6...76H}%
  \BibitemOpen
  \bibfield  {author} {\bibinfo {author} {\bibfnamefont {J.}~\bibnamefont
  {{Heyl}}}\ and\ \bibinfo {author} {\bibfnamefont {I.}~\bibnamefont
  {{Caiazzo}}},\ }\bibfield  {title} {\bibinfo {title} {{Strongly Magnetized
  Sources: QED and X-ray Polarization}},\ }\href
  {https://doi.org/10.3390/galaxies6030076} {\bibfield  {journal} {\bibinfo
  {journal} {Galaxies}\ }\textbf {\bibinfo {volume} {6}},\ \bibinfo {eid} {76}
  (\bibinfo {year} {2018})},\ \Eprint {https://arxiv.org/abs/1802.00358}
  {arXiv:1802.00358 [astro-ph.HE]} \BibitemShut {NoStop}%
\bibitem [{\citenamefont {{Caiazzo}}\ and\ \citenamefont
  {{Heyl}}(2018)}]{2018Galax...6...57C}%
  \BibitemOpen
  \bibfield  {author} {\bibinfo {author} {\bibfnamefont {I.}~\bibnamefont
  {{Caiazzo}}}\ and\ \bibinfo {author} {\bibfnamefont {J.}~\bibnamefont
  {{Heyl}}},\ }\bibfield  {title} {\bibinfo {title} {{Probing Black Hole
  Magnetic Fields with QED}},\ }\href {https://doi.org/10.3390/galaxies6020057}
  {\bibfield  {journal} {\bibinfo  {journal} {Galaxies}\ }\textbf {\bibinfo
  {volume} {6}},\ \bibinfo {eid} {57} (\bibinfo {year} {2018})},\ \Eprint
  {https://arxiv.org/abs/1805.11018} {arXiv:1805.11018 [astro-ph.HE]}
  \BibitemShut {NoStop}%
\bibitem [{\citenamefont {{Shakeri}}\ \emph {et~al.}(2017)\citenamefont
  {{Shakeri}}, \citenamefont {{Haghighat}},\ and\ \citenamefont
  {{Xue}}}]{2017JCAP...10..014S}%
  \BibitemOpen
  \bibfield  {author} {\bibinfo {author} {\bibfnamefont {S.}~\bibnamefont
  {{Shakeri}}}, \bibinfo {author} {\bibfnamefont {M.}~\bibnamefont
  {{Haghighat}}},\ and\ \bibinfo {author} {\bibfnamefont {S.-S.}\ \bibnamefont
  {{Xue}}},\ }\bibfield  {title} {\bibinfo {title} {{JCAP}},\ }\href
  {https://doi.org/10.1088/1475-7516/2017/10/014} {\bibfield  {journal}
  {\bibinfo  {journal} {jcap}\ }\textbf {\bibinfo {volume} {2017}},\ \bibinfo
  {eid} {014} (\bibinfo {year} {2017})},\ \Eprint
  {https://arxiv.org/abs/1704.04750} {arXiv:1704.04750 [astro-ph.HE]}
  \BibitemShut {NoStop}%
\bibitem [{\citenamefont {Harding}\ and\ \citenamefont
  {Lai}(2006)}]{Harding:2006qn}%
  \BibitemOpen
  \bibfield  {author} {\bibinfo {author} {\bibfnamefont {A.~K.}\ \bibnamefont
  {Harding}}\ and\ \bibinfo {author} {\bibfnamefont {D.}~\bibnamefont {Lai}},\
  }\bibfield  {title} {\bibinfo {title} {{Physics of Strongly Magnetized
  Neutron Stars}},\ }\href {https://doi.org/10.1088/0034-4885/69/9/R03}
  {\bibfield  {journal} {\bibinfo  {journal} {Rept. Prog. Phys.}\ }\textbf
  {\bibinfo {volume} {69}},\ \bibinfo {pages} {2631} (\bibinfo {year}
  {2006})},\ \Eprint {https://arxiv.org/abs/astro-ph/0606674}
  {arXiv:astro-ph/0606674} \BibitemShut {NoStop}%
\bibitem [{\citenamefont {{S.Heyl, Jeremy}}(2012)}]{Heyl}%
  \BibitemOpen
  \bibfield  {author} {\bibinfo {author} {\bibnamefont {{S.Heyl, Jeremy}}},\
  }\bibfield  {title} {\bibinfo {title} {Astrophysical implications of
  strong-field qed},\ }in\ \href {https://doi.org/10.1051/iesc/2012qed04002}
  {\emph {\bibinfo {booktitle} {IESC Proceedings}}},\ \bibinfo {editor} {edited
  by\ \bibinfo {editor} {\bibnamefont {Array}}}\ (\bibinfo {year} {2012})\ p.\
  \bibinfo {pages} {04002}\BibitemShut {NoStop}%
\bibitem [{\citenamefont {Bernicot}\ and\ \citenamefont
  {Guillet}(2008)}]{1loop1}%
  \BibitemOpen
  \bibfield  {author} {\bibinfo {author} {\bibfnamefont {C.}~\bibnamefont
  {Bernicot}}\ and\ \bibinfo {author} {\bibfnamefont {J.~P.}\ \bibnamefont
  {Guillet}},\ }\bibfield  {title} {\bibinfo {title} {{Six-Photon Amplitudes in
  Scalar QED}},\ }\href {https://doi.org/10.1088/1126-6708/2008/01/059}
  {\bibfield  {journal} {\bibinfo  {journal} {JHEP}\ }\textbf {\bibinfo
  {volume} {01}},\ \bibinfo {pages} {059}},\ \Eprint
  {https://arxiv.org/abs/0711.4713} {arXiv:0711.4713 [hep-ph]} \BibitemShut
  {NoStop}%
\bibitem [{\citenamefont {Binoth}\ \emph {et~al.}(2007)\citenamefont {Binoth},
  \citenamefont {Heinrich}, \citenamefont {Gehrmann},\ and\ \citenamefont
  {Mastrolia}}]{Binoth:2007ca}%
  \BibitemOpen
  \bibfield  {author} {\bibinfo {author} {\bibfnamefont {T.}~\bibnamefont
  {Binoth}}, \bibinfo {author} {\bibfnamefont {G.}~\bibnamefont {Heinrich}},
  \bibinfo {author} {\bibfnamefont {T.}~\bibnamefont {Gehrmann}},\ and\
  \bibinfo {author} {\bibfnamefont {P.}~\bibnamefont {Mastrolia}},\ }\bibfield
  {title} {\bibinfo {title} {{Six-Photon Amplitudes}},\ }\href
  {https://doi.org/10.1016/j.physletb.2007.04.032} {\bibfield  {journal}
  {\bibinfo  {journal} {Phys. Lett. B}\ }\textbf {\bibinfo {volume} {649}},\
  \bibinfo {pages} {422} (\bibinfo {year} {2007})},\ \Eprint
  {https://arxiv.org/abs/hep-ph/0703311} {arXiv:hep-ph/0703311} \BibitemShut
  {NoStop}%
\bibitem [{\citenamefont {Badger}\ \emph {et~al.}(2009)\citenamefont {Badger},
  \citenamefont {Bjerrum-Bohr},\ and\ \citenamefont {Vanhove}}]{Badger}%
  \BibitemOpen
  \bibfield  {author} {\bibinfo {author} {\bibfnamefont {S.}~\bibnamefont
  {Badger}}, \bibinfo {author} {\bibfnamefont {N.~E.~J.}\ \bibnamefont
  {Bjerrum-Bohr}},\ and\ \bibinfo {author} {\bibfnamefont {P.}~\bibnamefont
  {Vanhove}},\ }\bibfield  {title} {\bibinfo {title} {{Simplicity in the
  Structure of QED and Gravity Amplitudes}},\ }\href
  {https://doi.org/10.1088/1126-6708/2009/02/038} {\bibfield  {journal}
  {\bibinfo  {journal} {JHEP}\ }\textbf {\bibinfo {volume} {02}},\ \bibinfo
  {pages} {038}},\ \Eprint {https://arxiv.org/abs/0811.3405} {arXiv:0811.3405
  [hep-th]} \BibitemShut {NoStop}%
\bibitem [{\citenamefont {A~H}\ \emph {et~al.}(2024{\natexlab{a}})\citenamefont
  {A~H}, \citenamefont {Chaubey}, \citenamefont {Fraaije}, \citenamefont
  {Hirschi},\ and\ \citenamefont {Shao}}]{2loop1}%
  \BibitemOpen
  \bibfield  {author} {\bibinfo {author} {\bibfnamefont {A.}~\bibnamefont
  {A~H}}, \bibinfo {author} {\bibfnamefont {E.}~\bibnamefont {Chaubey}},
  \bibinfo {author} {\bibfnamefont {M.}~\bibnamefont {Fraaije}}, \bibinfo
  {author} {\bibfnamefont {V.}~\bibnamefont {Hirschi}},\ and\ \bibinfo {author}
  {\bibfnamefont {H.-S.}\ \bibnamefont {Shao}},\ }\bibfield  {title} {\bibinfo
  {title} {{Light-by-light scattering at next-to-leading order in QCD and
  QED}},\ }\href {https://doi.org/10.1016/j.physletb.2024.138555} {\bibfield
  {journal} {\bibinfo  {journal} {Phys. Lett. B}\ }\textbf {\bibinfo {volume}
  {851}},\ \bibinfo {pages} {138555} (\bibinfo {year} {2024}{\natexlab{a}})},\
  \Eprint {https://arxiv.org/abs/2312.16956} {arXiv:2312.16956 [hep-ph]}
  \BibitemShut {NoStop}%
\bibitem [{\citenamefont {A~H}\ \emph {et~al.}(2024{\natexlab{b}})\citenamefont
  {A~H}, \citenamefont {Chaubey},\ and\ \citenamefont {Shao}}]{2loop2}%
  \BibitemOpen
  \bibfield  {author} {\bibinfo {author} {\bibfnamefont {A.}~\bibnamefont
  {A~H}}, \bibinfo {author} {\bibfnamefont {E.}~\bibnamefont {Chaubey}},\ and\
  \bibinfo {author} {\bibfnamefont {H.-S.}\ \bibnamefont {Shao}},\ }\bibfield
  {title} {\bibinfo {title} {{Two-loop massive QCD and QED helicity amplitudes
  for light-by-light scattering}},\ }\href
  {https://doi.org/10.1007/JHEP03(2024)121} {\bibfield  {journal} {\bibinfo
  {journal} {JHEP}\ }\textbf {\bibinfo {volume} {03}},\ \bibinfo {pages}
  {121}},\ \Eprint {https://arxiv.org/abs/2312.16966} {arXiv:2312.16966
  [hep-ph]} \BibitemShut {NoStop}%
\bibitem [{\citenamefont {Henn}\ \emph
  {et~al.}(2024{\natexlab{a}})\citenamefont {Henn}, \citenamefont
  {Matija\v{s}i\'c}, \citenamefont {Miczajka}, \citenamefont {Peraro},
  \citenamefont {Xu},\ and\ \citenamefont {Zhang}}]{1loop2}%
  \BibitemOpen
  \bibfield  {author} {\bibinfo {author} {\bibfnamefont {J.~M.}\ \bibnamefont
  {Henn}}, \bibinfo {author} {\bibfnamefont {A.}~\bibnamefont
  {Matija\v{s}i\'c}}, \bibinfo {author} {\bibfnamefont {J.}~\bibnamefont
  {Miczajka}}, \bibinfo {author} {\bibfnamefont {T.}~\bibnamefont {Peraro}},
  \bibinfo {author} {\bibfnamefont {Y.}~\bibnamefont {Xu}},\ and\ \bibinfo
  {author} {\bibfnamefont {Y.}~\bibnamefont {Zhang}},\ }\bibfield  {title}
  {\bibinfo {title} {{A computation of two-loop six-point Feynman integrals in
  dimensional regularization}},\ }\href
  {https://doi.org/10.1007/JHEP08(2024)027} {\bibfield  {journal} {\bibinfo
  {journal} {JHEP}\ }\textbf {\bibinfo {volume} {08}},\ \bibinfo {pages}
  {027}},\ \Eprint {https://arxiv.org/abs/2403.19742} {arXiv:2403.19742
  [hep-ph]} \BibitemShut {NoStop}%
\bibitem [{\citenamefont {Fadin}\ and\ \citenamefont
  {Lee}(2023)}]{Fadin:2023phc}%
  \BibitemOpen
  \bibfield  {author} {\bibinfo {author} {\bibfnamefont {V.~S.}\ \bibnamefont
  {Fadin}}\ and\ \bibinfo {author} {\bibfnamefont {R.~N.}\ \bibnamefont
  {Lee}},\ }\bibfield  {title} {\bibinfo {title} {{Two-loop radiative
  corrections to e$^{+}$e$^{-}$ \textrightarrow{}
  \ensuremath{\gamma}\ensuremath{\gamma}$^{*}$ cross section}},\ }\href
  {https://doi.org/10.1007/JHEP11(2023)148} {\bibfield  {journal} {\bibinfo
  {journal} {JHEP}\ }\textbf {\bibinfo {volume} {11}},\ \bibinfo {pages}
  {148}},\ \Eprint {https://arxiv.org/abs/2308.09479} {arXiv:2308.09479
  [hep-ph]} \BibitemShut {NoStop}%
\bibitem [{\citenamefont {Anastasiou}\ \emph {et~al.}(2002)\citenamefont
  {Anastasiou}, \citenamefont {Glover},\ and\ \citenamefont
  {Tejeda-Yeomans}}]{Anastasiou:2002zn}%
  \BibitemOpen
  \bibfield  {author} {\bibinfo {author} {\bibfnamefont {C.}~\bibnamefont
  {Anastasiou}}, \bibinfo {author} {\bibfnamefont {E.~W.~N.}\ \bibnamefont
  {Glover}},\ and\ \bibinfo {author} {\bibfnamefont {M.~E.}\ \bibnamefont
  {Tejeda-Yeomans}},\ }\bibfield  {title} {\bibinfo {title} {{Two loop QED and
  QCD corrections to massless fermion boson scattering}},\ }\href
  {https://doi.org/10.1016/S0550-3213(02)00140-2} {\bibfield  {journal}
  {\bibinfo  {journal} {Nucl. Phys. B}\ }\textbf {\bibinfo {volume} {629}},\
  \bibinfo {pages} {255} (\bibinfo {year} {2002})},\ \Eprint
  {https://arxiv.org/abs/hep-ph/0201274} {arXiv:hep-ph/0201274} \BibitemShut
  {NoStop}%
\bibitem [{\citenamefont {Dave}\ and\ \citenamefont
  {Torres~Bobadilla}(2025)}]{Dave:2024ewl}%
  \BibitemOpen
  \bibfield  {author} {\bibinfo {author} {\bibfnamefont {T.}~\bibnamefont
  {Dave}}\ and\ \bibinfo {author} {\bibfnamefont {W.~J.}\ \bibnamefont
  {Torres~Bobadilla}},\ }\bibfield  {title} {\bibinfo {title} {Helicity
  amplitudes in massless qed to higher orders in the dimensional regulator},\
  }\href {https://doi.org/10.1103/PhysRevD.111.036024} {\bibfield  {journal}
  {\bibinfo  {journal} {Phys. Rev. D}\ }\textbf {\bibinfo {volume} {111}},\
  \bibinfo {pages} {036024} (\bibinfo {year} {2025})},\ \Eprint
  {https://arxiv.org/abs/2411.07063} {arXiv:2411.07063 [hep-ph]} \BibitemShut
  {NoStop}%
\bibitem [{\citenamefont {Mahlon}(1994)}]{Mahlon}%
  \BibitemOpen
  \bibfield  {author} {\bibinfo {author} {\bibfnamefont {G.}~\bibnamefont
  {Mahlon}},\ }\bibfield  {title} {\bibinfo {title} {{One loop multi - photon
  helicity amplitudes}},\ }\href {https://doi.org/10.1103/PhysRevD.49.2197}
  {\bibfield  {journal} {\bibinfo  {journal} {Phys. Rev. D}\ }\textbf {\bibinfo
  {volume} {49}},\ \bibinfo {pages} {2197} (\bibinfo {year} {1994})},\ \Eprint
  {https://arxiv.org/abs/hep-ph/9311213} {arXiv:hep-ph/9311213} \BibitemShut
  {NoStop}%
\bibitem [{\citenamefont {Henn}\ \emph
  {et~al.}(2024{\natexlab{b}})\citenamefont {Henn}, \citenamefont
  {Matija\v{s}i\'c}, \citenamefont {Miczajka}, \citenamefont {Peraro},
  \citenamefont {Xu},\ and\ \citenamefont {Zhang}}]{Henn:2024ngj}%
  \BibitemOpen
  \bibfield  {author} {\bibinfo {author} {\bibfnamefont {J.~M.}\ \bibnamefont
  {Henn}}, \bibinfo {author} {\bibfnamefont {A.}~\bibnamefont
  {Matija\v{s}i\'c}}, \bibinfo {author} {\bibfnamefont {J.}~\bibnamefont
  {Miczajka}}, \bibinfo {author} {\bibfnamefont {T.}~\bibnamefont {Peraro}},
  \bibinfo {author} {\bibfnamefont {Y.}~\bibnamefont {Xu}},\ and\ \bibinfo
  {author} {\bibfnamefont {Y.}~\bibnamefont {Zhang}},\ }\bibfield  {title}
  {\bibinfo {title} {{A computation of two-loop six-point Feynman integrals in
  dimensional regularization}},\ }\href
  {https://doi.org/10.1007/JHEP08(2024)027} {\bibfield  {journal} {\bibinfo
  {journal} {JHEP}\ }\textbf {\bibinfo {volume} {08}},\ \bibinfo {pages}
  {027}},\ \Eprint {https://arxiv.org/abs/2403.19742} {arXiv:2403.19742
  [hep-ph]} \BibitemShut {NoStop}%
\bibitem [{\citenamefont {Forner}\ \emph {et~al.}(2025)\citenamefont {Forner},
  \citenamefont {Nega},\ and\ \citenamefont {Tancredi}}]{Forner:2024ojj}%
  \BibitemOpen
  \bibfield  {author} {\bibinfo {author} {\bibfnamefont {F.}~\bibnamefont
  {Forner}}, \bibinfo {author} {\bibfnamefont {C.}~\bibnamefont {Nega}},\ and\
  \bibinfo {author} {\bibfnamefont {L.}~\bibnamefont {Tancredi}},\ }\bibfield
  {title} {\bibinfo {title} {{On the photon self-energy to three loops in
  QED}},\ }\href {https://doi.org/10.1007/JHEP03(2025)148} {\bibfield
  {journal} {\bibinfo  {journal} {JHEP}\ }\textbf {\bibinfo {volume} {03}},\
  \bibinfo {pages} {148}},\ \Eprint {https://arxiv.org/abs/2411.19042}
  {arXiv:2411.19042 [hep-th]} \BibitemShut {NoStop}%
\bibitem [{\citenamefont {Ole\u{i}nik}(1967)}]{Oleinik1}%
  \BibitemOpen
  \bibfield  {author} {\bibinfo {author} {\bibfnamefont {V.~P.}\ \bibnamefont
  {Ole\u{i}nik}},\ }\bibfield  {title} {\bibinfo {title} {Resonance effects in
  the field of an intense laser beam},\ }\href@noop {} {\bibfield  {journal}
  {\bibinfo  {journal} {Soviet Journal of Experimental and Theoretical
  Physics}\ }\textbf {\bibinfo {volume} {25}} (\bibinfo {year}
  {1967})}\BibitemShut {NoStop}%
\bibitem [{\citenamefont {Ole\u{i}nik}(1968)}]{Oleinik2}%
  \BibitemOpen
  \bibfield  {author} {\bibinfo {author} {\bibfnamefont {V.~P.}\ \bibnamefont
  {Ole\u{i}nik}},\ }\bibfield  {title} {\bibinfo {title} {Resonance effects in
  the field of an intense laser beam ii},\ }\href@noop {} {\bibfield  {journal}
  {\bibinfo  {journal} {Soviet Journal of Experimental and Theoretical
  Physics}\ }\textbf {\bibinfo {volume} {26}} (\bibinfo {year}
  {1968})}\BibitemShut {NoStop}%
\bibitem [{\citenamefont {Fedorov}\ \emph {et~al.}(2006)\citenamefont
  {Fedorov}, \citenamefont {Efremov},\ and\ \citenamefont
  {Volkov}}]{FEDOROV2006413}%
  \BibitemOpen
  \bibfield  {author} {\bibinfo {author} {\bibfnamefont {M.}~\bibnamefont
  {Fedorov}}, \bibinfo {author} {\bibfnamefont {M.}~\bibnamefont {Efremov}},\
  and\ \bibinfo {author} {\bibfnamefont {P.}~\bibnamefont {Volkov}},\
  }\bibfield  {title} {\bibinfo {title} {Double- and multi-photon pair
  production and electron–positron entanglement},\ }\href
  {https://doi.org/https://doi.org/10.1016/j.optcom.2006.01.063} {\bibfield
  {journal} {\bibinfo  {journal} {Optics Communications}\ }\textbf {\bibinfo
  {volume} {264}},\ \bibinfo {pages} {413} (\bibinfo {year} {2006})},\ \bibinfo
  {note} {quantum Control of Light and Matter}\BibitemShut {NoStop}%
\bibitem [{\citenamefont {Loetstedt}\ and\ \citenamefont
  {Jentschura}(2009)}]{Loetstedt:2009zz}%
  \BibitemOpen
  \bibfield  {author} {\bibinfo {author} {\bibfnamefont {E.}~\bibnamefont
  {Loetstedt}}\ and\ \bibinfo {author} {\bibfnamefont {U.~D.}\ \bibnamefont
  {Jentschura}},\ }\bibfield  {title} {\bibinfo {title} {{Correlated two-photon
  emission by transitions of Dirac-Volkov states in intense laser fields: QED
  predictions}},\ }\href {https://doi.org/10.1103/PhysRevA.80.053419}
  {\bibfield  {journal} {\bibinfo  {journal} {Phys. Rev. A}\ }\textbf {\bibinfo
  {volume} {80}},\ \bibinfo {pages} {053419} (\bibinfo {year} {2009})},\
  \Eprint {https://arxiv.org/abs/0911.4765} {arXiv:0911.4765 [quant-ph]}
  \BibitemShut {NoStop}%
\bibitem [{\citenamefont {Roshchupkin}\ \emph {et~al.}(2021)\citenamefont
  {Roshchupkin}, \citenamefont {Larin},\ and\ \citenamefont
  {Dubov}}]{Roshchupkin:2021yhd}%
  \BibitemOpen
  \bibfield  {author} {\bibinfo {author} {\bibfnamefont {S.~P.}\ \bibnamefont
  {Roshchupkin}}, \bibinfo {author} {\bibfnamefont {N.~R.}\ \bibnamefont
  {Larin}},\ and\ \bibinfo {author} {\bibfnamefont {V.~V.}\ \bibnamefont
  {Dubov}},\ }\bibfield  {title} {\bibinfo {title} {{Resonant photoproduction
  of ultrarelativistic electron-positron pairs on a nucleus in moderate and
  strong monochromatic light fields}},\ }\href
  {https://doi.org/10.1103/PhysRevD.104.116011} {\bibfield  {journal} {\bibinfo
   {journal} {Phys. Rev. D}\ }\textbf {\bibinfo {volume} {104}},\ \bibinfo
  {pages} {116011} (\bibinfo {year} {2021})},\ \Eprint
  {https://arxiv.org/abs/2108.04955} {arXiv:2108.04955 [hep-ph]} \BibitemShut
  {NoStop}%
\bibitem [{\citenamefont {Bragin}\ and\ \citenamefont
  {Di~Piazza}(2020)}]{PhysRevD.102.116012}%
  \BibitemOpen
  \bibfield  {author} {\bibinfo {author} {\bibfnamefont {S.}~\bibnamefont
  {Bragin}}\ and\ \bibinfo {author} {\bibfnamefont {A.}~\bibnamefont
  {Di~Piazza}},\ }\bibfield  {title} {\bibinfo {title} {Electron-positron
  annihilation into two photons in an intense plane-wave field},\ }\href
  {https://doi.org/10.1103/PhysRevD.102.116012} {\bibfield  {journal} {\bibinfo
   {journal} {Phys. Rev. D}\ }\textbf {\bibinfo {volume} {102}},\ \bibinfo
  {pages} {116012} (\bibinfo {year} {2020})},\ \Eprint
  {https://arxiv.org/abs/2003.02231} {arXiv:2003.02231 [hep-ph]} \BibitemShut
  {NoStop}%
\bibitem [{\citenamefont {King}(2015)}]{DC1}%
  \BibitemOpen
  \bibfield  {author} {\bibinfo {author} {\bibfnamefont {B.}~\bibnamefont
  {King}},\ }\bibfield  {title} {\bibinfo {title} {Double compton scattering in
  a constant crossed field},\ }\href
  {https://doi.org/10.1103/PhysRevA.91.033415} {\bibfield  {journal} {\bibinfo
  {journal} {Phys. Rev. A}\ }\textbf {\bibinfo {volume} {91}},\ \bibinfo
  {pages} {033415} (\bibinfo {year} {2015})},\ \Eprint
  {https://arxiv.org/abs/1410.5478} {arXiv:1410.5478 [hep-ph]} \BibitemShut
  {NoStop}%
\bibitem [{\citenamefont {Mackenroth}\ and\ \citenamefont
  {Di~Piazza}(2013)}]{DC2}%
  \BibitemOpen
  \bibfield  {author} {\bibinfo {author} {\bibfnamefont {F.}~\bibnamefont
  {Mackenroth}}\ and\ \bibinfo {author} {\bibfnamefont {A.}~\bibnamefont
  {Di~Piazza}},\ }\bibfield  {title} {\bibinfo {title} {Nonlinear double
  compton scattering in the ultrarelativistic quantum regime},\ }\href
  {https://doi.org/10.1103/PhysRevLett.110.070402} {\bibfield  {journal}
  {\bibinfo  {journal} {Phys. Rev. Lett.}\ }\textbf {\bibinfo {volume} {110}},\
  \bibinfo {pages} {070402} (\bibinfo {year} {2013})},\ \Eprint
  {https://arxiv.org/abs/1208.3424} {arXiv:1208.3424 [hep-ph]} \BibitemShut
  {NoStop}%
\bibitem [{\citenamefont {Hu}\ and\ \citenamefont {Huang}(2014)}]{Trident1}%
  \BibitemOpen
  \bibfield  {author} {\bibinfo {author} {\bibfnamefont {H.}~\bibnamefont
  {Hu}}\ and\ \bibinfo {author} {\bibfnamefont {J.}~\bibnamefont {Huang}},\
  }\bibfield  {title} {\bibinfo {title} {Trident pair production in colliding
  bright x-ray laser beams},\ }\href
  {https://doi.org/10.1103/PhysRevA.89.033411} {\bibfield  {journal} {\bibinfo
  {journal} {Phys. Rev. A}\ }\textbf {\bibinfo {volume} {89}},\ \bibinfo
  {pages} {033411} (\bibinfo {year} {2014})},\ \Eprint
  {https://arxiv.org/abs/1308.5324} {arXiv:1308.5324 [physics.atom-ph]}
  \BibitemShut {NoStop}%
\bibitem [{\citenamefont {Ilderton}(2011)}]{Trident2}%
  \BibitemOpen
  \bibfield  {author} {\bibinfo {author} {\bibfnamefont {A.}~\bibnamefont
  {Ilderton}},\ }\bibfield  {title} {\bibinfo {title} {Trident pair production
  in strong laser pulses},\ }\href
  {https://doi.org/10.1103/PhysRevLett.106.020404} {\bibfield  {journal}
  {\bibinfo  {journal} {Phys. Rev. Lett.}\ }\textbf {\bibinfo {volume} {106}},\
  \bibinfo {pages} {020404} (\bibinfo {year} {2011})},\ \Eprint
  {https://arxiv.org/abs/1011.4072} {arXiv:1011.4072 [hep-ph]} \BibitemShut
  {NoStop}%
\bibitem [{\citenamefont {King}\ and\ \citenamefont {Ruhl}(2013)}]{Trident3}%
  \BibitemOpen
  \bibfield  {author} {\bibinfo {author} {\bibfnamefont {B.}~\bibnamefont
  {King}}\ and\ \bibinfo {author} {\bibfnamefont {H.}~\bibnamefont {Ruhl}},\
  }\bibfield  {title} {\bibinfo {title} {Trident pair production in a constant
  crossed field},\ }\href {https://doi.org/10.1103/PhysRevD.88.013005}
  {\bibfield  {journal} {\bibinfo  {journal} {Phys. Rev. D}\ }\textbf {\bibinfo
  {volume} {88}},\ \bibinfo {pages} {013005} (\bibinfo {year} {2013})},\
  \Eprint {https://arxiv.org/abs/1303.1356} {arXiv:1303.1356 [hep-ph]}
  \BibitemShut {NoStop}%
\bibitem [{\citenamefont {Dinu}\ and\ \citenamefont
  {Torgrimsson}(2018)}]{Trident4}%
  \BibitemOpen
  \bibfield  {author} {\bibinfo {author} {\bibfnamefont {V.}~\bibnamefont
  {Dinu}}\ and\ \bibinfo {author} {\bibfnamefont {G.}~\bibnamefont
  {Torgrimsson}},\ }\bibfield  {title} {\bibinfo {title} {Trident pair
  production in plane waves: Coherence, exchange, and spacetime
  inhomogeneity},\ }\href {https://doi.org/10.1103/PhysRevD.97.036021}
  {\bibfield  {journal} {\bibinfo  {journal} {Phys. Rev. D}\ }\textbf {\bibinfo
  {volume} {97}},\ \bibinfo {pages} {036021} (\bibinfo {year} {2018})},\
  \Eprint {https://arxiv.org/abs/1711.04344} {arXiv:1711.04344 [hep-ph]}
  \BibitemShut {NoStop}%
\bibitem [{\citenamefont {King}\ and\ \citenamefont
  {Fedotov}(2018)}]{Trident5}%
  \BibitemOpen
  \bibfield  {author} {\bibinfo {author} {\bibfnamefont {B.}~\bibnamefont
  {King}}\ and\ \bibinfo {author} {\bibfnamefont {A.~M.}\ \bibnamefont
  {Fedotov}},\ }\bibfield  {title} {\bibinfo {title} {{Effect of interference
  on the trident process in a constant crossed field}},\ }\href
  {https://doi.org/10.1103/PhysRevD.98.016005} {\bibfield  {journal} {\bibinfo
  {journal} {Phys. Rev. D}\ }\textbf {\bibinfo {volume} {98}},\ \bibinfo
  {pages} {016005} (\bibinfo {year} {2018})},\ \Eprint
  {https://arxiv.org/abs/1801.07300} {arXiv:1801.07300 [hep-ph]} \BibitemShut
  {NoStop}%
\bibitem [{\citenamefont {Torgrimsson}(2020)}]{DCT}%
  \BibitemOpen
  \bibfield  {author} {\bibinfo {author} {\bibfnamefont {G.}~\bibnamefont
  {Torgrimsson}},\ }\bibfield  {title} {\bibinfo {title} {Nonlinear photon
  trident versus double compton scattering and resummation of one-step terms},\
  }\href {https://doi.org/10.1103/PhysRevD.102.116008} {\bibfield  {journal}
  {\bibinfo  {journal} {Phys. Rev. D}\ }\textbf {\bibinfo {volume} {102}},\
  \bibinfo {pages} {116008} (\bibinfo {year} {2020})},\ \Eprint
  {https://arxiv.org/abs/2010.02128} {arXiv:2010.02128 [hep-ph]} \BibitemShut
  {NoStop}%
\bibitem [{\citenamefont {de~Vos}\ \emph {et~al.}(2024)\citenamefont {de~Vos},
  \citenamefont {Postema}, \citenamefont {Schaap}, \citenamefont {Di~Piazza},\
  and\ \citenamefont {Luiten}}]{deVos:2023pen}%
  \BibitemOpen
  \bibfield  {author} {\bibinfo {author} {\bibfnamefont {T.~D.~C.}\
  \bibnamefont {de~Vos}}, \bibinfo {author} {\bibfnamefont {J.~J.}\
  \bibnamefont {Postema}}, \bibinfo {author} {\bibfnamefont {B.~H.}\
  \bibnamefont {Schaap}}, \bibinfo {author} {\bibfnamefont {A.}~\bibnamefont
  {Di~Piazza}},\ and\ \bibinfo {author} {\bibfnamefont {O.~J.}\ \bibnamefont
  {Luiten}},\ }\bibfield  {title} {\bibinfo {title} {{Production of entangled x
  rays through nonlinear double Compton scattering}},\ }\href
  {https://doi.org/10.1103/PhysRevA.110.043702} {\bibfield  {journal} {\bibinfo
   {journal} {Phys. Rev. A}\ }\textbf {\bibinfo {volume} {110}},\ \bibinfo
  {pages} {043702} (\bibinfo {year} {2024})},\ \Eprint
  {https://arxiv.org/abs/2311.17807} {arXiv:2311.17807 [quant-ph]} \BibitemShut
  {NoStop}%
\bibitem [{\citenamefont {{Ba{\v{i}}er}}\ \emph {et~al.}(1975)\citenamefont
  {{Ba{\v{i}}er}}, \citenamefont {{Mil'Shte{\v{i}}n}},\ and\ \citenamefont
  {{Strakhovenko}}}]{1975JETP...42..961B}%
  \BibitemOpen
  \bibfield  {author} {\bibinfo {author} {\bibfnamefont {V.~N.}\ \bibnamefont
  {{Ba{\v{i}}er}}}, \bibinfo {author} {\bibfnamefont {A.~I.}\ \bibnamefont
  {{Mil'Shte{\v{i}}n}}},\ and\ \bibinfo {author} {\bibfnamefont {V.~M.}\
  \bibnamefont {{Strakhovenko}}},\ }\bibfield  {title} {\bibinfo {title}
  {{Interaction between a photon and an intense electromagnetic wave}},\
  }\href@noop {} {\bibfield  {journal} {\bibinfo  {journal} {Soviet Journal of
  Experimental and Theoretical Physics}\ }\textbf {\bibinfo {volume} {42}},\
  \bibinfo {pages} {961} (\bibinfo {year} {1975})}\BibitemShut {NoStop}%
\bibitem [{\citenamefont {Becker}\ and\ \citenamefont
  {Mitter}(1975)}]{BeckerVP}%
  \BibitemOpen
  \bibfield  {author} {\bibinfo {author} {\bibfnamefont {W.}~\bibnamefont
  {Becker}}\ and\ \bibinfo {author} {\bibfnamefont {H.}~\bibnamefont
  {Mitter}},\ }\bibfield  {title} {\bibinfo {title} {Vacuum polarization in
  laser fields},\ }\href {https://doi.org/10.1088/0305-4470/8/10/017}
  {\bibfield  {journal} {\bibinfo  {journal} {Journal of Physics A:
  Mathematical and General}\ }\textbf {\bibinfo {volume} {8}},\ \bibinfo
  {pages} {1638} (\bibinfo {year} {1975})}\BibitemShut {NoStop}%
\bibitem [{\citenamefont {Meuren}\ \emph
  {et~al.}(2013{\natexlab{a}})\citenamefont {Meuren}, \citenamefont {Keitel},\
  and\ \citenamefont {Di~Piazza}}]{PhysRevD.88.013007}%
  \BibitemOpen
  \bibfield  {author} {\bibinfo {author} {\bibfnamefont {S.}~\bibnamefont
  {Meuren}}, \bibinfo {author} {\bibfnamefont {C.~H.}\ \bibnamefont {Keitel}},\
  and\ \bibinfo {author} {\bibfnamefont {A.}~\bibnamefont {Di~Piazza}},\
  }\bibfield  {title} {\bibinfo {title} {Polarization operator for plane-wave
  background fields},\ }\href {https://doi.org/10.1103/PhysRevD.88.013007}
  {\bibfield  {journal} {\bibinfo  {journal} {Phys. Rev. D}\ }\textbf {\bibinfo
  {volume} {88}},\ \bibinfo {pages} {013007} (\bibinfo {year}
  {2013}{\natexlab{a}})},\ \Eprint {https://arxiv.org/abs/1304.7672}
  {arXiv:1304.7672 [hep-ph]} \BibitemShut {NoStop}%
\bibitem [{\citenamefont {Ilderton}\ and\ \citenamefont
  {Torgrimsson}(2016)}]{Ilderton:2016qpj}%
  \BibitemOpen
  \bibfield  {author} {\bibinfo {author} {\bibfnamefont {A.}~\bibnamefont
  {Ilderton}}\ and\ \bibinfo {author} {\bibfnamefont {G.}~\bibnamefont
  {Torgrimsson}},\ }\bibfield  {title} {\bibinfo {title} {{Worldline approach
  to helicity flip in plane waves}},\ }\href
  {https://doi.org/10.1103/PhysRevD.93.085006} {\bibfield  {journal} {\bibinfo
  {journal} {Phys. Rev. D}\ }\textbf {\bibinfo {volume} {93}},\ \bibinfo
  {pages} {085006} (\bibinfo {year} {2016})},\ \Eprint
  {https://arxiv.org/abs/1601.05021} {arXiv:1601.05021 [hep-th]} \BibitemShut
  {NoStop}%
\bibitem [{\citenamefont {Narozhnyi}(1968)}]{VP}%
  \BibitemOpen
  \bibfield  {author} {\bibinfo {author} {\bibfnamefont {N.}~\bibnamefont
  {Narozhnyi}},\ }\href@noop {} {\bibfield  {journal} {\bibinfo  {journal} {Zh.
  Eksp. Teor. Fiz.}\ }\textbf {\bibinfo {volume} {55}} (\bibinfo {year}
  {1968})},\ \bibinfo {note} {[Sov. Phys. JETP 28, (1969)]}\BibitemShut
  {NoStop}%
\bibitem [{\citenamefont {Meuren}\ \emph
  {et~al.}(2013{\natexlab{b}})\citenamefont {Meuren}, \citenamefont {Keitel},\
  and\ \citenamefont {Di~Piazza}}]{Meuren:2013oya}%
  \BibitemOpen
  \bibfield  {author} {\bibinfo {author} {\bibfnamefont {S.}~\bibnamefont
  {Meuren}}, \bibinfo {author} {\bibfnamefont {C.~H.}\ \bibnamefont {Keitel}},\
  and\ \bibinfo {author} {\bibfnamefont {A.}~\bibnamefont {Di~Piazza}},\
  }\bibfield  {title} {\bibinfo {title} {{Polarization operator for plane-wave
  background fields}},\ }\href {https://doi.org/10.1103/PhysRevD.88.013007}
  {\bibfield  {journal} {\bibinfo  {journal} {Phys. Rev. D}\ }\textbf {\bibinfo
  {volume} {88}},\ \bibinfo {pages} {013007} (\bibinfo {year}
  {2013}{\natexlab{b}})},\ \Eprint {https://arxiv.org/abs/1304.7672}
  {arXiv:1304.7672 [hep-ph]} \BibitemShut {NoStop}%
\bibitem [{\citenamefont {Dinu}\ \emph {et~al.}(2014)\citenamefont {Dinu},
  \citenamefont {Heinzl}, \citenamefont {Ilderton}, \citenamefont {Marklund},\
  and\ \citenamefont {Torgrimsson}}]{Dinu:2013gaa}%
  \BibitemOpen
  \bibfield  {author} {\bibinfo {author} {\bibfnamefont {V.}~\bibnamefont
  {Dinu}}, \bibinfo {author} {\bibfnamefont {T.}~\bibnamefont {Heinzl}},
  \bibinfo {author} {\bibfnamefont {A.}~\bibnamefont {Ilderton}}, \bibinfo
  {author} {\bibfnamefont {M.}~\bibnamefont {Marklund}},\ and\ \bibinfo
  {author} {\bibfnamefont {G.}~\bibnamefont {Torgrimsson}},\ }\bibfield
  {title} {\bibinfo {title} {{Vacuum refractive indices and helicity flip in
  strong-field QED}},\ }\href {https://doi.org/10.1103/PhysRevD.89.125003}
  {\bibfield  {journal} {\bibinfo  {journal} {Phys. Rev. D}\ }\textbf {\bibinfo
  {volume} {89}},\ \bibinfo {pages} {125003} (\bibinfo {year} {2014})},\
  \Eprint {https://arxiv.org/abs/1312.6419} {arXiv:1312.6419 [hep-ph]}
  \BibitemShut {NoStop}%
\bibitem [{\citenamefont {Meuren}\ \emph {et~al.}(2015)\citenamefont {Meuren},
  \citenamefont {Hatsagortsyan}, \citenamefont {Keitel},\ and\ \citenamefont
  {Di~Piazza}}]{Meuren:2014uia}%
  \BibitemOpen
  \bibfield  {author} {\bibinfo {author} {\bibfnamefont {S.}~\bibnamefont
  {Meuren}}, \bibinfo {author} {\bibfnamefont {K.~Z.}\ \bibnamefont
  {Hatsagortsyan}}, \bibinfo {author} {\bibfnamefont {C.~H.}\ \bibnamefont
  {Keitel}},\ and\ \bibinfo {author} {\bibfnamefont {A.}~\bibnamefont
  {Di~Piazza}},\ }\bibfield  {title} {\bibinfo {title} {{Polarization operator
  approach to pair creation in short laser pulses}},\ }\href
  {https://doi.org/10.1103/PhysRevD.91.013009} {\bibfield  {journal} {\bibinfo
  {journal} {Phys. Rev. D}\ }\textbf {\bibinfo {volume} {91}},\ \bibinfo
  {pages} {013009} (\bibinfo {year} {2015})},\ \Eprint
  {https://arxiv.org/abs/1406.7235} {arXiv:1406.7235 [hep-ph]} \BibitemShut
  {NoStop}%
\bibitem [{\citenamefont {Di~Piazza}\ \emph {et~al.}(2007)\citenamefont
  {Di~Piazza}, \citenamefont {Milstein},\ and\ \citenamefont
  {Keitel}}]{DiPiazza:2007yx}%
  \BibitemOpen
  \bibfield  {author} {\bibinfo {author} {\bibfnamefont {A.}~\bibnamefont
  {Di~Piazza}}, \bibinfo {author} {\bibfnamefont {A.~I.}\ \bibnamefont
  {Milstein}},\ and\ \bibinfo {author} {\bibfnamefont {C.~H.}\ \bibnamefont
  {Keitel}},\ }\bibfield  {title} {\bibinfo {title} {{Photon splitting in a
  laser field}},\ }\href {https://doi.org/10.1103/PhysRevA.76.032103}
  {\bibfield  {journal} {\bibinfo  {journal} {Phys. Rev. A}\ }\textbf {\bibinfo
  {volume} {76}},\ \bibinfo {pages} {032103} (\bibinfo {year} {2007})},\
  \Eprint {https://arxiv.org/abs/0704.0695} {arXiv:0704.0695 [hep-ph]}
  \BibitemShut {NoStop}%
\bibitem [{\citenamefont {Adler}\ \emph {et~al.}(1970)\citenamefont {Adler},
  \citenamefont {Bahcall}, \citenamefont {Callan},\ and\ \citenamefont
  {Rosenbluth}}]{PhysRevLett.25.1061}%
  \BibitemOpen
  \bibfield  {author} {\bibinfo {author} {\bibfnamefont {S.~L.}\ \bibnamefont
  {Adler}}, \bibinfo {author} {\bibfnamefont {J.~N.}\ \bibnamefont {Bahcall}},
  \bibinfo {author} {\bibfnamefont {C.~G.}\ \bibnamefont {Callan}},\ and\
  \bibinfo {author} {\bibfnamefont {M.~N.}\ \bibnamefont {Rosenbluth}},\
  }\bibfield  {title} {\bibinfo {title} {Photon splitting in a strong magnetic
  field},\ }\href {https://doi.org/10.1103/PhysRevLett.25.1061} {\bibfield
  {journal} {\bibinfo  {journal} {Phys. Rev. Lett.}\ }\textbf {\bibinfo
  {volume} {25}},\ \bibinfo {pages} {1061} (\bibinfo {year}
  {1970})}\BibitemShut {NoStop}%
\bibitem [{\citenamefont {Bialynicka-Birula}\ and\ \citenamefont
  {Bialynicki-Birula}(1970)}]{PhysRevD.2.2341}%
  \BibitemOpen
  \bibfield  {author} {\bibinfo {author} {\bibfnamefont {Z.}~\bibnamefont
  {Bialynicka-Birula}}\ and\ \bibinfo {author} {\bibfnamefont {I.}~\bibnamefont
  {Bialynicki-Birula}},\ }\bibfield  {title} {\bibinfo {title} {Nonlinear
  effects in quantum electrodynamics. photon propagation and photon splitting
  in an external field},\ }\href {https://doi.org/10.1103/PhysRevD.2.2341}
  {\bibfield  {journal} {\bibinfo  {journal} {Phys. Rev. D}\ }\textbf {\bibinfo
  {volume} {2}},\ \bibinfo {pages} {2341} (\bibinfo {year} {1970})}\BibitemShut
  {NoStop}%
\bibitem [{\citenamefont {{Baier}}\ \emph {et~al.}(1997)\citenamefont
  {{Baier}}, \citenamefont {{Mil'shtein}},\ and\ \citenamefont
  {{Shaisultanov}}}]{Baier}%
  \BibitemOpen
  \bibfield  {author} {\bibinfo {author} {\bibfnamefont {V.~N.}\ \bibnamefont
  {{Baier}}}, \bibinfo {author} {\bibfnamefont {A.~I.}\ \bibnamefont
  {{Mil'shtein}}},\ and\ \bibinfo {author} {\bibfnamefont {R.~Z.}\ \bibnamefont
  {{Shaisultanov}}},\ }\bibfield  {title} {\bibinfo {title} {Photon splitting
  in an ultrastrong magnetic field},\ }\href
  {https://doi.org/https://doi.org/10.1134/1.558149} {\bibfield  {journal}
  {\bibinfo  {journal} {J. Exp. Theor. Phys.}\ }\textbf {\bibinfo {volume}
  {84}},\ \bibinfo {pages} {29–34} (\bibinfo {year} {1997})}\BibitemShut
  {NoStop}%
\bibitem [{\citenamefont {Baier}\ \emph {et~al.}(1996)\citenamefont {Baier},
  \citenamefont {Milstein},\ and\ \citenamefont
  {Shaisultanov}}]{PhysRevLett.77.1691}%
  \BibitemOpen
  \bibfield  {author} {\bibinfo {author} {\bibfnamefont {V.~N.}\ \bibnamefont
  {Baier}}, \bibinfo {author} {\bibfnamefont {A.~I.}\ \bibnamefont
  {Milstein}},\ and\ \bibinfo {author} {\bibfnamefont {R.~Z.}\ \bibnamefont
  {Shaisultanov}},\ }\bibfield  {title} {\bibinfo {title} {Photon splitting in
  a very strong magnetic field},\ }\href
  {https://doi.org/10.1103/PhysRevLett.77.1691} {\bibfield  {journal} {\bibinfo
   {journal} {Phys. Rev. Lett.}\ }\textbf {\bibinfo {volume} {77}},\ \bibinfo
  {pages} {1691} (\bibinfo {year} {1996})}\BibitemShut {NoStop}%
\bibitem [{\citenamefont {Adler}(1971)}]{ADLER1971599}%
  \BibitemOpen
  \bibfield  {author} {\bibinfo {author} {\bibfnamefont {S.~L.}\ \bibnamefont
  {Adler}},\ }\bibfield  {title} {\bibinfo {title} {Photon splitting and photon
  dispersion in a strong magnetic field},\ }\href
  {https://doi.org/https://doi.org/10.1016/0003-4916(71)90154-0} {\bibfield
  {journal} {\bibinfo  {journal} {Annals of Physics}\ }\textbf {\bibinfo
  {volume} {67}},\ \bibinfo {pages} {599} (\bibinfo {year} {1971})}\BibitemShut
  {NoStop}%
\bibitem [{\citenamefont {Adler}\ and\ \citenamefont
  {Schubert}(1996)}]{Adler:1996cja}%
  \BibitemOpen
  \bibfield  {author} {\bibinfo {author} {\bibfnamefont {S.~L.}\ \bibnamefont
  {Adler}}\ and\ \bibinfo {author} {\bibfnamefont {C.}~\bibnamefont
  {Schubert}},\ }\bibfield  {title} {\bibinfo {title} {{Photon splitting in a
  strong magnetic field: Recalculation and comparison with previous
  calculations}},\ }\href {https://doi.org/10.1103/PhysRevLett.77.1695}
  {\bibfield  {journal} {\bibinfo  {journal} {Phys. Rev. Lett.}\ }\textbf
  {\bibinfo {volume} {77}},\ \bibinfo {pages} {1695} (\bibinfo {year}
  {1996})},\ \Eprint {https://arxiv.org/abs/hep-th/9605035}
  {arXiv:hep-th/9605035} \BibitemShut {NoStop}%
\bibitem [{\citenamefont {Itzykson}\ and\ \citenamefont
  {Zuber}(1980)}]{Itzykson:1980rh}%
  \BibitemOpen
  \bibfield  {author} {\bibinfo {author} {\bibfnamefont {C.}~\bibnamefont
  {Itzykson}}\ and\ \bibinfo {author} {\bibfnamefont {J.~B.}\ \bibnamefont
  {Zuber}},\ }\href@noop {} {\emph {\bibinfo {title} {{Quantum Field
  Theory}}}},\ International Series In Pure and Applied Physics\ (\bibinfo
  {publisher} {McGraw-Hill},\ \bibinfo {address} {New York},\ \bibinfo {year}
  {1980})\BibitemShut {NoStop}%
\bibitem [{\citenamefont {Martin}\ \emph {et~al.}(2003)\citenamefont {Martin},
  \citenamefont {Schubert},\ and\ \citenamefont
  {Villanueva~Sandoval}}]{ChrisLow}%
  \BibitemOpen
  \bibfield  {author} {\bibinfo {author} {\bibfnamefont {L.~C.}\ \bibnamefont
  {Martin}}, \bibinfo {author} {\bibfnamefont {C.}~\bibnamefont {Schubert}},\
  and\ \bibinfo {author} {\bibfnamefont {V.~M.}\ \bibnamefont
  {Villanueva~Sandoval}},\ }\bibfield  {title} {\bibinfo {title} {{On the
  low-energy limit of the QED N photon amplitudes}},\ }\href
  {https://doi.org/10.1016/S0550-3213(03)00578-9} {\bibfield  {journal}
  {\bibinfo  {journal} {Nucl. Phys. B}\ }\textbf {\bibinfo {volume} {668}},\
  \bibinfo {pages} {335} (\bibinfo {year} {2003})},\ \Eprint
  {https://arxiv.org/abs/hep-th/0301022} {arXiv:hep-th/0301022} \BibitemShut
  {NoStop}%
\bibitem [{\citenamefont {Edwards}\ \emph {et~al.}(2018)\citenamefont
  {Edwards}, \citenamefont {Huet},\ and\ \citenamefont
  {Schubert}}]{Edwards:2018vjd}%
  \BibitemOpen
  \bibfield  {author} {\bibinfo {author} {\bibfnamefont {J.~P.}\ \bibnamefont
  {Edwards}}, \bibinfo {author} {\bibfnamefont {A.}~\bibnamefont {Huet}},\ and\
  \bibinfo {author} {\bibfnamefont {C.}~\bibnamefont {Schubert}},\ }\bibfield
  {title} {\bibinfo {title} {{On the low-energy limit of the QED N-photon
  amplitudes: part 2}},\ }\href
  {https://doi.org/10.1016/j.nuclphysb.2018.07.026} {\bibfield  {journal}
  {\bibinfo  {journal} {Nucl. Phys. B}\ }\textbf {\bibinfo {volume} {935}},\
  \bibinfo {pages} {198} (\bibinfo {year} {2018})},\ \Eprint
  {https://arxiv.org/abs/1807.10697} {arXiv:1807.10697 [hep-th]} \BibitemShut
  {NoStop}%
\bibitem [{\citenamefont {Dunne}\ and\ \citenamefont
  {Schubert}(2002{\natexlab{a}})}]{Dunne:2002qf}%
  \BibitemOpen
  \bibfield  {author} {\bibinfo {author} {\bibfnamefont {G.~V.}\ \bibnamefont
  {Dunne}}\ and\ \bibinfo {author} {\bibfnamefont {C.}~\bibnamefont
  {Schubert}},\ }\bibfield  {title} {\bibinfo {title} {{Two loop selfdual
  Euler-Heisenberg Lagrangians. 1. Real part and helicity amplitudes}},\ }\href
  {https://doi.org/10.1088/1126-6708/2002/08/053} {\bibfield  {journal}
  {\bibinfo  {journal} {JHEP}\ }\textbf {\bibinfo {volume} {08}},\ \bibinfo
  {pages} {053}},\ \Eprint {https://arxiv.org/abs/hep-th/0205004}
  {arXiv:hep-th/0205004} \BibitemShut {NoStop}%
\bibitem [{\citenamefont {Baier}\ \emph {et~al.}(2018)\citenamefont {Baier},
  \citenamefont {Rebhan},\ and\ \citenamefont {W{\"o}dlinger}}]{Baier:2018vso}%
  \BibitemOpen
  \bibfield  {author} {\bibinfo {author} {\bibfnamefont {R.}~\bibnamefont
  {Baier}}, \bibinfo {author} {\bibfnamefont {A.}~\bibnamefont {Rebhan}},\ and\
  \bibinfo {author} {\bibfnamefont {M.}~\bibnamefont {W{\"o}dlinger}},\
  }\bibfield  {title} {\bibinfo {title} {{Light-by-Light Scattering in the
  Presence of Magnetic Fields}},\ }\href
  {https://doi.org/10.1103/PhysRevD.98.056001} {\bibfield  {journal} {\bibinfo
  {journal} {Phys. Rev. D}\ }\textbf {\bibinfo {volume} {98}},\ \bibinfo
  {pages} {056001} (\bibinfo {year} {2018})},\ \Eprint
  {https://arxiv.org/abs/1804.06140} {arXiv:1804.06140 [hep-ph]} \BibitemShut
  {NoStop}%
\bibitem [{\citenamefont {Ahmadiniaz}\ \emph
  {et~al.}(2024{\natexlab{a}})\citenamefont {Ahmadiniaz}, \citenamefont
  {Lopez-Lopez},\ and\ \citenamefont {Schubert}}]{Ahmadiniaz:2023jwd}%
  \BibitemOpen
  \bibfield  {author} {\bibinfo {author} {\bibfnamefont {N.}~\bibnamefont
  {Ahmadiniaz}}, \bibinfo {author} {\bibfnamefont {M.~A.}\ \bibnamefont
  {Lopez-Lopez}},\ and\ \bibinfo {author} {\bibfnamefont {C.}~\bibnamefont
  {Schubert}},\ }\bibfield  {title} {\bibinfo {title} {{Low-energy limit of
  N-photon amplitudes in a constant field}},\ }\href
  {https://doi.org/10.1016/j.physletb.2024.138610} {\bibfield  {journal}
  {\bibinfo  {journal} {Phys. Lett. B}\ }\textbf {\bibinfo {volume} {852}},\
  \bibinfo {pages} {138610} (\bibinfo {year} {2024}{\natexlab{a}})},\ \Eprint
  {https://arxiv.org/abs/2312.07047} {arXiv:2312.07047 [hep-th]} \BibitemShut
  {NoStop}%
\bibitem [{\citenamefont {Lopez-Lopez}(2025)}]{MishaLow}%
  \BibitemOpen
  \bibfield  {author} {\bibinfo {author} {\bibfnamefont {M.~A.}\ \bibnamefont
  {Lopez-Lopez}},\ }\bibfield  {title} {\bibinfo {title} {{Low-energy limit of
  N-photon amplitudes in a constant field: Part II}},\ }\href
  {https://doi.org/10.1016/j.physletb.2024.139157} {\bibfield  {journal}
  {\bibinfo  {journal} {Phys. Lett. B}\ }\textbf {\bibinfo {volume} {860}},\
  \bibinfo {pages} {139157} (\bibinfo {year} {2025})},\ \Eprint
  {https://arxiv.org/abs/2408.16474} {arXiv:2408.16474 [hep-th]} \BibitemShut
  {NoStop}%
\bibitem [{\citenamefont {Edwards}\ and\ \citenamefont
  {Schubert}(2019)}]{UsRep}%
  \BibitemOpen
  \bibfield  {author} {\bibinfo {author} {\bibfnamefont {J.~P.}\ \bibnamefont
  {Edwards}}\ and\ \bibinfo {author} {\bibfnamefont {C.}~\bibnamefont
  {Schubert}},\ }\bibfield  {title} {\bibinfo {title} {{Quantum mechanical path
  integrals in the first quantised approach to quantum field theory}}\
  }(\bibinfo {year} {2019})\ \Eprint {https://arxiv.org/abs/1912.10004}
  {arXiv:1912.10004 [hep-th]} \BibitemShut {NoStop}%
\bibitem [{\citenamefont {Schubert}(2001)}]{ChrisRev}%
  \BibitemOpen
  \bibfield  {author} {\bibinfo {author} {\bibfnamefont {C.}~\bibnamefont
  {Schubert}},\ }\bibfield  {title} {\bibinfo {title} {{Perturbative quantum
  field theory in the string inspired formalism}},\ }\href
  {https://doi.org/10.1016/S0370-1573(01)00013-8} {\bibfield  {journal}
  {\bibinfo  {journal} {Phys. Rept.}\ }\textbf {\bibinfo {volume} {355}},\
  \bibinfo {pages} {73} (\bibinfo {year} {2001})},\ \Eprint
  {https://arxiv.org/abs/hep-th/0101036} {arXiv:hep-th/0101036} \BibitemShut
  {NoStop}%
\bibitem [{\citenamefont {Ahmadiniaz}\ \emph
  {et~al.}(2024{\natexlab{b}})\citenamefont {Ahmadiniaz}, \citenamefont
  {Banda~Guzm\'an}, \citenamefont {Edwards}, \citenamefont {Lopez-Lopez},
  \citenamefont {Mata}, \citenamefont {Rodriguez~Chac\'on}, \citenamefont
  {Schubert},\ and\ \citenamefont {Shaisultanov}}]{Ahmadiniaz:2024rvi}%
  \BibitemOpen
  \bibfield  {author} {\bibinfo {author} {\bibfnamefont {N.}~\bibnamefont
  {Ahmadiniaz}}, \bibinfo {author} {\bibfnamefont {V.~M.}\ \bibnamefont
  {Banda~Guzm\'an}}, \bibinfo {author} {\bibfnamefont {J.}~\bibnamefont
  {Edwards}}, \bibinfo {author} {\bibfnamefont {M.~A.}\ \bibnamefont
  {Lopez-Lopez}}, \bibinfo {author} {\bibfnamefont {C.~M.}\ \bibnamefont
  {Mata}}, \bibinfo {author} {\bibfnamefont {L.~A.}\ \bibnamefont
  {Rodriguez~Chac\'on}}, \bibinfo {author} {\bibfnamefont {C.}~\bibnamefont
  {Schubert}},\ and\ \bibinfo {author} {\bibfnamefont {R.}~\bibnamefont
  {Shaisultanov}},\ }\bibfield  {title} {\bibinfo {title} {{Worldline
  integration of photon amplitudes}},\ }\href
  {https://doi.org/10.22323/1.467.0011} {\bibfield  {journal} {\bibinfo
  {journal} {PoS}\ }\textbf {\bibinfo {volume} {LL2024}},\ \bibinfo {pages}
  {011} (\bibinfo {year} {2024}{\natexlab{b}})},\ \Eprint
  {https://arxiv.org/abs/2407.07383} {arXiv:2407.07383 [hep-th]} \BibitemShut
  {NoStop}%
\bibitem [{\citenamefont {Schubert}(2023)}]{Schubert:2023bed}%
  \BibitemOpen
  \bibfield  {author} {\bibinfo {author} {\bibfnamefont {C.}~\bibnamefont
  {Schubert}},\ }\bibfield  {title} {\bibinfo {title} {{The worldline formalism
  in strong-field QED}},\ }\href
  {https://doi.org/10.1088/1742-6596/2494/1/012020} {\bibfield  {journal}
  {\bibinfo  {journal} {J. Phys. Conf. Ser.}\ }\textbf {\bibinfo {volume}
  {2494}},\ \bibinfo {pages} {012020} (\bibinfo {year} {2023})},\ \Eprint
  {https://arxiv.org/abs/2304.07404} {arXiv:2304.07404 [hep-th]} \BibitemShut
  {NoStop}%
\bibitem [{\citenamefont {Edwards}\ and\ \citenamefont
  {Schubert}(2022)}]{Edwards:2021uif}%
  \BibitemOpen
  \bibfield  {author} {\bibinfo {author} {\bibfnamefont {J.~P.}\ \bibnamefont
  {Edwards}}\ and\ \bibinfo {author} {\bibfnamefont {C.}~\bibnamefont
  {Schubert}},\ }\bibfield  {title} {\bibinfo {title} {{Plane Wave Backgrounds
  in the Worldline Formalism}},\ }\href
  {https://doi.org/10.1088/1742-6596/2249/1/012019} {\bibfield  {journal}
  {\bibinfo  {journal} {J. Phys. Conf. Ser.}\ }\textbf {\bibinfo {volume}
  {2249}},\ \bibinfo {pages} {012019} (\bibinfo {year} {2022})},\ \Eprint
  {https://arxiv.org/abs/2112.13944} {arXiv:2112.13944 [hep-th]} \BibitemShut
  {NoStop}%
\bibitem [{\citenamefont {Ahmadiniaz}\ \emph
  {et~al.}(2023{\natexlab{b}})\citenamefont {Ahmadiniaz}, \citenamefont
  {Lopez-Arcos}, \citenamefont {Lopez-Lopez},\ and\ \citenamefont
  {Schubert}}]{Ahmadiniaz:2020jgo}%
  \BibitemOpen
  \bibfield  {author} {\bibinfo {author} {\bibfnamefont {N.}~\bibnamefont
  {Ahmadiniaz}}, \bibinfo {author} {\bibfnamefont {C.}~\bibnamefont
  {Lopez-Arcos}}, \bibinfo {author} {\bibfnamefont {M.~A.}\ \bibnamefont
  {Lopez-Lopez}},\ and\ \bibinfo {author} {\bibfnamefont {C.}~\bibnamefont
  {Schubert}},\ }\bibfield  {title} {\bibinfo {title} {{The QED four-photon
  amplitudes off-shell: Part 1}},\ }\href
  {https://doi.org/10.1016/j.nuclphysb.2023.116216} {\bibfield  {journal}
  {\bibinfo  {journal} {Nucl. Phys. B}\ }\textbf {\bibinfo {volume} {991}},\
  \bibinfo {pages} {116216} (\bibinfo {year} {2023}{\natexlab{b}})},\ \Eprint
  {https://arxiv.org/abs/2012.11791} {arXiv:2012.11791 [hep-th]} \BibitemShut
  {NoStop}%
\bibitem [{\citenamefont {Ahmadiniaz}\ \emph
  {et~al.}(2023{\natexlab{c}})\citenamefont {Ahmadiniaz}, \citenamefont
  {Lopez-Arcos}, \citenamefont {Lopez-Lopez},\ and\ \citenamefont
  {Schubert}}]{Ahmadiniaz:2023vrk}%
  \BibitemOpen
  \bibfield  {author} {\bibinfo {author} {\bibfnamefont {N.}~\bibnamefont
  {Ahmadiniaz}}, \bibinfo {author} {\bibfnamefont {C.}~\bibnamefont
  {Lopez-Arcos}}, \bibinfo {author} {\bibfnamefont {M.~A.}\ \bibnamefont
  {Lopez-Lopez}},\ and\ \bibinfo {author} {\bibfnamefont {C.}~\bibnamefont
  {Schubert}},\ }\bibfield  {title} {\bibinfo {title} {{The QED four-photon
  amplitudes off-shell: Part 2}},\ }\href
  {https://doi.org/10.1016/j.nuclphysb.2023.116217} {\bibfield  {journal}
  {\bibinfo  {journal} {Nucl. Phys. B}\ }\textbf {\bibinfo {volume} {991}},\
  \bibinfo {pages} {116217} (\bibinfo {year} {2023}{\natexlab{c}})},\ \Eprint
  {https://arxiv.org/abs/2303.12072} {arXiv:2303.12072 [hep-th]} \BibitemShut
  {NoStop}%
\bibitem [{\citenamefont {Edwards}\ \emph {et~al.}(2022)\citenamefont
  {Edwards}, \citenamefont {Mata},\ and\ \citenamefont
  {Schubert}}]{Edwards:2022dbd}%
  \BibitemOpen
  \bibfield  {author} {\bibinfo {author} {\bibfnamefont {J.~P.}\ \bibnamefont
  {Edwards}}, \bibinfo {author} {\bibfnamefont {C.~M.}\ \bibnamefont {Mata}},\
  and\ \bibinfo {author} {\bibfnamefont {C.}~\bibnamefont {Schubert}},\
  }\bibfield  {title} {\bibinfo {title} {{One-loop amplitudes in the worldline
  formalism}},\ }\href {https://doi.org/10.1088/1402-4896/ac6a1e} {\bibfield
  {journal} {\bibinfo  {journal} {Phys. Scripta}\ }\textbf {\bibinfo {volume}
  {97}},\ \bibinfo {pages} {064002} (\bibinfo {year} {2022})},\ \Eprint
  {https://arxiv.org/abs/2201.12457} {arXiv:2201.12457 [hep-th]} \BibitemShut
  {NoStop}%
\bibitem [{\citenamefont {Reuter}\ \emph {et~al.}(1997)\citenamefont {Reuter},
  \citenamefont {Schmidt},\ and\ \citenamefont {Schubert}}]{Reuter:1996zm}%
  \BibitemOpen
  \bibfield  {author} {\bibinfo {author} {\bibfnamefont {M.}~\bibnamefont
  {Reuter}}, \bibinfo {author} {\bibfnamefont {M.~G.}\ \bibnamefont
  {Schmidt}},\ and\ \bibinfo {author} {\bibfnamefont {C.}~\bibnamefont
  {Schubert}},\ }\bibfield  {title} {\bibinfo {title} {{Constant external
  fields in gauge theory and the spin 0, 1/2, 1 path integrals}},\ }\href
  {https://doi.org/10.1006/aphy.1997.5716} {\bibfield  {journal} {\bibinfo
  {journal} {Annals Phys.}\ }\textbf {\bibinfo {volume} {259}},\ \bibinfo
  {pages} {313} (\bibinfo {year} {1997})},\ \Eprint
  {https://arxiv.org/abs/hep-th/9610191} {arXiv:hep-th/9610191} \BibitemShut
  {NoStop}%
\bibitem [{\citenamefont {Schmidt}\ and\ \citenamefont
  {Schubert}(1993)}]{Schmidt:1993rk}%
  \BibitemOpen
  \bibfield  {author} {\bibinfo {author} {\bibfnamefont {M.~G.}\ \bibnamefont
  {Schmidt}}\ and\ \bibinfo {author} {\bibfnamefont {C.}~\bibnamefont
  {Schubert}},\ }\bibfield  {title} {\bibinfo {title} {{On the calculation of
  effective actions by string methods}},\ }\href
  {https://doi.org/10.1016/0370-2693(93)91537-W} {\bibfield  {journal}
  {\bibinfo  {journal} {Phys. Lett. B}\ }\textbf {\bibinfo {volume} {318}},\
  \bibinfo {pages} {438} (\bibinfo {year} {1993})},\ \Eprint
  {https://arxiv.org/abs/hep-th/9309055} {arXiv:hep-th/9309055} \BibitemShut
  {NoStop}%
\bibitem [{\citenamefont {Shaisultanov}(1996)}]{Shaisultanov:1995tm}%
  \BibitemOpen
  \bibfield  {author} {\bibinfo {author} {\bibfnamefont {R.}~\bibnamefont
  {Shaisultanov}},\ }\bibfield  {title} {\bibinfo {title} {{On the string
  inspired approach to QED in external field}},\ }\href
  {https://doi.org/10.1016/0370-2693(96)00359-0} {\bibfield  {journal}
  {\bibinfo  {journal} {Phys. Lett. B}\ }\textbf {\bibinfo {volume} {378}},\
  \bibinfo {pages} {354} (\bibinfo {year} {1996})},\ \Eprint
  {https://arxiv.org/abs/hep-th/9512142} {arXiv:hep-th/9512142} \BibitemShut
  {NoStop}%
\bibitem [{\citenamefont {Kors}\ and\ \citenamefont
  {Schmidt}(1999)}]{Kors:1998ew}%
  \BibitemOpen
  \bibfield  {author} {\bibinfo {author} {\bibfnamefont {B.}~\bibnamefont
  {Kors}}\ and\ \bibinfo {author} {\bibfnamefont {M.~G.}\ \bibnamefont
  {Schmidt}},\ }\bibfield  {title} {\bibinfo {title} {{The Effective two loop
  Euler-Heisenberg action for scalar and spinor QED in a general constant
  background field}},\ }\href {https://doi.org/10.1007/s100520050331}
  {\bibfield  {journal} {\bibinfo  {journal} {Eur. Phys. J. C}\ }\textbf
  {\bibinfo {volume} {6}},\ \bibinfo {pages} {175} (\bibinfo {year} {1999})},\
  \Eprint {https://arxiv.org/abs/hep-th/9803144} {arXiv:hep-th/9803144}
  \BibitemShut {NoStop}%
\bibitem [{\citenamefont {Fliegner}\ \emph {et~al.}(1997)\citenamefont
  {Fliegner}, \citenamefont {Reuter}, \citenamefont {Schmidt},\ and\
  \citenamefont {Schubert}}]{Fliegner:1997ra}%
  \BibitemOpen
  \bibfield  {author} {\bibinfo {author} {\bibfnamefont {D.}~\bibnamefont
  {Fliegner}}, \bibinfo {author} {\bibfnamefont {M.}~\bibnamefont {Reuter}},
  \bibinfo {author} {\bibfnamefont {M.~G.}\ \bibnamefont {Schmidt}},\ and\
  \bibinfo {author} {\bibfnamefont {C.}~\bibnamefont {Schubert}},\ }\bibfield
  {title} {\bibinfo {title} {{The Two loop Euler-Heisenberg Lagrangian in
  dimensional renormalization}},\ }\href {https://doi.org/10.1007/BF02634170}
  {\bibfield  {journal} {\bibinfo  {journal} {Theor. Math. Phys.}\ }\textbf
  {\bibinfo {volume} {113}},\ \bibinfo {pages} {1442} (\bibinfo {year}
  {1997})},\ \Eprint {https://arxiv.org/abs/hep-th/9704194}
  {arXiv:hep-th/9704194} \BibitemShut {NoStop}%
\bibitem [{\citenamefont {Dunne}\ and\ \citenamefont
  {Schubert}(2002{\natexlab{b}})}]{Dunne:2002qg}%
  \BibitemOpen
  \bibfield  {author} {\bibinfo {author} {\bibfnamefont {G.~V.}\ \bibnamefont
  {Dunne}}\ and\ \bibinfo {author} {\bibfnamefont {C.}~\bibnamefont
  {Schubert}},\ }\bibfield  {title} {\bibinfo {title} {{Two loop selfdual
  Euler-Heisenberg Lagrangians. 2. Imaginary part and Borel analysis}},\ }\href
  {https://doi.org/10.1088/1126-6708/2002/06/042} {\bibfield  {journal}
  {\bibinfo  {journal} {JHEP}\ }\textbf {\bibinfo {volume} {06}},\ \bibinfo
  {pages} {042}},\ \Eprint {https://arxiv.org/abs/hep-th/0205005}
  {arXiv:hep-th/0205005} \BibitemShut {NoStop}%
\bibitem [{\citenamefont {Alexandrou}\ \emph {et~al.}(1999)\citenamefont
  {Alexandrou}, \citenamefont {Rosenfelder},\ and\ \citenamefont
  {Schreiber}}]{Alexandrou:1998ia}%
  \BibitemOpen
  \bibfield  {author} {\bibinfo {author} {\bibfnamefont {C.}~\bibnamefont
  {Alexandrou}}, \bibinfo {author} {\bibfnamefont {R.}~\bibnamefont
  {Rosenfelder}},\ and\ \bibinfo {author} {\bibfnamefont {A.~W.}\ \bibnamefont
  {Schreiber}},\ }\bibfield  {title} {\bibinfo {title} {{Worldline path
  integral for the massive Dirac propagator: A four-dimensional approach}},\
  }\href {https://doi.org/10.1103/PhysRevA.59.1762} {\bibfield  {journal}
  {\bibinfo  {journal} {Phys. Rev. A}\ }\textbf {\bibinfo {volume} {59}},\
  \bibinfo {pages} {1762} (\bibinfo {year} {1999})},\ \Eprint
  {https://arxiv.org/abs/hep-th/9809101} {arXiv:hep-th/9809101} \BibitemShut
  {NoStop}%
\bibitem [{\citenamefont {Ahmad}\ \emph {et~al.}(2017)\citenamefont {Ahmad},
  \citenamefont {Ahmadiniaz}, \citenamefont {Corradini}, \citenamefont {Kim},\
  and\ \citenamefont {Schubert}}]{Ahmad_2017}%
  \BibitemOpen
  \bibfield  {author} {\bibinfo {author} {\bibfnamefont {A.}~\bibnamefont
  {Ahmad}}, \bibinfo {author} {\bibfnamefont {N.}~\bibnamefont {Ahmadiniaz}},
  \bibinfo {author} {\bibfnamefont {O.}~\bibnamefont {Corradini}}, \bibinfo
  {author} {\bibfnamefont {S.~P.}\ \bibnamefont {Kim}},\ and\ \bibinfo {author}
  {\bibfnamefont {C.}~\bibnamefont {Schubert}},\ }\bibfield  {title} {\bibinfo
  {title} {{Master formulas for the dressed scalar propagator in a constant
  field}},\ }\href {https://doi.org/10.1016/j.nuclphysb.2017.03.007} {\bibfield
   {journal} {\bibinfo  {journal} {Nucl. Phys. B}\ }\textbf {\bibinfo {volume}
  {919}},\ \bibinfo {pages} {9} (\bibinfo {year} {2017})},\ \Eprint
  {https://arxiv.org/abs/1612.02944} {arXiv:1612.02944 [hep-ph]} \BibitemShut
  {NoStop}%
\bibitem [{\citenamefont {Ahmadiniaz}\ \emph {et~al.}()\citenamefont
  {Ahmadiniaz}, \citenamefont {Banda~Guzm\'an}, \citenamefont {Bastianelli},
  \citenamefont {Corradini}, \citenamefont {Edwards},\ and\ \citenamefont
  {Schubert}}]{fppaper3}%
  \BibitemOpen
  \bibfield  {author} {\bibinfo {author} {\bibfnamefont {N.}~\bibnamefont
  {Ahmadiniaz}}, \bibinfo {author} {\bibfnamefont {V.}~\bibnamefont
  {Banda~Guzm\'an}}, \bibinfo {author} {\bibfnamefont {F.}~\bibnamefont
  {Bastianelli}}, \bibinfo {author} {\bibfnamefont {O.}~\bibnamefont
  {Corradini}}, \bibinfo {author} {\bibfnamefont {J.}~\bibnamefont {Edwards}},\
  and\ \bibinfo {author} {\bibfnamefont {C.}~\bibnamefont {Schubert}},\
  }\bibfield  {title} {\bibinfo {title} {{Worldline master formulas for the
  dressed electron propagator. Part 3. Constant external fields}},\ }\href@noop
  {} {\ }\bibinfo {note} {In preparation}\BibitemShut {NoStop}%
\bibitem [{\citenamefont {Edwards}\ and\ \citenamefont
  {Schubert}(2017)}]{Edwards:2017bte}%
  \BibitemOpen
  \bibfield  {author} {\bibinfo {author} {\bibfnamefont {J.~P.}\ \bibnamefont
  {Edwards}}\ and\ \bibinfo {author} {\bibfnamefont {C.}~\bibnamefont
  {Schubert}},\ }\bibfield  {title} {\bibinfo {title} {{One-particle reducible
  contribution to the one-loop scalar propagator in a constant field}},\ }\href
  {https://doi.org/10.1016/j.nuclphysb.2017.08.002} {\bibfield  {journal}
  {\bibinfo  {journal} {Nucl. Phys. B}\ }\textbf {\bibinfo {volume} {923}},\
  \bibinfo {pages} {339} (\bibinfo {year} {2017})},\ \Eprint
  {https://arxiv.org/abs/1704.00482} {arXiv:1704.00482 [hep-th]} \BibitemShut
  {NoStop}%
\bibitem [{\citenamefont {Ahmadiniaz}\ \emph
  {et~al.}(2017{\natexlab{a}})\citenamefont {Ahmadiniaz}, \citenamefont
  {Bastianelli}, \citenamefont {Corradini}, \citenamefont {Edwards},\ and\
  \citenamefont {Schubert}}]{Ahmadiniaz:2017rrk}%
  \BibitemOpen
  \bibfield  {author} {\bibinfo {author} {\bibfnamefont {N.}~\bibnamefont
  {Ahmadiniaz}}, \bibinfo {author} {\bibfnamefont {F.}~\bibnamefont
  {Bastianelli}}, \bibinfo {author} {\bibfnamefont {O.}~\bibnamefont
  {Corradini}}, \bibinfo {author} {\bibfnamefont {J.~P.}\ \bibnamefont
  {Edwards}},\ and\ \bibinfo {author} {\bibfnamefont {C.}~\bibnamefont
  {Schubert}},\ }\bibfield  {title} {\bibinfo {title} {{One-particle reducible
  contribution to the one-loop spinor propagator in a constant field}},\ }\href
  {https://doi.org/10.1016/j.nuclphysb.2017.09.012} {\bibfield  {journal}
  {\bibinfo  {journal} {Nucl. Phys. B}\ }\textbf {\bibinfo {volume} {924}},\
  \bibinfo {pages} {377} (\bibinfo {year} {2017}{\natexlab{a}})},\ \Eprint
  {https://arxiv.org/abs/1704.05040} {arXiv:1704.05040 [hep-th]} \BibitemShut
  {NoStop}%
\bibitem [{\citenamefont {Ahmadiniaz}\ \emph {et~al.}(2019)\citenamefont
  {Ahmadiniaz}, \citenamefont {Edwards},\ and\ \citenamefont
  {Ilderton}}]{Ahmadiniaz:2019nhk}%
  \BibitemOpen
  \bibfield  {author} {\bibinfo {author} {\bibfnamefont {N.}~\bibnamefont
  {Ahmadiniaz}}, \bibinfo {author} {\bibfnamefont {J.~P.}\ \bibnamefont
  {Edwards}},\ and\ \bibinfo {author} {\bibfnamefont {A.}~\bibnamefont
  {Ilderton}},\ }\bibfield  {title} {\bibinfo {title} {{Reducible contributions
  to quantum electrodynamics in external fields}},\ }\href
  {https://doi.org/10.1007/JHEP05(2019)038} {\bibfield  {journal} {\bibinfo
  {journal} {JHEP}\ }\textbf {\bibinfo {volume} {05}},\ \bibinfo {pages}
  {038}},\ \Eprint {https://arxiv.org/abs/1901.09416} {arXiv:1901.09416
  [hep-th]} \BibitemShut {NoStop}%
\bibitem [{\citenamefont {Dittrich}\ and\ \citenamefont
  {Shaisultanov}(2000)}]{Dittrich:2000wz}%
  \BibitemOpen
  \bibfield  {author} {\bibinfo {author} {\bibfnamefont {W.}~\bibnamefont
  {Dittrich}}\ and\ \bibinfo {author} {\bibfnamefont {R.}~\bibnamefont
  {Shaisultanov}},\ }\bibfield  {title} {\bibinfo {title} {{Vacuum polarization
  in QED with worldline methods}},\ }\href
  {https://doi.org/10.1103/PhysRevD.62.045024} {\bibfield  {journal} {\bibinfo
  {journal} {Phys. Rev. D}\ }\textbf {\bibinfo {volume} {62}},\ \bibinfo
  {pages} {045024} (\bibinfo {year} {2000})},\ \Eprint
  {https://arxiv.org/abs/hep-th/0001171} {arXiv:hep-th/0001171} \BibitemShut
  {NoStop}%
\bibitem [{\citenamefont {Schubert}(2000)}]{Schubert:2000yt}%
  \BibitemOpen
  \bibfield  {author} {\bibinfo {author} {\bibfnamefont {C.}~\bibnamefont
  {Schubert}},\ }\bibfield  {title} {\bibinfo {title} {{Vacuum polarization
  tensors in constant electromagnetic fields. Part 1.}},\ }\href
  {https://doi.org/10.1016/S0550-3213(00)00423-5} {\bibfield  {journal}
  {\bibinfo  {journal} {Nucl. Phys. B}\ }\textbf {\bibinfo {volume} {585}},\
  \bibinfo {pages} {407} (\bibinfo {year} {2000})},\ \Eprint
  {https://arxiv.org/abs/hep-ph/0001288} {arXiv:hep-ph/0001288} \BibitemShut
  {NoStop}%
\bibitem [{\citenamefont {McKeon}\ and\ \citenamefont
  {Sherry}(1994)}]{McKeon:1994hd}%
  \BibitemOpen
  \bibfield  {author} {\bibinfo {author} {\bibfnamefont {D.~G.~C.}\
  \bibnamefont {McKeon}}\ and\ \bibinfo {author} {\bibfnamefont {T.~N.}\
  \bibnamefont {Sherry}},\ }\bibfield  {title} {\bibinfo {title} {{Radiative
  effects in a constant magnetic field using the quantum mechanical path
  integral}},\ }\href {https://doi.org/10.1142/S0217732394002021} {\bibfield
  {journal} {\bibinfo  {journal} {Mod. Phys. Lett. A}\ }\textbf {\bibinfo
  {volume} {9}},\ \bibinfo {pages} {2167} (\bibinfo {year} {1994})}\BibitemShut
  {NoStop}%
\bibitem [{\citenamefont {Franchino-Vi\~nas}\ \emph {et~al.}(2024)\citenamefont
  {Franchino-Vi\~nas}, \citenamefont {Garc\'\i{}a-P\'erez}, \citenamefont
  {Mazzitelli}, \citenamefont {Vitagliano},\ and\ \citenamefont
  {Haimovichi}}]{Franchino-Vinas:2023wea}%
  \BibitemOpen
  \bibfield  {author} {\bibinfo {author} {\bibfnamefont {S.~A.}\ \bibnamefont
  {Franchino-Vi\~nas}}, \bibinfo {author} {\bibfnamefont {C.}~\bibnamefont
  {Garc\'\i{}a-P\'erez}}, \bibinfo {author} {\bibfnamefont {F.~D.}\
  \bibnamefont {Mazzitelli}}, \bibinfo {author} {\bibfnamefont
  {V.}~\bibnamefont {Vitagliano}},\ and\ \bibinfo {author} {\bibfnamefont
  {U.~W.}\ \bibnamefont {Haimovichi}},\ }\bibfield  {title} {\bibinfo {title}
  {{Resummed heat kernel and effective action for Yukawa and QED}},\ }\href
  {https://doi.org/10.1016/j.physletb.2024.138684} {\bibfield  {journal}
  {\bibinfo  {journal} {Phys. Lett. B}\ }\textbf {\bibinfo {volume} {854}},\
  \bibinfo {pages} {138684} (\bibinfo {year} {2024})},\ \Eprint
  {https://arxiv.org/abs/2312.16303} {arXiv:2312.16303 [hep-th]} \BibitemShut
  {NoStop}%
\bibitem [{\citenamefont {Fecit}\ \emph {et~al.}(2025)\citenamefont {Fecit},
  \citenamefont {Franchino-Vi\~nas},\ and\ \citenamefont
  {Mazzitelli}}]{Fecit:2025kqb}%
  \BibitemOpen
  \bibfield  {author} {\bibinfo {author} {\bibfnamefont {F.}~\bibnamefont
  {Fecit}}, \bibinfo {author} {\bibfnamefont {S.}~\bibnamefont
  {Franchino-Vi\~nas}},\ and\ \bibinfo {author} {\bibfnamefont {F.~D.}\
  \bibnamefont {Mazzitelli}},\ }\bibfield  {title} {\bibinfo {title} {{Resummed
  effective actions and heat kernels: the Worldline approach and Yukawa
  assisted pair creation}},\ }\href@noop {} {\  (\bibinfo {year} {2025})},\
  \Eprint {https://arxiv.org/abs/2501.17094} {arXiv:2501.17094 [hep-th]}
  \BibitemShut {NoStop}%
\bibitem [{\citenamefont {Edwards}\ and\ \citenamefont
  {Schubert}(2021)}]{Edwards:2021vhg}%
  \BibitemOpen
  \bibfield  {author} {\bibinfo {author} {\bibfnamefont {J.~P.}\ \bibnamefont
  {Edwards}}\ and\ \bibinfo {author} {\bibfnamefont {C.}~\bibnamefont
  {Schubert}},\ }\bibfield  {title} {\bibinfo {title} {{N-photon amplitudes in
  a plane-wave background}},\ }\href
  {https://doi.org/10.1016/j.physletb.2021.136696} {\bibfield  {journal}
  {\bibinfo  {journal} {Phys. Lett. B}\ }\textbf {\bibinfo {volume} {822}},\
  \bibinfo {pages} {136696} (\bibinfo {year} {2021})},\ \Eprint
  {https://arxiv.org/abs/2105.08173} {arXiv:2105.08173 [hep-th]} \BibitemShut
  {NoStop}%
\bibitem [{\citenamefont {Schubert}\ and\ \citenamefont
  {Shaisultanov}(2023)}]{Schubert:2023gsl}%
  \BibitemOpen
  \bibfield  {author} {\bibinfo {author} {\bibfnamefont {C.}~\bibnamefont
  {Schubert}}\ and\ \bibinfo {author} {\bibfnamefont {R.}~\bibnamefont
  {Shaisultanov}},\ }\bibfield  {title} {\bibinfo {title} {{Master formulas for
  photon amplitudes in a combined constant and plane-wave background field}},\
  }\href {https://doi.org/10.1016/j.physletb.2023.137969} {\bibfield  {journal}
  {\bibinfo  {journal} {Phys. Lett. B}\ }\textbf {\bibinfo {volume} {843}},\
  \bibinfo {pages} {137969} (\bibinfo {year} {2023})},\ \Eprint
  {https://arxiv.org/abs/2303.08907} {arXiv:2303.08907 [hep-th]} \BibitemShut
  {NoStop}%
\bibitem [{\citenamefont {Copinger}\ \emph
  {et~al.}(2024{\natexlab{a}})\citenamefont {Copinger}, \citenamefont
  {Edwards}, \citenamefont {Ilderton},\ and\ \citenamefont
  {Rajeev}}]{Copinger:2024twl}%
  \BibitemOpen
  \bibfield  {author} {\bibinfo {author} {\bibfnamefont {P.}~\bibnamefont
  {Copinger}}, \bibinfo {author} {\bibfnamefont {J.~P.}\ \bibnamefont
  {Edwards}}, \bibinfo {author} {\bibfnamefont {A.}~\bibnamefont {Ilderton}},\
  and\ \bibinfo {author} {\bibfnamefont {K.}~\bibnamefont {Rajeev}},\
  }\bibfield  {title} {\bibinfo {title} {{All-multiplicity amplitudes in
  impulsive PP-waves from the worldline formalism}},\ }\href
  {https://doi.org/10.1007/JHEP09(2024)148} {\bibfield  {journal} {\bibinfo
  {journal} {JHEP}\ }\textbf {\bibinfo {volume} {09}},\ \bibinfo {pages}
  {148}},\ \Eprint {https://arxiv.org/abs/2405.07385} {arXiv:2405.07385
  [hep-th]} \BibitemShut {NoStop}%
\bibitem [{\citenamefont {Tarasov}\ and\ \citenamefont
  {Venugopalan}(2019)}]{Tarasov:2019rfp}%
  \BibitemOpen
  \bibfield  {author} {\bibinfo {author} {\bibfnamefont {A.}~\bibnamefont
  {Tarasov}}\ and\ \bibinfo {author} {\bibfnamefont {R.}~\bibnamefont
  {Venugopalan}},\ }\bibfield  {title} {\bibinfo {title} {{Structure functions
  at small x from worldlines: Unpolarized distributions}},\ }\href
  {https://doi.org/10.1103/PhysRevD.100.054007} {\bibfield  {journal} {\bibinfo
   {journal} {Phys. Rev. D}\ }\textbf {\bibinfo {volume} {100}},\ \bibinfo
  {pages} {054007} (\bibinfo {year} {2019})},\ \Eprint
  {https://arxiv.org/abs/1903.11624} {arXiv:1903.11624 [hep-ph]} \BibitemShut
  {NoStop}%
\bibitem [{\citenamefont {Corradini}\ \emph {et~al.}(2015)\citenamefont
  {Corradini}, \citenamefont {Schubert}, \citenamefont {Edwards},\ and\
  \citenamefont {Ahmadiniaz}}]{Corradini:2015tik}%
  \BibitemOpen
  \bibfield  {author} {\bibinfo {author} {\bibfnamefont {O.}~\bibnamefont
  {Corradini}}, \bibinfo {author} {\bibfnamefont {C.}~\bibnamefont {Schubert}},
  \bibinfo {author} {\bibfnamefont {J.~P.}\ \bibnamefont {Edwards}},\ and\
  \bibinfo {author} {\bibfnamefont {N.}~\bibnamefont {Ahmadiniaz}},\ }\bibfield
   {title} {\bibinfo {title} {{Spinning Particles in Quantum Mechanics and
  Quantum Field Theory}}\ }(\bibinfo {year} {2015})\ \Eprint
  {https://arxiv.org/abs/1512.08694} {arXiv:1512.08694 [hep-th]} \BibitemShut
  {NoStop}%
\bibitem [{\citenamefont {Copinger}\ and\ \citenamefont
  {Hidaka}(2023)}]{Copinger:2022gfz}%
  \BibitemOpen
  \bibfield  {author} {\bibinfo {author} {\bibfnamefont {P.}~\bibnamefont
  {Copinger}}\ and\ \bibinfo {author} {\bibfnamefont {Y.}~\bibnamefont
  {Hidaka}},\ }\bibfield  {title} {\bibinfo {title} {{Angular momentum
  inheritance from the Schwinger effect in (chromo)electromagnetic fields}},\
  }\href {https://doi.org/10.1093/ptep/ptad015} {\bibfield  {journal} {\bibinfo
   {journal} {PTEP}\ }\textbf {\bibinfo {volume} {2023}},\ \bibinfo {pages}
  {023B08} (\bibinfo {year} {2023})},\ \Eprint
  {https://arxiv.org/abs/2203.10917} {arXiv:2203.10917 [hep-ph]} \BibitemShut
  {NoStop}%
\bibitem [{\citenamefont {Feynman}(1951)}]{Feyn2}%
  \BibitemOpen
  \bibfield  {author} {\bibinfo {author} {\bibfnamefont {R.~P.}\ \bibnamefont
  {Feynman}},\ }\bibfield  {title} {\bibinfo {title} {{An Operator calculus
  having applications in quantum electrodynamics}},\ }\href
  {https://doi.org/10.1103/PhysRev.84.108} {\bibfield  {journal} {\bibinfo
  {journal} {Phys. Rev.}\ }\textbf {\bibinfo {volume} {84}},\ \bibinfo {pages}
  {108} (\bibinfo {year} {1951})}\BibitemShut {NoStop}%
\bibitem [{\citenamefont {Strassler}(1992)}]{Strass1}%
  \BibitemOpen
  \bibfield  {author} {\bibinfo {author} {\bibfnamefont {M.~J.}\ \bibnamefont
  {Strassler}},\ }\bibfield  {title} {\bibinfo {title} {{Field theory without
  Feynman diagrams: One loop effective actions}},\ }\href
  {https://doi.org/10.1016/0550-3213(92)90098-V} {\bibfield  {journal}
  {\bibinfo  {journal} {Nucl. Phys. B}\ }\textbf {\bibinfo {volume} {385}},\
  \bibinfo {pages} {145} (\bibinfo {year} {1992})},\ \Eprint
  {https://arxiv.org/abs/hep-ph/9205205} {arXiv:hep-ph/9205205} \BibitemShut
  {NoStop}%
\bibitem [{\citenamefont {Copinger}\ \emph
  {et~al.}(2024{\natexlab{b}})\citenamefont {Copinger}, \citenamefont
  {Edwards}, \citenamefont {Ilderton},\ and\ \citenamefont
  {Rajeev}}]{Copinger:2023ctz}%
  \BibitemOpen
  \bibfield  {author} {\bibinfo {author} {\bibfnamefont {P.}~\bibnamefont
  {Copinger}}, \bibinfo {author} {\bibfnamefont {J.~P.}\ \bibnamefont
  {Edwards}}, \bibinfo {author} {\bibfnamefont {A.}~\bibnamefont {Ilderton}},\
  and\ \bibinfo {author} {\bibfnamefont {K.}~\bibnamefont {Rajeev}},\
  }\bibfield  {title} {\bibinfo {title} {{Master formulas for N-photon tree
  level amplitudes in plane wave backgrounds}},\ }\href
  {https://doi.org/10.1103/PhysRevD.109.065003} {\bibfield  {journal} {\bibinfo
   {journal} {Phys. Rev. D}\ }\textbf {\bibinfo {volume} {109}},\ \bibinfo
  {pages} {065003} (\bibinfo {year} {2024}{\natexlab{b}})},\ \Eprint
  {https://arxiv.org/abs/2311.14638} {arXiv:2311.14638 [hep-th]} \BibitemShut
  {NoStop}%
\bibitem [{\citenamefont {Ahmadiniaz}\ \emph
  {et~al.}(2017{\natexlab{b}})\citenamefont {Ahmadiniaz}, \citenamefont
  {Bastianelli}, \citenamefont {Corradini}, \citenamefont {Edwards},\ and\
  \citenamefont {Schubert}}]{ReducibleSpinor}%
  \BibitemOpen
  \bibfield  {author} {\bibinfo {author} {\bibfnamefont {N.}~\bibnamefont
  {Ahmadiniaz}}, \bibinfo {author} {\bibfnamefont {F.}~\bibnamefont
  {Bastianelli}}, \bibinfo {author} {\bibfnamefont {O.}~\bibnamefont
  {Corradini}}, \bibinfo {author} {\bibfnamefont {J.~P.}\ \bibnamefont
  {Edwards}},\ and\ \bibinfo {author} {\bibfnamefont {C.}~\bibnamefont
  {Schubert}},\ }\bibfield  {title} {\bibinfo {title} {{One-particle reducible
  contribution to the one-loop spinor propagator in a constant field}},\ }\href
  {https://doi.org/10.1016/j.nuclphysb.2017.09.012} {\bibfield  {journal}
  {\bibinfo  {journal} {Nucl. Phys. B}\ }\textbf {\bibinfo {volume} {924}},\
  \bibinfo {pages} {377} (\bibinfo {year} {2017}{\natexlab{b}})},\ \Eprint
  {https://arxiv.org/abs/1704.05040} {arXiv:1704.05040 [hep-th]} \BibitemShut
  {NoStop}%
\bibitem [{\citenamefont {Ahmadiniaz}\ \emph
  {et~al.}(2022{\natexlab{b}})\citenamefont {Ahmadiniaz}, \citenamefont
  {Guzman}, \citenamefont {Bastianelli}, \citenamefont {Corradini},
  \citenamefont {Edwards},\ and\ \citenamefont {Schubert}}]{fppaper1}%
  \BibitemOpen
  \bibfield  {author} {\bibinfo {author} {\bibfnamefont {N.}~\bibnamefont
  {Ahmadiniaz}}, \bibinfo {author} {\bibfnamefont {V.~M.~B.}\ \bibnamefont
  {Guzman}}, \bibinfo {author} {\bibfnamefont {F.}~\bibnamefont {Bastianelli}},
  \bibinfo {author} {\bibfnamefont {O.}~\bibnamefont {Corradini}}, \bibinfo
  {author} {\bibfnamefont {J.~P.}\ \bibnamefont {Edwards}},\ and\ \bibinfo
  {author} {\bibfnamefont {C.}~\bibnamefont {Schubert}},\ }\bibfield  {title}
  {\bibinfo {title} {{Worldline master formulas for the dressed electron
  propagator. Part 2. On-shell amplitudes}},\ }\href
  {https://doi.org/10.1007/JHEP01(2022)050} {\bibfield  {journal} {\bibinfo
  {journal} {JHEP}\ }\textbf {\bibinfo {volume} {01}},\ \bibinfo {pages}
  {050}},\ \Eprint {https://arxiv.org/abs/2107.00199} {arXiv:2107.00199
  [hep-th]} \BibitemShut {NoStop}%
\bibitem [{\citenamefont {Ahmadiniaz}\ \emph {et~al.}(2020)\citenamefont
  {Ahmadiniaz}, \citenamefont {Banda~Guzm\'an}, \citenamefont {Bastianelli},
  \citenamefont {Corradini}, \citenamefont {Edwards},\ and\ \citenamefont
  {Schubert}}]{fppaper2}%
  \BibitemOpen
  \bibfield  {author} {\bibinfo {author} {\bibfnamefont {N.}~\bibnamefont
  {Ahmadiniaz}}, \bibinfo {author} {\bibfnamefont {V.~M.}\ \bibnamefont
  {Banda~Guzm\'an}}, \bibinfo {author} {\bibfnamefont {F.}~\bibnamefont
  {Bastianelli}}, \bibinfo {author} {\bibfnamefont {O.}~\bibnamefont
  {Corradini}}, \bibinfo {author} {\bibfnamefont {J.~P.}\ \bibnamefont
  {Edwards}},\ and\ \bibinfo {author} {\bibfnamefont {C.}~\bibnamefont
  {Schubert}},\ }\bibfield  {title} {\bibinfo {title} {{Worldline master
  formulas for the dressed electron propagator. Part I. Off-shell
  amplitudes}},\ }\href {https://doi.org/10.1007/JHEP08(2020)018} {\bibfield
  {journal} {\bibinfo  {journal} {JHEP}\ }\textbf {\bibinfo {volume}
  {08}}\bibfield  {number} {\bibinfo  {number} { (08)},\ \bibinfo {pages}
  {049}},\ }\Eprint {https://arxiv.org/abs/2004.01391} {arXiv:2004.01391
  [hep-th]} \BibitemShut {NoStop}%
\bibitem [{\citenamefont {Bonezzi}\ and\ \citenamefont
  {Kallimani}(2025)}]{Bonezzi:2025iza}%
  \BibitemOpen
  \bibfield  {author} {\bibinfo {author} {\bibfnamefont {R.}~\bibnamefont
  {Bonezzi}}\ and\ \bibinfo {author} {\bibfnamefont {M.~F.}\ \bibnamefont
  {Kallimani}},\ }\bibfield  {title} {\bibinfo {title} {{Worldline geometries
  for scattering amplitudes}},\ }\href
  {https://doi.org/10.1007/JHEP06(2025)167} {\bibfield  {journal} {\bibinfo
  {journal} {JHEP}\ }\textbf {\bibinfo {volume} {06}},\ \bibinfo {pages}
  {167}},\ \Eprint {https://arxiv.org/abs/2502.18030} {arXiv:2502.18030
  [hep-th]} \BibitemShut {NoStop}%
\bibitem [{\citenamefont {Bastianelli}\ \emph {et~al.}(2024)\citenamefont
  {Bastianelli}, \citenamefont {Corradini}, \citenamefont {Edwards},
  \citenamefont {McKeon},\ and\ \citenamefont {Schubert}}]{Yukawa}%
  \BibitemOpen
  \bibfield  {author} {\bibinfo {author} {\bibfnamefont {F.}~\bibnamefont
  {Bastianelli}}, \bibinfo {author} {\bibfnamefont {O.}~\bibnamefont
  {Corradini}}, \bibinfo {author} {\bibfnamefont {J.~P.}\ \bibnamefont
  {Edwards}}, \bibinfo {author} {\bibfnamefont {D.~G.~C.}\ \bibnamefont
  {McKeon}},\ and\ \bibinfo {author} {\bibfnamefont {C.}~\bibnamefont
  {Schubert}},\ }\bibfield  {title} {\bibinfo {title} {{Unified worldline
  treatment of Yukawa and axial couplings}},\ }\href
  {https://doi.org/10.1007/JHEP11(2024)152} {\bibfield  {journal} {\bibinfo
  {journal} {JHEP}\ }\textbf {\bibinfo {volume} {11}},\ \bibinfo {pages}
  {152}},\ \Eprint {https://arxiv.org/abs/2406.19988} {arXiv:2406.19988
  [hep-th]} \BibitemShut {NoStop}%
\bibitem [{\citenamefont {Ahmadiniaz}\ \emph
  {et~al.}(2022{\natexlab{c}})\citenamefont {Ahmadiniaz}, \citenamefont
  {Edwards}, \citenamefont {Lopez-Arcos}, \citenamefont {Lopez-Lopez},
  \citenamefont {Mata}, \citenamefont {Nicasio},\ and\ \citenamefont
  {Schubert}}]{Ahmadiniaz:2022yam}%
  \BibitemOpen
  \bibfield  {author} {\bibinfo {author} {\bibfnamefont {N.}~\bibnamefont
  {Ahmadiniaz}}, \bibinfo {author} {\bibfnamefont {J.~P.}\ \bibnamefont
  {Edwards}}, \bibinfo {author} {\bibfnamefont {C.}~\bibnamefont
  {Lopez-Arcos}}, \bibinfo {author} {\bibfnamefont {M.~A.}\ \bibnamefont
  {Lopez-Lopez}}, \bibinfo {author} {\bibfnamefont {C.~M.}\ \bibnamefont
  {Mata}}, \bibinfo {author} {\bibfnamefont {J.}~\bibnamefont {Nicasio}},\ and\
  \bibinfo {author} {\bibfnamefont {C.}~\bibnamefont {Schubert}},\ }\bibfield
  {title} {\bibinfo {title} {{Summing Feynman diagrams in the worldline
  formalism}},\ }\href {https://doi.org/10.22323/1.416.0052} {\bibfield
  {journal} {\bibinfo  {journal} {PoS}\ }\textbf {\bibinfo {volume} {LL2022}},\
  \bibinfo {pages} {052} (\bibinfo {year} {2022}{\natexlab{c}})},\ \Eprint
  {https://arxiv.org/abs/2208.06585} {arXiv:2208.06585 [hep-th]} \BibitemShut
  {NoStop}%
\bibitem [{\citenamefont {Ritus}(1985)}]{RN1}%
  \BibitemOpen
  \bibfield  {author} {\bibinfo {author} {\bibfnamefont {V.~I.}\ \bibnamefont
  {Ritus}},\ }\bibfield  {title} {\bibinfo {title} {{Quantum effects of the
  interaction of elementary particles with an intense electromagnetic field}},\
  }\href {https://doi.org/10.1007/BF01120220} {\bibfield  {journal} {\bibinfo
  {journal} {J. Russ. Laser Res.}\ }\textbf {\bibinfo {volume} {6}},\ \bibinfo
  {pages} {497} (\bibinfo {year} {1985})}\BibitemShut {NoStop}%
\bibitem [{\citenamefont {Fedotov}(2017)}]{RN2}%
  \BibitemOpen
  \bibfield  {author} {\bibinfo {author} {\bibfnamefont {A.}~\bibnamefont
  {Fedotov}},\ }\bibfield  {title} {\bibinfo {title} {Conjecture of
  perturbative qed breakdown at $\alpha \chi^{2/3} \geq 1$},\ }\href
  {https://doi.org/10.1088/1742-6596/826/1/012027} {\bibfield  {journal}
  {\bibinfo  {journal} {Journal of Physics: Conference Series}\ }\textbf
  {\bibinfo {volume} {826}},\ \bibinfo {pages} {012027} (\bibinfo {year}
  {2017})}\BibitemShut {NoStop}%
\bibitem [{\citenamefont {Fliegner}\ \emph {et~al.}(1998)\citenamefont
  {Fliegner}, \citenamefont {Haberl}, \citenamefont {Schmidt},\ and\
  \citenamefont {Schubert}}]{Fliegner_1998}%
  \BibitemOpen
  \bibfield  {author} {\bibinfo {author} {\bibfnamefont {D.}~\bibnamefont
  {Fliegner}}, \bibinfo {author} {\bibfnamefont {P.}~\bibnamefont {Haberl}},
  \bibinfo {author} {\bibfnamefont {M.}~\bibnamefont {Schmidt}},\ and\ \bibinfo
  {author} {\bibfnamefont {C.}~\bibnamefont {Schubert}},\ }\bibfield  {title}
  {\bibinfo {title} {The higher derivative expansion of the effective action by
  the string inspired method, ii},\ }\href
  {https://doi.org/10.1006/aphy.1997.5778} {\bibfield  {journal} {\bibinfo
  {journal} {Annals of Physics}\ }\textbf {\bibinfo {volume} {264}},\ \bibinfo
  {pages} {51–74} (\bibinfo {year} {1998})},\ \Eprint
  {https://arxiv.org/abs/hep-th/9707189} {arXiv:hep-th/9707189} \BibitemShut
  {NoStop}%
\bibitem [{\citenamefont {Auer}\ \emph {et~al.}(2004)\citenamefont {Auer},
  \citenamefont {Schmidt},\ and\ \citenamefont {Zahlten}}]{Auer:2003rq}%
  \BibitemOpen
  \bibfield  {author} {\bibinfo {author} {\bibfnamefont {T.}~\bibnamefont
  {Auer}}, \bibinfo {author} {\bibfnamefont {M.~G.}\ \bibnamefont {Schmidt}},\
  and\ \bibinfo {author} {\bibfnamefont {C.}~\bibnamefont {Zahlten}},\
  }\bibfield  {title} {\bibinfo {title} {{Resummed effective action in the
  worldline formalism}},\ }\href
  {https://doi.org/10.1016/j.nuclphysb.2003.10.040} {\bibfield  {journal}
  {\bibinfo  {journal} {Nucl. Phys. B}\ }\textbf {\bibinfo {volume} {677}},\
  \bibinfo {pages} {430} (\bibinfo {year} {2004})},\ \Eprint
  {https://arxiv.org/abs/hep-th/0306243} {arXiv:hep-th/0306243} \BibitemShut
  {NoStop}%
\bibitem [{\citenamefont {Wintucky}(1971)}]{wintucky1971formulas}%
  \BibitemOpen
  \bibfield  {author} {\bibinfo {author} {\bibfnamefont {E.~G.}\ \bibnamefont
  {Wintucky}},\ }\href@noop {} {\emph {\bibinfo {title} {Formulas for nth order
  derivatives of hyperbolic and trigonometric functions}}}\ (\bibinfo
  {publisher} {National Aeronautics and Space Administration},\ \bibinfo {year}
  {1971})\BibitemShut {NoStop}%
\bibitem [{\citenamefont {Cvijović}(2009)}]{CVIJOVIC20093002}%
  \BibitemOpen
  \bibfield  {author} {\bibinfo {author} {\bibfnamefont {D.}~\bibnamefont
  {Cvijović}},\ }\bibfield  {title} {\bibinfo {title} {Derivative polynomials
  and closed-form higher derivative formulae},\ }\href
  {https://doi.org/https://doi.org/10.1016/j.amc.2009.09.047} {\bibfield
  {journal} {\bibinfo  {journal} {Applied Mathematics and Computation}\
  }\textbf {\bibinfo {volume} {215}},\ \bibinfo {pages} {3002} (\bibinfo {year}
  {2009})}\BibitemShut {NoStop}%
\bibitem [{\citenamefont {Boyadzhiev}(2007)}]{Derivs}%
  \BibitemOpen
  \bibfield  {author} {\bibinfo {author} {\bibfnamefont {K.~N.}\ \bibnamefont
  {Boyadzhiev}},\ }\bibfield  {title} {\bibinfo {title} {Derivative polynomials
  for tanh, tan, sech and sec in explicit form},\ }\href@noop {} {\bibfield
  {journal} {\bibinfo  {journal} {Fibonacci Quart.}\ }\textbf {\bibinfo
  {volume} {45}},\ \bibinfo {pages} {291–303} (\bibinfo {year}
  {2007})}\BibitemShut {NoStop}%
\end{thebibliography}%

\end{document}